\documentclass[a4paper,fleqn,usenatbib,useAMS]{mnras}

\usepackage{graphicx}	
\usepackage{amsmath}	

\newcommand{\kmsMpc}{\ifmmode  \,\rm km\,s^{-1}\,Mpc^{-1} \else $\,\rm km\,s^{-1}\,Mpc^{-1}  $ \fi }
\newcommand{\Mpc}{\ifmmode  {\rm~Mpc}  \else ${\rm~Mpc}$\fi}  
\newcommand{\kpc}{\ifmmode  {\rm~kpc}  \else ${\rm~kpc}$\fi}  
\newcommand{\Gyr}{\ifmmode  {\rm~Gyr}  \else ${\rm~Gyr}$\fi}
\newcommand{\Msun}{\ifmmode {\rm M}_{\odot} \else ${\rm M}_{\odot}$ \fi} 
\newcommand{\Msunpyr}{\ifmmode M_{\odot}{\rm~yr}^{-1} \else $M_{\odot}{\rm~yr}^{-1}$ \fi} 
\newcommand{\LCDM}{\ifmmode \Lambda{\rm CDM} \else $\Lambda{\rm CDM}$ \fi}
\newcommand{\Omegam}{\ifmmode \Omega_{\rm m} \else $\Omega_{\rm m}$\fi} 
\newcommand{\OmegaL}{\ifmmode \Omega_{\rm \Lambda} \else $\Omega_{\rm \Lambda}$\fi} 
\newcommand{\Omegab}{\ifmmode \Omega_{\rm b} \else $\Omega_{\rm b}$\fi} 
\newcommand{\csfrd}{\ifmmode \dot \rho^* \else $\dot \rho^*$ \fi} 
\newcommand{\fq}{\ifmmode {f_\mathrm{q}} \else ${f_\mathrm{q}}$ \fi}
\newcommand{\wpp}{\ifmmode {w_\mathrm{p}} \else ${w_\mathrm{p}}$ \fi}
\newcommand{\rpp}{\ifmmode {r_\mathrm{p}} \else ${r_\mathrm{p}}$ \fi}

\usepackage[T1]{fontenc}
\usepackage{ae,aecompl}
\usepackage{newtxtext,newtxmath}

\title[The empirical galaxy formation model {\sc Emerge}] 
{{\sc Emerge} -- An empirical model for the formation of galaxies since $z\sim10$}

\author[B. P. Moster et al.] {Benjamin P. Moster$^{1,2,3,}$\thanks{moster@usm.lmu.de}, Thorsten Naab$^2$, Simon D. M. White$^2$\\ 
  $^1$Universit\"ats-Sternwarte, Ludwig-Maximilians-Universit\"at M\"unchen, Scheinerstr. 1, 81679 M\"unchen, Germany\\
  $^2$Max-Planck Institut f\"ur Astrophysik, Karl-Schwarzschild Stra\ss e 1, 85748 Garching, Germany\\
  $^3$Kavli Institute for Cosmology and Institute of Astronomy, University of Cambridge, Madingley Rd, Cambridge CB3 0HA, UK\\
} 
 
\date{Last updated 2017 January 1; in original form 2017 January 1}

\pubyear{2017}

\begin{document}
\label{firstpage}
\pagerange{\pageref{firstpage}--\pageref{lastpage}}
\maketitle

\begin{abstract}
We present {\sc Emerge}, an \textbf{E}mpirical \textbf{M}od\textbf{E}l for the fo\textbf{R}mation of \textbf{G}alaxi\textbf{E}s, describing
the evolution of individual galaxies in large volumes from $z \sim 10$ to the present day.
We assign a star formation rate to each dark matter halo based on its growth rate, which specifies how much baryonic material becomes
available, and the instantaneous baryon conversion efficiency, which determines how efficiently this material is converted to stars, thereby
capturing the baryonic physics. 
Satellites are quenched following the delayed-then-rapid model, and they are tidally disrupted once their subhalo has lost a significant fraction of its mass.
The model is constrained with observed data extending out to high redshift.
The empirical relations are very flexible, and the model complexity is increased only if required by the data, assessed by several model selection statistics.
We find that for the same final halo mass galaxies can have very different star formation histories.
Nevertheless, the average star formation and accretion rates are in good agreement with models following an abundance matching strategy.
Galaxies that are quenched at $z=0$ typically have a higher peak star formation rate compared to their star-forming counterparts.
The accretion of stars can dominate the total mass of massive galaxies, but is insignificant for low-mass systems, independent of star-formation activity.
{\sc Emerge} predicts stellar-to-halo mass ratios for individual galaxies and introduces scatter self-consistently.
We find that at fixed halo mass, passive galaxies have a higher stellar mass on average.
The intra-cluster-mass in massive haloes can be up to 8 times larger than the mass of the central galaxy.
Clustering for star-forming and quenched galaxies is in good agreement with observational constraints, indicating a realistic assignment of galaxies to haloes.
\end{abstract}



\section{Introduction}
\label{sec:intro}

In the standard model of cosmology, only a small fraction of the present energy density of the Universe is in the
form of baryonic matter. The remaining dark components are the dynamically cold and collisionless dark matter
\citep{Zwicky:1933aa,Davis:1985aa}, and a near-uniform dark energy field which can be described by a cosmological
constant \citep*{Riess:1998aa,Perlmutter:1999aa,Perlmutter:1999ab}. Together they form the foundation of the \LCDM
theory in which structure formation proceeds through gravitationally driven hierarchical collapse and merging. In the
standard picture, galaxies form by the cooling and condensation of gas in the centres of virialised dark matter halos
\citep{White:1978aa,Fall:1980aa,Blumenthal:1984aa}, which results in a tight correlation between the properties of
haloes and those of the galaxies they host.

The formation and evolution of dark matter haloes has been studied extensively with large cosmological $N$-body simulations
(\citealt{Springel:2005ab,Boylan-Kolchin:2009aa}; \citealt*{Klypin:2011aa}; \citealt{,Angulo:2012aa,Klypin:2016aa}). As this process
only depends on gravity and the initial conditions have been measured very accurately \citep{Planck-Collaboration:2016aa},
these simulations have converged and make accurate and definite predictions for the properties of dark matter haloes at all
cosmic epochs \citep{Frenk:2012aa,Knebe:2013aa}. The halo mass function (HMF) found in the simulations is very steep and
the dark matter is distributed over many orders of magnitude. If galaxies were forming with the same efficiency in haloes of
different masses, we would expect the galaxy stellar mass function (SMF) to have the same shape as the HMF. However,
the observed local SMF has a very different shape, with a much shallower slope at the low-mass end and an exponential
cut-off at much smaller masses \citep{Li:2009aa,Bernardi:2013aa}. This tension indicates the complexity of the baryonic
physics regulating galaxy formation, such as gas cooling, star formation, and feedback processes. 

There are several pathways to learn about the formation and evolution of galaxies. The most popular method are {\it `ab initio'
models}, where an initial distribution of gas and dark matter is evolved according to a specified set of relevant physical processes,
including all the various baryonic physics that one thinks is important (see the reviews by \citealt{Somerville:2015aa} and
\citealt{Naab:2016aa} for more details). In {\it hydrodynamical simulations} the baryonic component
is discretised and evolved hydrodynamically \citep{Hirschmann:2014aa,Vogelsberger:2014aa,Dubois:2014aa,
Hopkins:2014aa,Khandai:2015aa,Schaye:2015aa}, while in {\it semi-analytic models} (SAMs) it is separated from the
dark-matter-dominated growth of structure by post-processing halo merger trees with a series of physically motivated recipes
(\citealt{White:1991aa}; \citealt*{Kauffmann:1993aa}; \citealt{Cole:1994aa}; \citealt{Somerville:1999aa}; \citealt{Kang:2005aa};
\citealt*{Monaco:2007aa}; \citealt{Benson:2012aa}).
The advantage of ab initio models is that they track galaxies and haloes self-consistently through cosmic time, and can test the
impact of different physical processes on galaxy properties. If the models disagree with observations the model is changed
either by implementing the physical processes differently, or by including new physical processes. However, these methods
can only achieve a limited resolution, so that simplified and highly uncertain `sub-grid' models have to be used to treat the
unresolved physical processes, such as star and black hole formation and the related feedback. The effects of these processes
then become tunable via free parameters, such that the models typically need to be calibrated with observations and are
effectively phenomenological. Due to the complex interaction of different physical prescriptions, the model parameters
can be degenerate and difficult to interpret if the model is not constrained well by data \citep{Lu:2012aa}. Moreover, there is considerable
uncertainty about whether the physics of galaxy formation is reliably represented.

A different option has emerged with the advent of data sets from large galaxy surveys \citep{York:2000aa,Colless:2001aa,
Lilly:2007aa,Driver:2011aa,Grogin:2011aa,McCracken:2012aa} and avoids explicitly modelling the baryonic physics.
Instead, {\it empirical models} of galaxy formation use relations with adjustable parameters to statistically link observed
galaxy properties to simulated dark matter haloes. In this way they can model galaxy formation unbiased by assumptions
on poorly understood baryonic physics, instead `marginalising' over these uncertainties. Predictions by empirical models
are therefore very useful for planning future surveys and for the interpretation of the observations.
Moreover, they provide a framework for ab initio models, and can thus help to constrain the
relevant physical processes. Empirical results have been widely used to fix unconstrained parameters in the sub-grid
models of hydrodynamic simulations, e.g. by requiring that simulated galaxies reproduce the empirically determined
stellar-to-halo mass (SHM) ratio. Hence empirical and ab initio models are complementary methods that can be applied to study
the physics that drives galaxy formation.

In the {\it halo occupation distribution} and the related {\it conditional luminosity function} formalisms, the distribution of galaxies
having specified intrinsic properties within main haloes of a given mass is constrained using galaxy abundance and clustering
statistics (\citealt*{Peacock:2000aa,Seljak:2000aa,White:2001aa,Berlind:2002aa,Yang:2003aa,Zehavi:2004aa,Zheng:2005aa};
\citealt{Tinker:2005aa,Brown:2008aa,Leauthaud:2012aa}). These approaches have typically been used at low redshift, as reliable
galaxy clustering measurements are not available at high redshift.
This problem can be circumvent by directly connecting galaxies to the underlying substructure. The {\it subhalo abundance matching}
method links the luminosity or stellar mass of a galaxy to the dark matter halo mass by matching the cumulative abundance
of galaxies to those of haloes and subhaloes (\citealt{Mo:1996aa,Wechsler:1998aa,Vale:2004aa}; \citealt*{Conroy:2006aa};
\citealt{Wang:2006aa,Wang:2007aa};
\citealt{Moster:2010aa}; \citealt*{Behroozi:2010aa}; \citealt{Guo:2010aa,Trujillo-Gomez:2011aa}). The clustering statistics of
galaxies can then be readily derived from the clustering of haloes in the simulation resulting in a remarkable agreement with
observed correlation functions. There are two ways to derive the relation between stellar and halo mass in this context. In the
backward modelling approach, observed galaxies and simulated haloes in an equal volume are rank ordered by mass and then
matched one by one. The SHM relation can then be described by a fitting function. In the forward modelling
approach, simulated dark matter haloes are populated with galaxies using a parameterised relation between stellar and halo
mass. The free parameters are then constrained by requiring that the observed SMF be reproduced. This method has the
advantage that scatter in the relation can easily be added to account for possible differences in the formation histories of haloes
at a fixed mass.

\begin{figure*}
\begin{tabular}{cc}
\includegraphics[height=70mm]{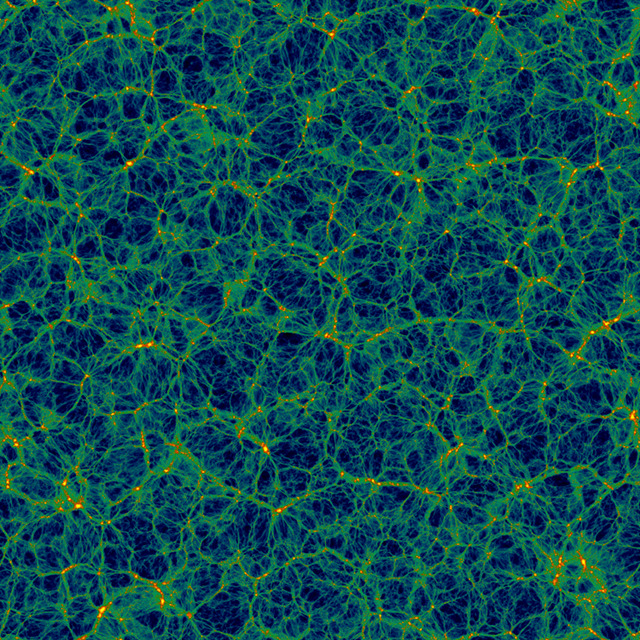} ~~~ & 
\includegraphics[height=70mm]{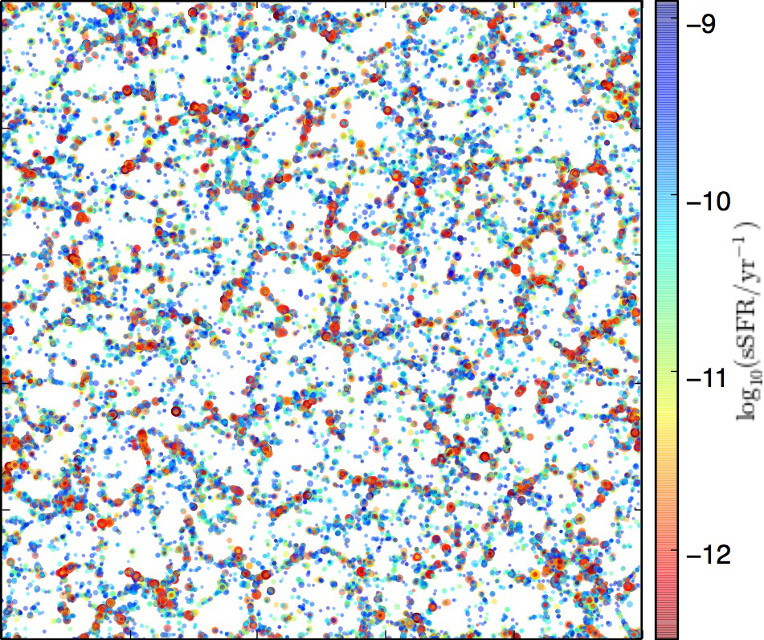} \\
\end{tabular}
\caption{\textit{Left panel}: Slice through the cosmological dark
  matter simulation with $6\Mpc$ thickness and $750\Mpc$
  width. The colour indicates dark matter density (red: high, blue:
  low).
  \textit{Right panel}: The same slice
  populated with galaxies using the new empirical model {\sc Emerge}, presented in
  this paper. Each dot corresponds to a galaxy -- the size scales with stellar mass and the colour
  corresponds to the specific star formation rate as indicated by the
  colour bar. The model makes definite predictions about stellar masses and
  star formation activity of central and satellite galaxies in
  individual dark matter haloes and subhaloes.}
\label{fig:box}
\end{figure*}

As the evolution of dark matter haloes is determined by the cosmological model, the link between galaxies and haloes can
be employed to infer the evolution of galaxy properties from the growth histories of haloes through cosmic time
(\citealt*{Conroy:2007aa}; \citealt{White:2007aa}; \citealt*{Zheng:2007ab}; \citealt{Firmani:2010aa,Wang:2013aa}).
In particular \citet{Conroy:2009aa} show how this method can be used to empirically constrain the average star formation
histories (SFHs) and stellar mass growth of galaxies in haloes with a given mass since $z=2$. This approach has been
extended by \citet*{Moster:2013aa} and \citet*{Behroozi:2013aa} to $z\sim8$ using halo merger trees that have been
extracted from cosmological simulations. These multi-epoch abundance matching models have been very successful in
describing the average evolution of galaxy properties in dark matter haloes. However, they do not self-consistently track
the growth history of individual galaxies. Inferred galaxy properties such as clustering then only depend on halo mass,
although it is well established that the spatial distribution of dark matter haloes depends on their formation time
\citep*[e.g.][]{Gao:2005aa}. Hence galaxy properties should also depend on the formation history of the halo.

A simple way to design an empirical model for the evolution of individual galaxies is presented by \citet*{Mutch:2013aa}.
Instead of linking integrated properties such as the present stellar mass to the halo mass, they connect instantaneous
properties. In this model the star formation rate (SFR) of a galaxy is given by the product of the halo growth rate
which determines how much material becomes available for star formation, and a parameterised baryon conversion
efficiency which only depends on halo mass and determines how effectively this material is converted into stars.
The stellar mass of a galaxy is then computed by integrating the SFHs of each galaxy through cosmic time taking
into account galaxy mergers. In this way a complete formation history for every galaxy is provided. The connection
between star formation and halo growth is also found observationally \citep{Tinker:2012aa} and in hydrodynamical
simulations \citep{Feldmann:2016aa}. An alternative method is to directly relate the SFH to the mass of the halo at
any given time \citep{Lu:2014aa,Lu:2015aa}.

In this work we present the novel empirical model {\sc Emerge}\footnote{\textbf{E}mpirical \textbf{M}od\textbf{E}l for the
fo\textbf{R}mation of \textbf{G}alaxi\textbf{E}s} that describes the formation of individual galaxies in dark matter
haloes. We follow \citet{Mutch:2013aa} and compute the SFR from the halo growth rate and the instantaneous conversion
efficiency. However, we use a more realistic parameterisation that depends on redshift and allows for different slopes
at the low and high-mass ends. Moreover, instead of setting the SFR to zero for all satellite galaxies, we include
empirical treatments for star formation after infall based on the `delayed-then-rapid' mechanism found by
\citet*{Wetzel:2012aa}, and for tidal stripping. To construct this new empirical model we follow the philosophy laid out in
\citet{Lu:2014aa}, and attempt to let the data speak for themselves in a way that is as independent as possible of any
model assumptions. Specifically, we assume that structure formation is determined by a \LCDM cosmology, that the SFR
of a galaxy depends on its halo's mass and growth rate and the redshift, and that once a galaxy becomes a satellite
it continues to form stars until it is rapidly quenched, its stars are stripped to the background when its halo has lost a
significant fraction of its mass, and it merges with the central galaxy on a dynamical friction timescale otherwise.
We constrain each of these processes with a specific data set minimising the correlation between parameters. The
complexity of the model is increased stepwise if the data require it, which we assess with a number of different
model selection statistics. In this way we aim to construct the simplest model that is in agreement with the data.

This paper is organised as follows. In Section \ref{sec:simobs} we describe the cosmological dark matter simulations and
the observational data sets we use to constrain our model. The methodology of our new empirical model is presented in
Section \ref{sec:method}. We further discuss how the model is constrained and compare our best-fit model to the
data. In Section \ref{sec:results} we present our main results for the growth of the stellar component and the
integrated conversion efficiency. We discuss the model in Section \ref{sec:disc} and provide a summary and an outlook
in Section \ref{sec:sum}. In Appendix \ref{sec:bayes} the model selection process is explained in detail, and in
Appendix \ref{sec:cov} we discuss the correlation between model parameters.

Throughout this work we assume a Planck \LCDM cosmology with (\Omegam, \OmegaL, \Omegab, $h$, $n_\mathrm{s}$,
$\sigma_8$) = (0.308, 0.692, 0.0484, 0.6781, 0.9677, 0.8149). We employ a \citet{Chabrier:2003aa} initial mass
function (IMF) and we convert all stellar masses and SFRs to this IMF. All virial masses are computed according to the
overdensity criterion by \citet{Bryan:1998aa}. In order to simplify the notation, we will use the capital $M$ to denote
dark matter halo masses and the lower case $m$ to denote galaxy stellar masses.

\section{Simulations and Observations}
\label{sec:simobs}

Empirical models connect observed galaxy properties to simulated dark matter haloes. The main
pillars of each empirical model are therefore a cosmological $N$-body simulation from which dark
matter haloes and merger trees are extracted, and observed data. In this work we used two dark
matter simulations with side lengths of $150\Mpc$ and $200\Mpc$, respectively.
We further used five different observational constraints taken from the literature. In this section
we provide more details on the simulations and observations.

\subsection{Dark matter simulations}
\label{sec:sims}

\begin{table}
 \caption{Observed cosmic star formation rate densities}
 \label{tab:obscsfrd}
 \begin{tabular}{lcccc}
  \hline
  Publication & $z$ & Area & IMF & $\lambda$\\
  \hline
  \citet{Robotham:2011aa} & 0.0~-~0.1 & 833.13 & S & UV\\
  \citet{Salim:2007aa} & 0.0~-~0.2 & 741 & C & UV\\
  \citet{Gunawardhana:2015aa} & 0.0~-~0.4 & 144 & C & H$\alpha$\\
  \citet{Ly:2011aa} & 0.8 & 0.82 & S & H$\alpha$\\
  \citet{Zheng:2007aa} & 0.2~-~1.0 & 0.458 & C & UV/IR\\
  \citet{Rujopakarn:2010aa} & 0.0~-~1.2 & $\le9$ & S & FIR\\
  \citet{Smolcic:2009aa} & 0.1~-~1.3 & 2 & S & 1.4 GHz\\
  \citet{Shim:2009aa} & 0.7~-~1.9 & 0.029 & S & H$\alpha$\\
  \citet{Tadaki:2011aa} & 0.0~-~0.2 & 0.016 & S & H$\alpha$\\
  \citet{Sobral:2013aa} & 2.2 & $\le1.68$ & S & H$\alpha$\\
  \citet{Magnelli:2011aa} & 1.3~-~2.3 & 0.079 & S & IR\\
  \citet{Hayes:2010aa} & 2.2 & 0.016 & S & H$\alpha$\\
  \citet{Karim:2011aa} & 0.2~-~3.0 & 1.72 & C & 1.4 GHz\\
  \citet{Ly:2011ab} & 1~-~3 & 0.242 & S & UV\\
  \citet{Kajisawa:2010aa} & 0.5~-~3.5 & 0.029 & S & UV/IR\\
  \citet{Reddy:2009aa} & 1.9~-~3.4 & 0.906 & K & UV\\
  \citet{Burgarella:2013aa} & 0~-~4 & $\le0.6$ & S & UV/IR\\
  \citet{Cucciati:2012aa} & 0~-~4.5 & 0.611 & S & UV\\
  \citet{Dunne:2009aa} & 0~-~5 & 0.8 & S & 1.4 GHz\\
  \citet{Le-Borgne:2009aa} & 1~-~5 & 0.07 & S & IR/mm\\
  \citet{van-der-Burg:2010aa} & 3~-~5 & 4 & S & UV\\
  \citet{Bourne:2016aa} & 0.5~-~6 & 0.064 & C & UV/IR\\
  \citet{Duncan:2014aa} & 4~-~7 & $\le0.017$ & C & UV\\  
  \citet{Oesch:2013aa} & 3.8~-~11 & 0.045 & S & UV\\
  \citet{McLure:2013aa} & 6~-~10 & $\le0.05$ & S & UV\\
    \hline
  \end{tabular}
 \medskip\\
  \textbf{Notes:} Columns are publication (1), redshift range $z$ (2), survey area in deg$^2$ (3),
  IMF (4): C \citep{Chabrier:2003aa}, S \citep{Salpeter:1955aa}, K \citep{Kroupa:2001aa}, and spectral range used to convert fluxes into SFRs (5).
\end{table}

The empirical model presented in this paper follows the growth of dark matter haloes and assigns a
SFR to the galaxy at its centre. For this, we have extracted the halo merger trees from two
$N$-body simulations. The first simulation has a smaller volume and fewer particles, and consequently
results in fewer halo merger trees. It is therefore well suited to run the empirical model multiple times,
as done for parameter space exploration. The second simulation has a larger volume and more particles
and hence more merger trees. We only use this simulation in single runs with previously determined
parameters to cover a larger mass range. As we will show, the simulations lead to identical results for
the galaxy populations. For both simulations we adopted cosmological parameters consistent with the
latest results by the \citet{Planck-Collaboration:2016aa}. Specifically, we chose $\Omegam=0.308$,
$\OmegaL=0.692$, $\Omegab=0.0484$, $H_0=67.81\kmsMpc$,
$n_\mathrm{s}=0.9677$, and $\sigma_8=0.8149$.

The initial conditions for both simulations were generated with the {\sc Music} code \citep{Hahn:2011aa}
using a power spectrum computed with the \texttt{CAMB} code \citep*{Lewis:2000aa}. The first simulation
has a side length of $150\Mpc$ and contains $512^3$ dark matter particles corresponding to a particle
mass of $9.88\times10^8\Msun$. The second simulation has a side length of $200\Mpc$ and contains
$1024^3$ dark matter particles corresponding to a particle mass of $2.92\times10^8\Msun$. Both simulations
were run with periodic boundary conditions from redshift $z=63$ to 0 using the TreePM code {\sc Gadget3}
\citep{Springel:2005aa} creating 94 snapshots, equally spaced in scale factor ($\Delta a=0.01$). The
gravitational softening was $6\kpc$ for the $150\Mpc$ simulation, and $3.3\kpc$ for the $200\Mpc$
simulation. In the left panel of Figure \ref{fig:box} we show a density map of the first simulation, remapped
into a sheet with $6\Mpc$ thickness and $750\Mpc$ width using the method presented by \citet{Carlson:2010aa}.

Dark matter haloes and subhaloes in the simulations were identified with the seven-dimensional halo finder
{\sc Rockstar} \citep*{Behroozi:2013ac} for each snapshot. Halo masses were calculated using spherical
overdensities, according to the criterion for a spherical collapse model of a tophat perturbation in a
\LCDM cosmology by \citet{Bryan:1998aa}. Given a minimal particle number of 100 for each halo the
minimally resolved halo mass is $M_\mathrm{min}=10^{11}\Msun$ for the $150\Mpc$ box, and
$M_\mathrm{min}=10^{10.5}\Msun$ for the $200\Mpc$ box. Merger trees were generated using the
{\sc ConsistentTrees} algorithm \citep{Behroozi:2013ad}, providing a physically consistent evolution of
halo properties across time steps. We use the term `main halo' to refer to distinct haloes that are not
located within a larger halo, while the term `subhalo' refers to all other haloes. We further assume that
both main haloes and subhaloes host galaxies at their centres. The galaxy at the centre of a main halo
is referred to as `central galaxy', and all galaxies within subhaloes are referred to as `satellite galaxy'.

\begin{table}
 \caption{Observed specific star formation rates}
 \label{tab:obsssfr}
 \begin{tabular}{lcccc}
  \hline
  Publication & $z$ & Area & IMF & $\lambda$\\
  \hline
  \citet{Salim:2007aa} & 0.0~-~0.2 & 741 & C & UV\\
  \citet{Zheng:2007aa} & 0.2~-~1.0 & 0.458 & C & UV/IR\\
  \citet{Twite:2012aa} & 1.0 & 1.4 & C & H$\alpha$\\
  \citet{Noeske:2007aa} & 0.2~-~1.1 & 0.5 & K & UV/IR\\
  \citet{Tadaki:2011aa} & 2.2 & 0.016 & S & H$\alpha$\\
  \citet{Whitaker:2012aa} & 0.0~-~2.5 & 0.4 & C & UV/IR\\
  \citet{Daddi:2007aa} & 1.4~-~2.5 & 0.06 & S & UV-1.4 GHz\\
  \citet{Salmi:2012aa} & 0.9~-~1.3 & 0.06 & C & UV\\
  \citet{Karim:2011aa} & 0.2~-~3.0 & 1.72 & C & 1.4 GHz\\
  \citet{Kajisawa:2010aa} & 0.5~-~3.5 & 0.029 & S & UV/IR\\
  \citet{Reddy:2012aa} & 1.4~-~3.7 & 0.44 & S & UV/IR\\
  \citet{Feulner:2005aa} & 0.4~-~5.0 & 0.014 & S & UV/IR\\  
  \citet{Lee:2012aa} & 3.3~-~4.3 & 5.3 & C & UV/IR\\    
  \citet{Gonzalez:2011aa} & 4~-~6 & 0.015 & S & UV/IR\\ 
  \citet{Schaerer:2010aa} & 6~-~8 & 2 & S & UV\\    
  \citet{Labbe:2013aa} & 8 & 0.04 & S & UV/IR\\    
  \citet{McLure:2011aa} & 6~-~8.7 & 0.013 & C & UV\\  
  \citet{Duncan:2014aa} & 4~-~7 & 0.017 & C & UV\\   
    \hline
  \end{tabular}
 \medskip\\
  \textbf{Notes:} Columns are publication (1), redshift range $z$ (2), survey area in deg$^2$ (3), IMF (4), and spectral range used
  to convert fluxes into SFRs (5).
\end{table}

\subsection{Observations}
\label{sec:observations}

The empirical model assigns a SFR to each galaxy based on the growth rate of its halo, and
the stellar masses are computed by integrating these. To constrain the star formation we use five different
observational constraints: SMFs, cosmic SFR densities (CSFRDs), specific SFRs (sSFRs), fractions of
quenched galaxies, and projected galaxy correlation functions. We convert all units to
physical units using $h=0.6781$. All stellar masses and SFRs are converted to be consistent with
a \citet{Chabrier:2003aa} IMF. An evolving or non-universal IMF is not considered.

We specifically do not take into consideration any other systematic effects, such as different stellar populations
synthesis models, dust models, spectral energy distribution fitting methods, assumed SFHs,
metallicities, photometry, redshift measurements, and cosmic variance. Instead, we consider these effects to be
sources of error for the model SFRs and stellar masses. As observational studies typically do not take these
sources into account when quoting the errors, we calculate the variance between different observations and add
the result quadratically to each data point to estimate the true systematic errors. In this way, the uncertainties in the
observations, i.e. scatter between different data sets, will get translated into model uncertainties. The resulting confidence
levels in modelled galaxy properties will thus reflect our lack of knowledge on systematic observational effects.
We do not combine different measurements into an average sample (e.g. at various redshifts), but compute a
corresponding model prediction for each measured data point.

\begin{table}
 \caption{Observed stellar mass functions}
 \label{tab:obssmf}
 \begin{tabular}{llccc}
  \hline
  Publication & Abb. & $z$ & Area & IMF   \\
  \hline
  \citet{Baldry:2012aa} & Bal12 & 0.0~-~0.1 & 143 & C\\
  \citet{Li:2009aa} & LW09 & 0.0~-~0.2 & 6,437 & C\\
  \citet{Bernardi:2013aa} & Ber13 & 0.0~-~0.2 & 4,681 & C\\
  \citet{Perez-Gonzalez:2008aa} & Per08 & 0.0~-~4.0 & 0.184 & S\\
  \citet{Ilbert:2010aa} & Ilb10 & 0.0~-~2.0 & 2 & C\\
  \citet{Pozzetti:2010aa} & Poz10 & 0.1~-~1.0 & 1.4 & C\\
  \citet{Ilbert:2013aa} & Ilb13 & 0.2~-~4.0 & 1.52 & C\\
  \citet{Moustakas:2013aa} & Mou13 & 0.2~-~1.0 & 9 & C\\
  \citet{Muzzin:2013aa} & Muz13 & 0.2~-~4.0 & 1.62 & K\\
  \citet{Kajisawa:2009aa} & Kaj09 & 0.5~-~3.5 & 0.0364 & S\\
  \citet{Santini:2012aa} & San12 & 0.6~-~4.5 & 0.0092 & S\\
  \citet{Mortlock:2011aa} & Mor11 & 1.0~-~3.5 & 0.0121 & S\\
  \citet{Marchesini:2009aa} & Mar09 & 1.3~-~4.0 & 0.1620 & K\\
  \citet{Caputi:2011aa} & Cap11 & 3.0~-~5.0 & 0.6 & S\\
  \citet{Grazian:2015aa} & Gra15 & 3.5~-~7.5 & 0.1025 & S\\
  \citet{Lee:2012aa} & Lee12 & 3.7~-~5.0 & 0.0889 & C\\
  \citet{Gonzalez:2011aa} & Gon11 & 4.0~-~7.0 & 0.0150 & S\\
  \citet{Duncan:2014aa} & Dun14 & 4.0~-~7.0 & 0.0167 & C\\
  \citet{Song:2016aa} & Son16 & 4.0~-~8.0 & 0.0778 & S\\
    \hline
  \end{tabular}
 \medskip\\
  \textbf{Notes:} Columns are publication (1), abbreviation (2), redshift range $z$ (3),
  survey area in deg$^2$(4), and IMF (5).
\end{table}

The normalisation of the instantaneous baryon conversion efficiency can be constrained with the observed
CSFRD. In Table \ref{tab:obscsfrd} and in Figure \ref{fig:csfrd} we summarise all data sets that we have used in
this work. A broad range of techniques has been used to convert fluxes into SFRs using narrowband (H$\alpha$),
broadband (UV-IR), and radio (1.4 GHz) surveys up to $z=11$. Recent observations find that the CSFRD
falls off rather steeply for $z>3$. Although there is generally a very good agreement between different data sets
we find a variance of 0.1 dex which we include in the errors. To constrain the low and high-mass
ends of the instantaneous conversion efficiency we use observed sSFRs for different stellar masses and redshifts
up to $z=8$. The data sets are summarised in Table \ref{tab:obsssfr} and in Figure \ref{fig:ssfr}. As the data
sets are not fully consistent with each other, we find a variance of 0.15 dex independent of stellar mass and
redshift, and correct the errors to account for this.

To constrain the overall evolution of galaxies and specifically contribution of galaxy mergers, we use the SMFs
from $z=0$ to $z=8$ presented in Table \ref{tab:obssmf} and in Figure \ref{fig:smf}. This includes data
from wide surveys that capture many massive galaxies, and from deep surveys that can well constrain the
low-mass tail. Most recent observations find rather steep low-mass slopes, reconciling the tension
between the integrated CSFRD and the integrated SMFs. At low redshift there is generally a good agreement
between different data sets. However, at higher redshift some of the data sets are not consistent with each other
showing that the systematic errors have not been fully considered. We find a variance of
0.015 dex at $z=0$ and 0.3 dex beyond $z=2$ and modify the errors accordingly. Additionally,
the limited photometric information means that stellar populations cannot be fully constrained, leading to
intrinsic scatter in the mass estimates relative to the true mass. As this effects becomes stronger at high
redshift, we estimate the scatter as $\sigma(z)=0.08+0.06z$, based on the results of \citet{Li:2009aa} at $z\sim0.1$
and \citet{Perez-Gonzalez:2008aa} at $z\lesssim4$. Thus, when comparing to observed stellar masses
we draw the stellar mass from a lognormal distribution with a mean value given by the model mass
and a scatter of $\sigma(z)$.

\begin{table}
 \caption{Observed quenched fractions}
 \label{tab:obsfq}
 \begin{tabular}{llcccc}
  \hline
  Publication & $z$ & Area & IMF & A/P  \\
  \hline
  \citet{Wetzel:2012aa} & 0.0~-~0.1 & 7,97 & C & sSFR\\
  \citet{Drory:2009aa} & 0.2~-~1.0 & 1.73 & C & NUV-$R$-$J$\\
  \citet{Ilbert:2013aa} & 0.2~-~4.0 & 1.52 & C & NUV-$r^+$-$J$\\
  \citet{Moustakas:2013aa} & 0.2~-~1.0 & 9 & C & sSFR\\
  \citet{Lin:2014aa} & 0.2~-~0.8 & 70 & S & sSFR\\
  \citet{Muzzin:2013aa} & 0.2~-~4.0 & 1.62 & K & $U$-$V$-$J$\\
    \hline
  \end{tabular}
 \medskip\\
  \textbf{Notes:} Columns are publication (1), redshift range $z$ (2), total survey area in deg$^2$ (3), IMF (4), and method to separate
  active from passive galaxies (5): active-to-passive cut based on the sSFR, or based on location in rest-frame colour-colour diagram (e.g. $U$-$V$-$J$:
  $U-V$ vs. $V-J$).
\end{table}

\begin{table}
 \caption{Observed projected correlation functions}
 \label{tab:obswp}
 \begin{tabular}{lccc}
  \hline
  Publication & Survey & Stellar mass & IMF \\
  \hline
  \citet{Li:2006aa} & ~~~SDSS/DR2~~~ & ~~~K03~~~ & ~~~C~~~\\
  \citet{Guo:2011aa} & SDSS/DR7 & K03  & ~~~C~~~\\
  \citet{Yang:2012aa} & SDSS/DR7 & B03  & ~~~C~~~\\
    \hline
  \end{tabular}
 \medskip\\
    \textbf{Notes:} Columns are publication (1), survey data (2), method to derive stellar mass (3): K03 \citep{Kauffmann:2003aa}, B03 \citep{Bell:2003aa},
    and IMF (4).
\end{table}

As in our model satellite galaxies keep forming stars for a specific timescale after their halo reaches its peak mass, we
need an observational constraint to determine if it is still forming stars or has been quenched. Several authors
have measured the SMF for both active and passive galaxies up to $z=0$, and can thus provide the fraction
of quenched galaxies as a function of stellar mass. These measurements are summarised in Table \ref{tab:obsfq}
and in Figure \ref{fig:fq}. To distinguish quenched from star forming galaxies observational studies use two different
techniques. One is to also measure the SFR of each object and then applying a cut at a certain sSFR. If the SFR
cannot be obtained for all galaxies because of limited depth, the classification can also be based on a
colour-colour diagram that is easier to obtain. Although it is clear that these techniques do not correspond perfectly
for each galaxy, it has been shown that they correlate well \citep{Williams:2009aa}. The variance between the
different data sets is 10 per cent and we update the errors accordingly.

In our model satellite galaxies get tidally stripped once their halo has fallen below a specific fraction of its peak mass.
This affects galaxy clustering on small scales, so we use the projected galaxy auto-correlation functions summarised
in Table \ref{tab:obswp} and Figure \ref{fig:wp} to constrain this fraction. All data sets are are based on the Sloan
Digital Sky Survery \citep{York:2000aa}, but while \citet{Li:2006aa} use the Data Release 2 \citep[DR2;][]{Abazajian:2004aa},
 \citet{Guo:2011aa} and \citet{Yang:2012aa} use the Data Release 7 \citep[DR7;][]{Abazajian:2009aa}. 
 Stellar masses in \citet{Li:2006aa} and  \citet{Guo:2011aa} are computed following \citet{Kauffmann:2003aa}, while
\citet{Yang:2012aa} follow \citet{Bell:2003aa}. Although the data sets are in good agreement we find a variance of 0.15 dex
which we add to the errors.

\section{Connecting galaxies and haloes} 
\label{sec:method}

In this section we describe how {\sc Emerge} relates galaxies to their
dark matter haloes. To populate the dark matter halo merger trees that
have been extracted from the simulations presented in section
\ref{sec:sims} with galaxies, we first compute the SFR
based on the growth rate of the halo, and then integrate this to
derive the stellar mass of each galaxy. We further consider the
relevant effects for satellite galaxies, i.e. quenching, stripping and
merging. For any set of model parameters we then compute a number of
mock observations and compare them to the data presented in section
\ref{sec:observations}. Finally, the model parameters are fitted to
reproduce the observations using a Markov chain Monte Carlo method.

\subsection{Star formation in dark matter haloes}

While in previous empirical models the stellar mass of a galaxy has
been linked to its halo mass, {\sc Emerge} connects the time
derivatives  of these two quantities, i.e. the SFR of
a galaxy is linked to growth rate of its halo. We assume that the
baryonic mass in every dark matter halo is given by the universal
fraction $f_\mathrm{b} = \Omegab/\Omegam = 0.156$,
and that the rate of infalling baryonic mass $\dot m_\mathrm{b}$ is
proportional to the halo growth rate $\dot M$. The SFR
is then given by the product of the baryonic growth rate and the
instantaneous baryon conversion efficiency $\epsilon$ which may depend
on halo mass and redshift:
\begin{equation}
\label{eqn:sfrcen}
\frac{{\rm d}m_* (M,z)}{{\rm d}t} = \frac{{\rm d}m_\mathrm{b}}{{\rm d}t} \cdot \, \epsilon(M,z) = f_\mathrm{b} \frac{{\rm d}M}{{\rm
    d}t} \cdot \, \epsilon(M,z)\;. 
\end{equation}
The baryonic growth rate at a given redshift $z$ thus determines the
amount of material that is available for galaxy formation, while the
conversion efficiency denotes how effectively this material
is converted into stars. In the following sections we show how we compute these two quantities.  

\subsubsection{The baryonic growth rate}
\label{sec:mdotbary}

The baryonic growth rate can be extracted from the dark matter
simulations by assuming that the fraction of infalling baryons is
equal to the universal fraction:
\begin{equation}
\frac{{\rm d}m_\mathrm{b}}{{\rm d}t} = f_\mathrm{b} \frac{{\rm d}M}{{\rm d}t}\;.
\end{equation}
All baryonic material that falls into the dark matter halo becomes
available for star formation within one dynamical time of the halo
$t_\mathrm{dyn} = (R_\mathrm{v}^3/\mathrm{G}M)^{1/2}$,
where $\mathrm{G}$ is the gravitational constant and
$R_\mathrm{v}$ is the virial radius. We therefore  let the
baryonic growth rate only depend on the halo growth rate averaged over
the dynamical time, defined as 
\begin{equation}
\left\langle\frac{{\rm d}M}{{\rm d}t}\right\rangle_\mathrm{dyn} = \frac{M(t) - M(t-t_\mathrm{dyn})}{t_\mathrm{dyn}} \; ,
\end{equation}
Consequently, the model becomes independent of the time resolution of
the simulation as long as the evaluation timesteps are smaller than
$t_\mathrm{dyn}$. 

This growth rate can be split into two contributions \citep*{Diemer:2013aa}: the
physical growth of the halo due to accretion (i.e. the mass growth inside a fixed radius), and
pseudo evolution due to the growing virial radius over cosmic time as the background density decreases.
The growth rate due to accretion can be combuted as
\begin{equation}
\frac{{\rm d}M}{{\rm d}t} = \left\langle\frac{{\rm d}M}{{\rm d}t}\right\rangle_\mathrm{dyn} - 4\pi \, R_\mathrm{v}^2 \, \rho(R_\mathrm{v}) \,
\left\langle\frac{{\rm d}R_\mathrm{v}}{{\rm d}t}\right\rangle_\mathrm{dyn} \; ,
\end{equation}
where $R_\mathrm{v}$ is the virial radius of the main progenitor and $\rho(R_\mathrm{v})$ is its density at the virial radius
\citep[c.f. eqn. 9 in][]{Diemer:2013aa}. The density for a \citet*{Navarro:1997aa} profile can be computed using the halo scale length
extracted from the simulation.

Here we are only interested in the infalling baryons that will be available for galaxy formation.
However, unlike dark matter, baryons do not splash back beyond the virial radius, but typically remain inside.
The rate of baryons becoming available for galaxy formation at the virial radius can thus be slightly higher than $f_\mathrm{b} \dot M$
in massive systems, more closely following the halo growth rate without the pseudo evolution correction. We have tested our model for 
standard growth rates and growth rates corrected for pseudo evolution and find that the results are identical except for a small shift (10
per cent) in the normalisation of the conversion efficiency. All results reported in this work are based on growth rates corrected for pseudo evolution.

The baryonic growth rate describes how much material is becoming available for star formation, so we require it to be non-negative.
Negative $\dot M$ will occur when haloes are stripped. In this case we set the growth rate to zero.

\subsubsection{The instantaneous baryon conversion efficiency}
\label{sec:epsilon}

The instantaneous baryon conversion efficiency $\epsilon$ describes how efficiently gas that is becoming available according to the
baryonic growth function are converted into stars. This efficiency parametrises the effects of physical processes that determine how gas 
is converted into stars, i.e. gas cooling, star formation, and various feedback processes. Therefore, it can be a function of many parameters 
and can become arbitrarily complex. Following the philosophy of our approach to find the simplest model that is able to reproduce a large
number of observations, we assume that the efficiency only depends on halo mass $M$ and redshift $z$. 

Comparing the observed SMF and the computed HMF, simple empirical models find that the integrated baryon
conversion efficiency (i.e. the ratio between all stellar mass and baryonic mass in a halo $m/m_\mathrm{b}$) is a strong function of halo
mass at any redshift. While peaking at a halo mass of $\log_{10}(M/\Msun)\approx 12$, the integrated efficiency declines
towards both lower and higher halo masses. In the currently favoured picture, feedback from supernovae can drive massive galactic winds
that are able to eject gas from the shallow potential wells of low mass haloes, reducing the availability of fuel for star formation. In
massive haloes it is believed that feedback from active galactic nuclei (AGN) can heat the gas in the halo and prevent it from
cooling. Also gravitational heating may play a role in further preventing the gas from cooling. The halo mass where galaxy formation
is most efficient can then be understood as the mass where the combination of the various physical processes that reduce the
efficiency is minimal integrated over time. However, the integrated efficiency is still several steps away from the instantaneous
efficiency as it includes the effects of stellar mass loss and galaxy mergers, so it may have a different dependence on halo mass and
redshift. 

\begin{figure}
\includegraphics[width=0.48\textwidth]{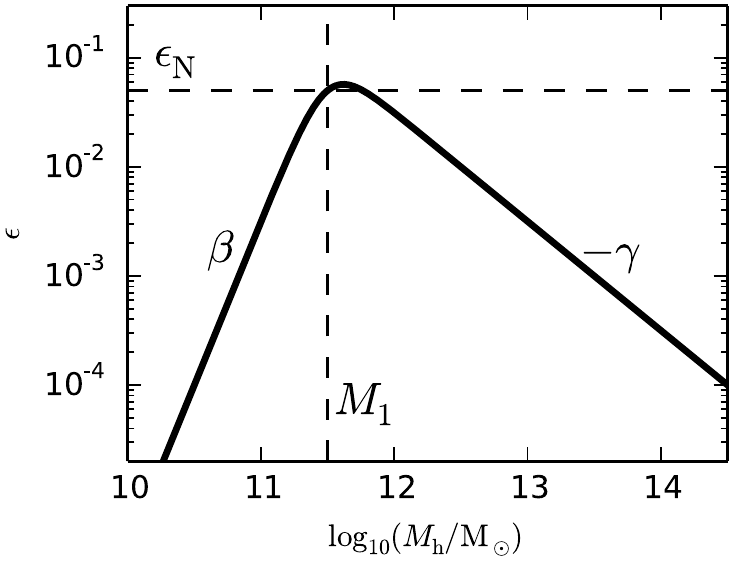}
\caption{Illustration of the
    dependence of the instantaneous baryon conversion efficiency
    $\epsilon$ on halo mass $M$. The efficiency peaks close to the
    characteristic halo mass $M_1$ where it has the normalisation
    $\epsilon_\mathrm{N}$. Towards low masses the efficiency decreases
    with the slope $\beta$ while towards high masses it decreases with
    the slope $\gamma$. } 
\label{fig:epsilon}
\end{figure}

To find a suitable parameterisation of the instantaneous baryon conversion efficiency, we used the predictions by
\citet{Moster:2013aa} and find that at every redshift the dependence of the instantaneous efficiency on halo mass can be described by a
double power law as is typically used for the integrated efficiency. This is consistent with the results of \citet*[][c.f. their
  figure 2]{Behroozi:2013ab}. Therefore we adopt the parameterisation used for the integrated efficiency introduced in \citet{Moster:2010aa}:
\begin{equation}
\label{eqn:epsilon}
\epsilon(M,z) = 2 \;\epsilon_\mathrm{N} \left[ \left(\frac{M}{M_1}\right)^{-\beta} + \left(\frac{M}{M_1}\right)^{\gamma}\right]^{-1} \; .
\end{equation}
It is governed by four free parameters: the normalisation $\epsilon_\mathrm{N}$, the characteristic mass $M_1$ where the
efficiency is equal to its normalisation, and the two slopes $\beta$ and $\gamma$ that determine how the efficiency decreases at low and
high mass, respectively. An illustration of the dependence of the efficiency on halo mass is shown in Figure \ref{fig:epsilon}. Baryon
conversion is most efficient at a halo mass close to the characteristic one: 
\begin{equation}
\label{eqn:effmax}
M_\mathrm{max}=M_1 \left(\frac{\beta}{\gamma}\right)^{1/(\beta+\gamma)} \; .
\end{equation}
We typically expect the values of $\beta$ and $\gamma$ to be positive, i.e. at low masses the efficiency increases with increasing mass,
while at high masses the efficiency decreases with increasing mass. However, we impose no a-priori restrictions on the values of the
slopes, such that the parameterisation is very flexible and can adopt to a wide range of observational constraints. 

As the integrated baryon conversion efficiency depends on redshift \citep[e.g.][]{Moster:2013aa}, we allow the parameters of the
instantaneous baryon conversion efficiency to vary with redshift. As before we try to find a parameterisation that is as simple as
possible. For each parameter we adopt a linear dependence on the scale factor $a=(z+1)^{-1}$, and use model selection statistics to determine
if this constitutes an improvement over a redshift-independent parameterisation (see Appendix \ref{sec:bayes}). We find that the
simplest model that can reproduce the data has  redshift-dependence in $M1$, $\epsilon_\mathrm{N}$, and $\beta$ while the high mass
slope $\gamma$ is constant over cosmic time. We hence adopt the following parameterisations:
\begin{align}
\log_{10} M_1(z)& = M_0 + M_\mathrm{z}(1-a) = M_0 + M_\mathrm{z}\frac{z}{z+1} \; ,\\
\epsilon_\mathrm{N}(z)& = \epsilon_0 + \epsilon_\mathrm{z}(1-a) = \epsilon_0 + \epsilon_\mathrm{z}\frac{z}{z+1} \; ,\\
\beta(z)& = \beta_0 + \beta_\mathrm{z}(1-a) = \beta_0 + \beta_\mathrm{z}\frac{z}{z+1} \; ,\\
\gamma(z)& = \gamma_0 \; .
\end{align}
We do not impose any a priori constraints on these parameters, except for physical boundary conditions, e.g. $\epsilon_o+\epsilon_\mathrm{z}>0$.
The instantaneous baryon conversion efficiency of any halo thus only depends on its mass and redshift, and no artificial scatter is added to it.
This means that two haloes with the same mass at a given redshift will have the same instantaneous efficiency. However, the integrated efficiency
and hence the stellar mass will depend on the full formation history of the halo, such that scatter is introduced self-consistently.

\subsubsection{The build-up of stellar mass}
\label{sec:buildup}

Having specified how we compute the baryonic growth rate $\dot m_{\mathrm b}$ and the instantaneous baryon conversion efficiency
$\epsilon(M,z)$, we can now calculate the SFR of the central galaxy $\dot m_*(M,\dot M,z)$ in each halo using equation
($\ref{eqn:sfrcen}$). The stellar mass $m_*$ of each galaxy will then grow by both star formation given by $\dot m_*$ (in-situ) and by the
assembly of stars that formed outside the galaxy and are being accreted given by $\dot m_\mathrm{acc}$ (ex-situ). Moreover, we have
to take into account the stellar mass that is being lost  as a consequence of dying stars $\dot m_\mathrm{loss}$. The stellar mass at
any time $t$ can then be calculated as 
\begin{align}
\label{eqn:mstar}
m_*(t)& = \int_0^t \mathrm{d}t^\prime \;\left[ \dot m_*(t^\prime) - \dot m_\mathrm{loss}(t^\prime)
+ \dot m_\mathrm{acc}(t^\prime)\right] \nonumber\\
& =  \int_0^t \mathrm{d}t^\prime \;\left[ \dot m_*(t^\prime) \cdot [1-f_\mathrm{loss}(t-t^\prime)]
+ \dot m_\mathrm{acc}(t^\prime)\right] \nonumber\\
& = m_\mathrm{SF} (t)+ m_\mathrm{acc} (t) \, ,
\end{align}
where $f_\mathrm{loss}(t)$ is the fraction of mass lost by a single stellar population with an age t. Using the FSPS package
\citep*{Conroy:2010aa,Conroy:2010ab} to calculate the rate of stellar mass loss for a \citet{Chabrier:2003aa} IMF, the fraction of lost
stars is well-fit by 
\begin{equation}
\label{eqn:massloss}
f_\mathrm{loss}(t) = 0.05 \ln \left(1+ \frac{t}{1.4~\mathrm{Myr}} \right) \; .
\end{equation}
Note that we do not employ the instantaneous recycling approximation (where $f_\mathrm{loss}$ is constant), but compute the mass loss as a
function of time.

The first term in equation (\ref{eqn:mstar}) can be integrated as $m_\mathrm{SF}(t) = \sum_{t_i<t} \dot m_{*,i} \, \Delta t_i \,
[1-f_\mathrm{loss} (t-t_i)]$. The second term  depend on how accreted stars are added to the central galaxy during a merger. We discuss the
details of merging galaxies in section \ref{sec:merging}.

\subsubsection{Star formation in growing haloes}
\label{sec:modelcentral}

\begin{figure}
\includegraphics[width=0.47\textwidth]{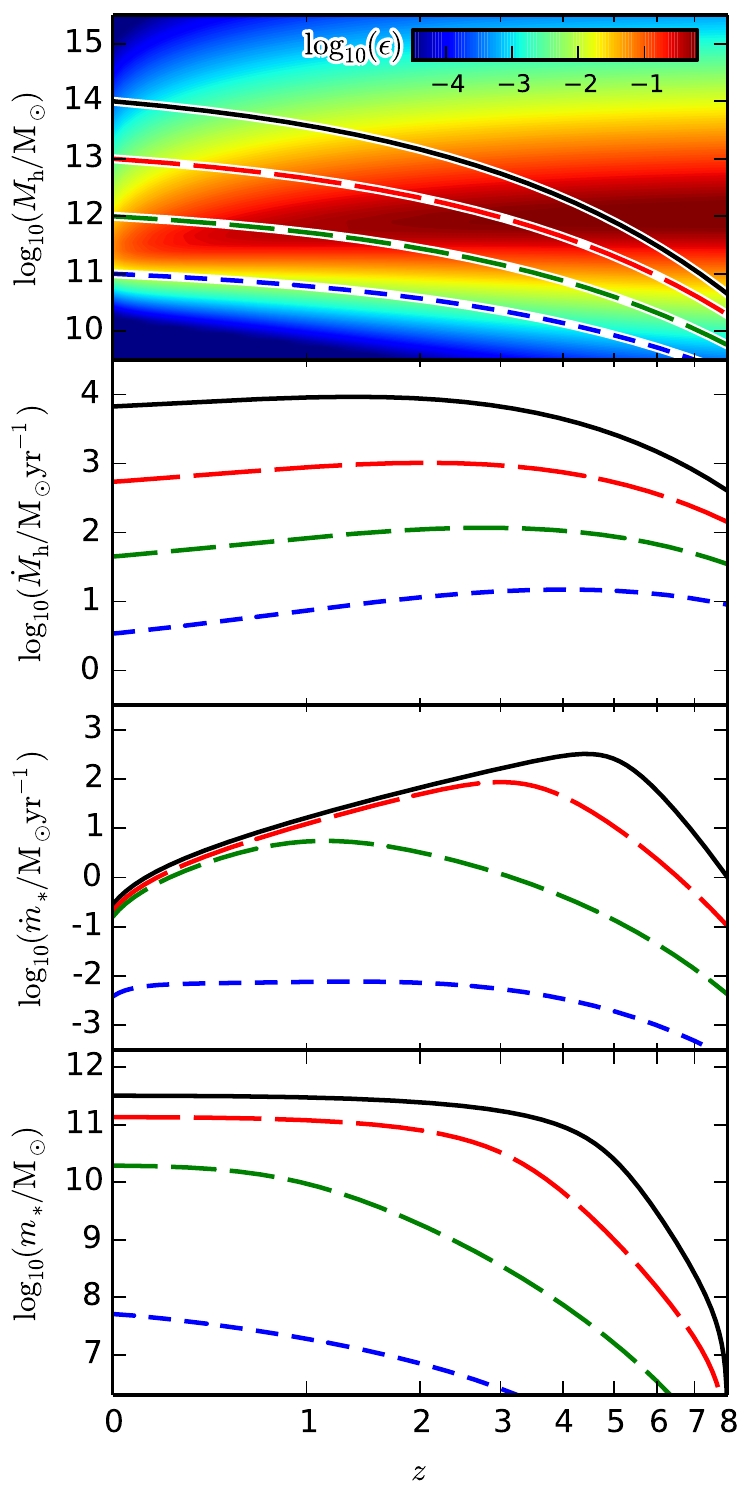}
\caption{Illustration of the growth of haloes and their galaxies. The
  first panel shows the evolution of the halo mass for four idealised
  haloes with $z=0$  masses of $\log_{10}(M/\Msun)=11,12,13$ and
  $14$. The background colour denotes the instantaneous baryon
  conversion efficiency $\epsilon(M,z)$ for a given halo mass and
  redshift. The second panel shows the corresponding growth rates. In
  the third panel the star formation rate for the galaxies within the
  four haloes is shown as computed with equation
  (\ref{eqn:sfrcen}). It peaks close to the redshift where the halo
  mass passes through the maximum efficiency. The fourth panel gives
  the time integrated stellar mass, assuming a mass loss rate of 40
  per cent. }
\label{fig:model}
\end{figure}

The model specified so far can be used to explain the growth of isolated central galaxies living in monotonically growing dark matter
haloes. Figure \ref{fig:model} illustrates this for four dark matter haloes with idealised average growth histories for different $z=0$
virial masses. For this illustration we have used the best-fit parameters that will be derived later, but a departure from their exact values
will not change the qualitative behaviour of the model. In the top panel the redshift evolution of the halo mass is indicated by the
lines of different colours. The background colour gives the instantaneous baryon conversion efficiency $\epsilon(M_{\rm h},z)$ for
a given halo mass and redshift. The most massive halo with a $z=0$ virial mass of $10^{14}\Msun$ (solid black line) has a path that
crosses the peak of the conversion efficiency already at $z\approx 4$, while less massive haloes will reach their peak efficiency later,
e.g. a halo with a $z=0$ virial mass of $10^{11}\Msun$ (dashed blue line) reaches its peak efficiency only around $z\approx 0$. 

The second panel gives the halo growth rates $\dot M_\mathrm{h}$ for each system. Typically the growth rates of massive haloes peak late,
i.e. they assemble most of their mass at low redshift, while haloes with lower mass have growth rates that peak earlier. However, the
growth histories are rather flat once the peak has been reached, i.e for $z\lesssim5$. The third panel shows the SFR of each
halo's central galaxy $\dot m_\mathrm{*}$, as computed with equation (\ref{eqn:sfrcen}), i.e. the product of the universal baryon
fraction $f_\mathrm{b}$, the halo's growth rates $\dot M_\mathrm{h}$ (from panel 2), and the instantaneous baryon conversion efficiency
$\epsilon(M_{\rm h},z)$ at that halo mass and redshift (from top panel). Since the variation in the halo growth rate is typically much 
smaller than the variation in the conversion efficiency, the change in the SFR is mostly dominated by the change in the
conversion efficiency. At high redshift, when all haloes still  have low mass, their conversion efficiency is rather low, leading to a low
SFR for the central galaxies. As the halo mass grows, the conversion efficiency grows as well leading to higher SFRs.
Once the halo mass has reached the value where the conversion efficiency peaks, the SFR reaches its
maximum (with a slight modulation due the non-constant growth rate). As the halo becomes more massive, the conversion efficiency
decreases again, leading to  lower SFRs, so that the central galaxies stop forming stars. As massive haloes pass through
the peak of the conversion efficiency earlier, their central galaxies' SFR reaches its maximum at a higher redshift than
systems with a lower mass. As a result, massive galaxies typically stop star formation earlier than low mass galaxies.

Finally, the bottom panel gives the stellar mass of each halo's central galaxy $m_\mathrm{*}$, which has been computed by integrating
the SFH. For this illustration only, mergers have been neglected, and a constant stellar mass loss of 40 per cent has
been assumed. Typically, central galaxies in massive haloes have formed most of their in-situ mass at high redshift (e.g. $z\approx 3$
for $M_\mathrm{h}=10^{14}\Msun$), and can then only grow by mergers. On the other side, central galaxies in low mass haloes have
formed most of their stellar mass relatively recently.

\subsection{Satellite galaxies}
\label{sec:modelsatellites}

Having specified the model for isolated, monotonously growing haloes (i.e. for central galaxies), we now have to consider situations where
the halo is losing mass, and eventually merging with a larger halo (i.e. satellite galaxies). In {\sc Emerge}, we make no formal distinction
between haloes and subhaloes (or central and satellite galaxies) based any specific halo radius. Instead, the only input in our model is
whether a halo grows in mass or not. We consider three effects that can impact the growth of galaxies in haloes that have stopped growing:
quenching after the halo has not grown for some time, stripping once the halo has lost a significant fraction of its mass, and merging once
a subhalo has lost its orbital energy. As with the model for growing haloes, we have chosen the simplest model that is able to reproduce
all observational data. However, it would be straight forward to increase the model complexity if new observational data require it.

\subsubsection{Quenching}
\label{sec:quenching}

When a halo starts being accreted by a larger halo its own growth rate begins to decline. Eventually the halo reaches its peak mass,
typically still far outside the virial radius of the larger halo, after which tidal forces strip mass from the halo. The galaxy in such
a halo then experiences a reduced gas infall rate, such that the cold gas reservoir will eventually be used up by star formation, and the
galaxy will be quenched. 

Using a galaxy group catalog from the Sloan Digital Sky Survey, together with cosmological simulation, \citet{Wetzel:2013aa} study the
SFHs and quenching timescales of satellite galaxies. They find a `delayed-then-rapid' quenching scenario in which
the SFRs evolve unaffected for a few Gyr after infall, after which star formation quenches rapidly. In this way, satellites can grow
significantly in stellar mass after infall, nearly identical to central galaxies. Moreover, they find that quenching time-scales are 
shorter for more massive satellites but do not depend on host halo mass.

Using these empirical results, we construct our quenching model as follows: at each time step in the formation history of a halo, we
record its previous maximum virial mass $M_\mathrm{peak}(t)$, and the time at which this mass was reached $t_\mathrm{peak}(t)$. If the
current mass is lower than the peak mass ($M<M_\mathrm{peak}$), we keep the SFR of its galaxy constant, i.e. we use $\dot m
(t_\mathrm{peak}$). After a time $\tau$ has elapsed and the halo mass is still below its previous peak mass, the SFR of the galaxy is set to
0. We parameterise this quenching time with respect to the halo's dynamical time $t_\mathrm{dyn}$ (i.e. proportional to the current
Hubble time), and allow for longer quenching times for low mass satellites: 
\begin{equation}
\label{eqn:satquenching}
\tau = t_\mathrm{dyn} \cdot \tau_0 \left(\frac{m_*}{10^{10}\Msun}\right)^{-\tau_\mathrm{s}}  \; .
\end{equation}
We further require a minimum quenching time of $\tau_0 \cdot t_\mathrm{dyn}$. The satellite quenching model thus has two free
parameters, the normalisation $\tau_0$ describing the quenching time for massive galaxies with a stellar mass of $m_* \ge 10^{10}\Msun$,
and the slope $\tau_\mathrm{s}$ that describes how the quenching time of low mass galaxies changes with stellar mass. Both parameters are
mostly constrained by the observed fraction of quenched galaxies as a function of stellar mass at different redshifts. For a positive
$\tau_\mathrm{s}$, we get longer quenching times for less massive satellites \citep[as found by][]{Wetzel:2013aa}. Using a satellite
mass dependent parameterisation leads to a significantly better fit to the quenched fractions of low mass galaxies (see Appendix
\ref{sec:bayes} for more details). We also tried a model in which we do not keep the SFR constant after the peak halo mass has been
reached, but let it decay exponentially until the quenching time has elapsed (when it is set to 0). However, we did not find any
significant improvement over our standard model indicating that the data do not require this decay (see Appendix \ref{sec:bayes}).
If the halo mass becomes larger than the previous peak mass, the SFR is again determined by the halo growth rate and the instantaneous
conversion efficiency (eqn. $\ref{eqn:sfrcen}$).

\subsubsection{Stripping}
\label{sec:stripping}

While a subhalo orbits within its host halo, strong gravitational tidal forces strip mass from the outer regions of the subhalo,
lowering its own gravitational potential. As the stars in its galaxy are centrally concentrated and more tightly bound than the dark
matter, the stellar mass of the satellite changes only slightly until most of the dark matter has been stripped off. However, at some point,
the halo mass has been lowered enough that it becomes comparable to the mass of the galaxy, and the gravitational potential of the halo is
not strong enough to protect the galaxy from becoming stripped as well. Consequently, once the halo has lost enough mass, the stars in
its centre get tidally stripped to the host halo, where they are added to the stellar halo, or equivalently to the intra-cluster mass (ICM). 

We implement this process by comparing the halo mass at each time to the peak mass of the halo through its history. If the current halo
mass $M$ is smaller than a fraction $f_\mathrm{s}$ of the peak mass $M_\mathrm{peak}$, 
\begin{equation}
\label{eqn:satstripping}
M < f_\mathrm{s} \cdot M_\mathrm{peak}
\end{equation}
we move all stellar mass $m$ within this halo to the ICM, $m_\mathrm{ICM,new}=m_\mathrm{ICM,old}+m$, and then set $m=\dot
m=0$. The stripping parameter is mostly constrained by small-scale clustering, as early (late) stripping leads to less (more) satellites
close to the central, such that the one-halo term of the galaxy two-point correlation function is lowered (enhanced). We also tried an
alternative stripping model in which the galaxy is stripped, once the halo mass has been reduced below a given factor of the stellar mass of
the galaxy in the centre of the halo ($M < f_\mathrm{s} \cdot m$), but this model resulted in a worse fit to the data (see Appendix
\ref{sec:bayes}). 

This is one of two channels by which the ICM can grow (the other being merging). The gravitational potential of a subhalo is too weak to
support its own stellar halo. Thus, once a halo becomes a subhalo, we move all the mass in its ICM to the ICM of its host halo.

\subsubsection{Merging}
\label{sec:merging}

Eventually a subhalo will have lost all of its orbital energy due to dynamical friction and the satellite galaxy will merge with the
central galaxy. We assume that during such a merger a fraction of satellite stars $f_\mathrm{esc}$ can escape from the central galaxy
and ends up in the halo as diffuse stellar material not detected in standard surveys. The merger remnant will then have stellar mass of 
\begin{equation}
\label{eqn:satmerging}
m_\mathrm{rem} = m_\mathrm{cen} + m_\mathrm{sat} \cdot (1-f_\mathrm{esc}) \; ,
\end{equation}
while the ICM of the host halo will grow by $m_\mathrm{ICM,new}=m_\mathrm{ICM,old}+f_\mathrm{esc}
m_\mathrm{sat}$. We treat the escape fraction $f_\mathrm{esc}$ as a free parameter. It is mostly constrained by the evolution of the
massive end of the SMF at low redshift. A low escape fraction will lead to a strong evolution, while a high escape
fraction will lead to little growth in the massive end. Note that the exact value of $f_\mathrm{esc}$ will depend on how the stellar mass
function has been derived. If stellar masses are computed from Petrosian magnitudes, most of the mass in the outskirts of the galaxy
will be seen as belonging to the ICM, resulting in a higher value for the escape fraction. If instead the stellar masses are computed from
fits to the light profiles, less mass will be classified as ICM, so that the escape fraction is slower (see section \ref{sec:smfz0} 
for more discussion).

Due to the finite mass resolution of the simulations, sub-haloes can no longer be identified once tidally stripped below the resolution
limit. Since mass loss can be substantial, this is important even for fairly massive subhaloes, and a special treatment of these so-called
`orphans' becomes necessary. We determine the orbital parameters at  the last moment when a subhalo is identified in the simulation and use
them in the dynamical friction estimate given by \citet*{Boylan-Kolchin:2008aa}, which is applicable at radii
$R<R_\mathrm{v}$. We keep the disrupted subhalo and the associated satellite until the dynamical friction time $t_\mathrm{df}$ has
elapsed, and assume that only then does it truly merge with the main halo. While orbiting, the mean distance between the satellite and the
central decays proportional to $f_\mathrm{dec}=\sqrt{1-\Delta t/t_\mathrm{df}}$, where $\Delta t$ is the time since the subhalo
was last identified \citep[see section 8.1.1 of][]{Binney:1987aa}. We therefore place the orphan galaxy randomly on a sphere around the
central galaxy with a radius of $r=r_0f_\mathrm{dec}$, where $r_0$ is the distance when the subhalo was last identified. To compute if the
galaxy gets tidally stripped, we extrapolate the halo mass linearly. Finally, if a host halo of an orphan merges with a larger
halo and becomes a subhalo itself, we reset the merger clock, i.e. we recompute the dynamical friction time with respect to the new central
galaxy, and assume the orphan will merge with it once this new time has elapsed.

\subsection{Obtaining mock observations}
\label{sec:mocks}

\begin{table}
 \caption{Fitting results from MCMC}
 \label{tab:bestfit}
 \begin{tabular}{@{}lccc@{}}
  \hline
  Parameter & Best-fit & Upper 1$\sigma$ & Lower 1$\sigma$\\
  \hline
  $M_0$ & 11.339 & +0.005 & -0.008\\
  $M_z$ & ~0.692 & +0.010 & -0.009\\
  $\epsilon_0$ & ~0.005 & +0.001 & -0.001\\
  $\epsilon_z$ & ~0.689 & +0.003 & -0.003\\
  $\beta_0$ & ~3.344 & +0.084 & -0.101\\
  $\beta_z$ & -2.079 & +0.127 & -0.134\\
  $\gamma_0$ & ~0.966 & +0.002 & -0.003\\
  \hline
  $f_\mathrm{esc}$ & ~0.388 & +0.002 & -0.002\\
  $f_\mathrm{s}$ & ~0.122 & +0.001 & -0.001\\
  $\tau_0$ & ~4.282 & +0.015 & -0.020\\
  $\tau_\mathrm{s}$ & ~0.363 & +0.014 & -0.014\\
  \hline
  \end{tabular}
 \medskip\\
  \textbf{Notes:} All masses are in units of \Msun.
\end{table}

Having specified the model and its free parameters, we are now able to populate all merger trees of the simulation with galaxies. To
determine the values and uncertainties of our parameters, we need to compare the predictions of the model to observations. As we aim to
avoid degeneracies between model parameters, we choose a set of observations that have a different constraining power for each
parameter. The idea here is to have a subset of observations that mostly constrains one given parameter, while another subset of
observations constrains another parameter, and so forth. 

To constrain the overall evolution of the instantaneous conversion efficiency $\epsilon$ and specifically its characteristic mass ($M_0$
and $M_z$), we use the SMFs $\Phi$ up to $z=8$. The CSFRD \csfrd up to $z=11$ mostly constrains the evolution of the normalisation of $\epsilon$
($\epsilon_0$ and $\epsilon_z$). The sSFRs $\Psi$ for galaxies of different stellar masses up to $z=9$ can constrain the evolution of the slopes of
$\epsilon$ ($\beta_0$, $\beta_z$, and $\gamma_0$). The low redshift evolution of the massive end of the SMF together with the sSFR
of massive galaxies at low redshift put strong constraints on the escape fraction $f_\mathrm{esc}$. The quenching
parameters $\tau_0$ and $\tau_\mathrm{s}$ can be constrained with the fraction of quenched galaxies \fq as a function of stellar mass up to
$z=4$. Finally, galaxy clustering on small scales, i.e. the projected two-point correlation function \wpp for different stellar masses at
small radii, put strong constraints on the stripping parameter $f_\mathrm{s}$. 

We construct our mock SMFs at the same redshift and mass bins as the observations, so that we can directly compare each
individual data point $\Phi_i(m_i,z_i)$ . The CSFRD is computed at each simulation redshift. We interpolate
between these redshifts using cubic splines to get the data points $\csfrd_i(z_i)$ which we compare to an observation at a specific
redshift $z_i$. To compute the sSFR, we bin in stellar mass and redshift, and calculate the mean sSFR
for each grid point. We use a 2d cubic spline interpolation on this grid to get $\Psi_i(m_i,z_i)$, which can be
compared to a specific observed data point at a given redshift and stellar mass. 

We compute the fraction of quenched galaxies $\fq_i(m_i,z)$ directly for the same stellar mass and redshift bins as the used
observations. For that we divide the number of all quenched galaxies that lie in a specific bin by the total number of galaxies in the same
bin. We define a galaxy to be quenched if its sSFR is below a redshift dependent threshold given by
$\Psi<0.3t_\mathrm{H}^{-1}$, where $t_\mathrm{H}$ is the Hubble time at that redshift \citep[see e.g.][]{Franx:2008aa}.  

Finally, we calculate the $z=0$ galaxy two-point correlation function in a given stellar mass bin using kd-trees following
\citet{Moore:2001aa}, which we found to be very effective. Once the tree is constructed, we count the number of pairs in a distance bin
$dd(r)$, and compute the average number of pairs for a random distribution $N_\mathrm{p}(r) = 2 \pi N^2 r^2 \Delta r
L_\mathrm{box}^{-3}$, where $N$ is the total number of galaxies in a stellar mass bin. The real space correlation function is then given by
$\xi(r)=dd(r)/N_\mathrm{p}(r)-1$, and we can compute the projected correlation function $\wpp_i(\rpp_i,m_i)$ at the same projected radii
\rpp as the observations, by integrating $\xi(r)$ along the line of sight:
\begin{equation}
\label{eqn:xi2wp}
\wpp (\rpp,m) = 2 \int_{\rpp}^\infty {\rm d}r \, r \,\xi(r,m) \, \left(r^2-\rpp^2\right)^{-1/2} \;.
\end{equation}

\subsection{Fitting the model parameters}
\label{sec:modelfitting}

\begin{figure}
\includegraphics[width=0.48\textwidth]{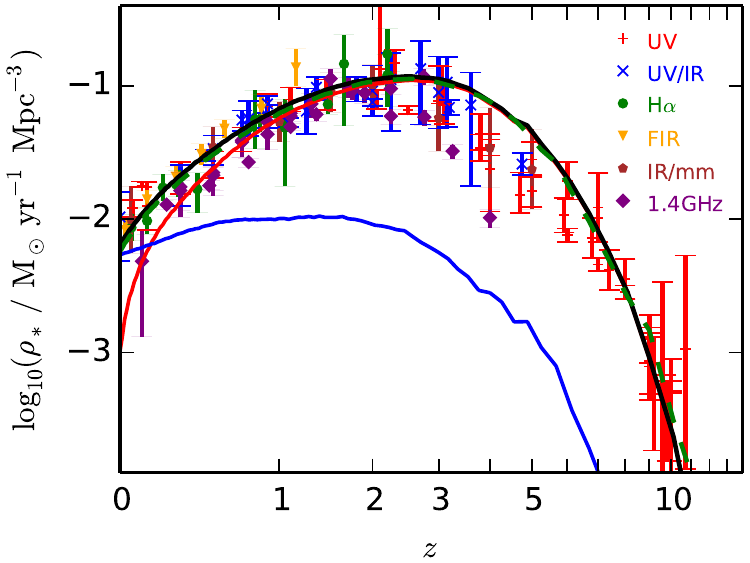}
\caption{Cosmic star formation rate density as a function of
  redshift. The symbols of different colours denote observational
  estimates derived with different methods (c.f. Table \ref{tab:obscsfrd}).
  The black solid line shows the
  model prediction for the best-fit parameters for the $150\Mpc$
  box, while the green dashed line shows the model with the same
  parameters for the $200\Mpc$ box.
  The red/blue solid lines show the contribution from galaxies
  and their progenitors that are quenched/star-forming at $z=0$.
}
\label{fig:csfrd}
\end{figure}

\begin{figure*}
\includegraphics[width=0.99\textwidth]{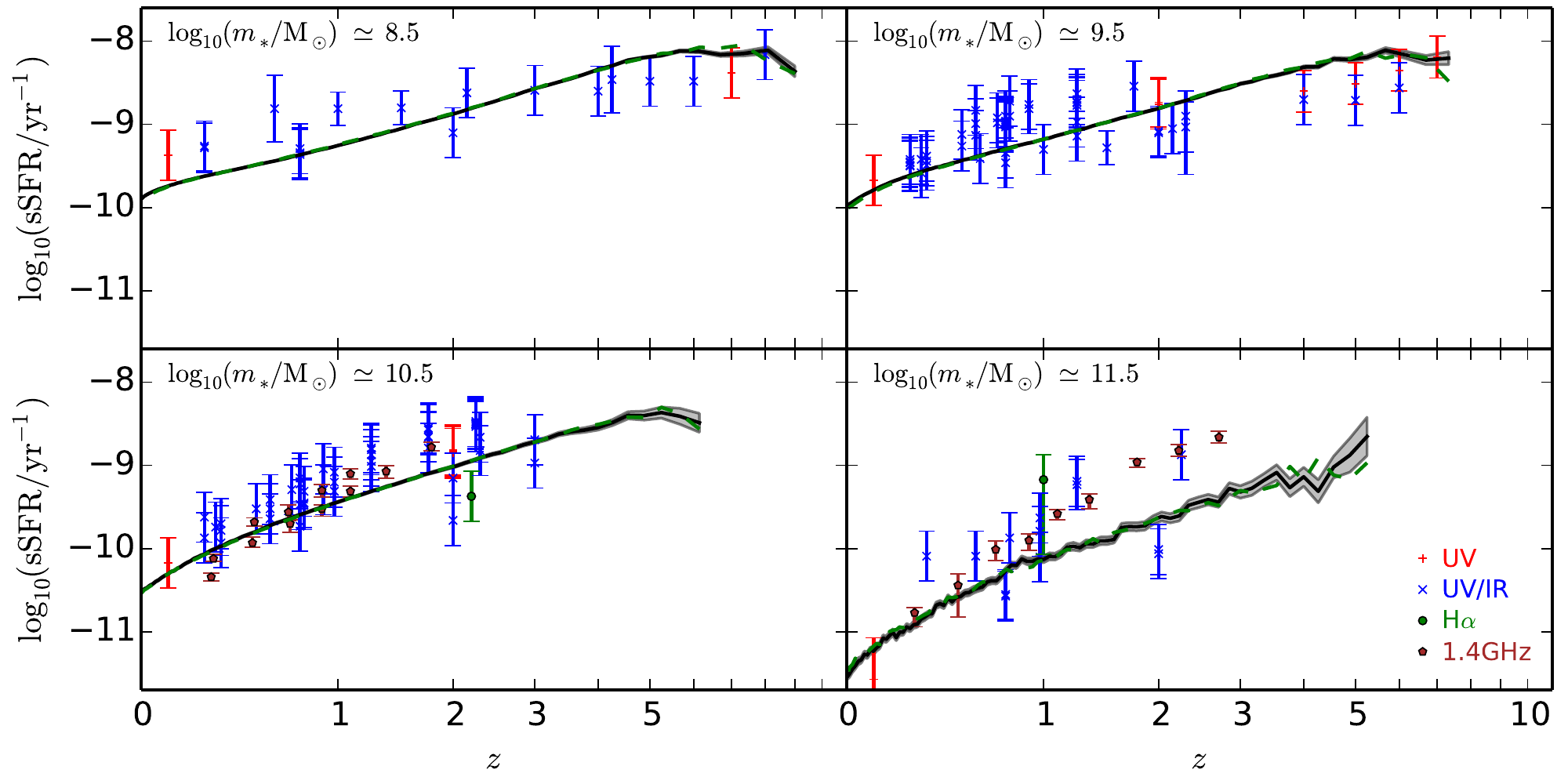}
\caption{Specific star formation rates as a function of redshift for
  galaxies with a given stellar mass (at the corresponding
  redshift). The symbols of different colours denote observational
  estimates derived with different methods (c.f. Table \ref{tab:obsssfr}).
  The black solid line shows the model prediction for the best fit
  parameters for the box with a side length of $150\Mpc$, while the
  green dashed line shows the model with the same parameters for
  the box with $200\Mpc$. The shaded regions correspond to poisson noise.
}
\label{fig:ssfr}
\end{figure*}

For a specific model, i.e. a given set of parameters $\vec \theta$, we can compute the mock observations $\vec \mu (\vec \theta)$ and compare
them to the observations $\vec \omega$ to get the difference $\vec \Delta$: 
\begin{align}
\vec \mu (\vec \theta)& = (\log_{10}\Phi_\mathrm{m},\log_{10}\csfrd_\mathrm{m},\log_{10}\Psi_\mathrm{m},\fq_\mathrm{m},\wpp_\mathrm{m})\\
\vec \omega& = (\log_{10}\Phi_\mathrm{o},\log_{10}\csfrd_\mathrm{o},\log_{10}\Psi_\mathrm{o},\fq_\mathrm{o},\wpp_\mathrm{o})\\
\vec \Delta& = \vec \omega - \vec \mu (\vec \theta) \; .
\end{align}
We can then compute
\begin{equation}
\label{eqn:chi2}
\chi^2 = \vec\Delta^T \, C^{-1} \, \vec\Delta \; ,
\end{equation}
where $C$ is the covariance matrix of the observed data, and assign a likelihood to the model
\begin{equation}
\label{eqn:likelihood}
\mathcal{L} = \exp \left( -\chi^2/2\right) \; .
\end{equation}
If available, we use the full covariance matrix, otherwise we calculate it as $C = \mathrm{diag}(\sigma^2_1,...,\sigma^2_N)$, where $\sigma_i$
is the uncertainty of the $i$th data point.

Having assigned a probability to each possible model, we can now try to find the best-fit model, i.e. the model that maximises the
likelihood (and minimises $\chi^2$), and derive the model uncertainty, i.e. the 1$\sigma$ errors of the parameters. Before sampling the
posterior probability distribution, we found it very effective to first find the best-fit model using a dedicated method. We employed
the {\sc Hybrid} method presented in \citet{Elson:2007aa}, which combines elements of simulated annealing, Markov-Chain Monte Carlo,
and particle-swarm methods. In {\sc Hybrid} a set of chains is run according to the Metropolis-Hastings algorithm, but the step size of
each walker is adjusted based on the ratio of the $\chi^2$ of a walker and the average $\chi^2$ at this step (i.e. how well the walker does
compared to the others). Furthermore, the step size of all walkers is adjusted based on the ratio of the average $\chi^2$ at this step and
the initial average $\chi^2$ (i.e. how well the walkers do compared to their starting position). To find the maximum likelihood, we ran 14
sets of chains starting from different regions in parameter space, each with 30 walkers and 2,000 steps (so 60,000 models each). This was
sufficient, so that all sets of chains found the same minimum. 

Once we have identified the global minimum, we are interested in the posterior probability distribution around this point to derive the
parameter uncertainties. We sample the posterior using the affine invariant ensemble sampler for MCMC presented by
\citet{Goodman:2010aa}. In this method an ensemble of walkers is evolved by proposing a new position for a walker $k$ stretching along
the line to another random walker $j$: $\vec\theta_{k,\mathrm{new}} = \vec\theta_j + Z(\vec\theta_k-\vec\theta_j)$, with the random number
$Z$ drawn from the distribution $g(z)=1/\sqrt{z}$. The new position is accepted with the probability
$q=Z^{N-1}\mathcal{L}(\vec\theta_{k,\mathrm{new}})/\mathcal{L}(\vec\theta_k)$, where $N$ is the number of parameters. After a few autocorrelation
times the current walker positions represent an independent sample of the  posterior distribution. We use 10 ensembles of 100 walkers each,
which we initialise in a tight sphere around the previously found best-fit model, and evolve each walker for 1,000 steps, i.e. 1,000,000
models are computed. At this point the 1$\sigma$ parameter errors derived from the walker distribution have converged. We extract the
errors from the final ensembles.

\subsection{Fitting results}
\label{sec:fittingresults}

The best-fit parameters and their 1$\sigma$ uncertainties are presented in Table \ref{tab:bestfit}. The characteristic halo mass of
the instantaneous conversion efficiency $M_1$ decreases from $1.1\times10^{12}\Msun$ at high redshift to $2.2\times10^{11}\Msun$ at
$z=0$. The peak efficiency decreases from 70 per cent to lower than one per cent at $z=0$. The low mass slope of the efficiency steepens
drastically from 1.3 to 3.3, leading to much lower efficiencies for low mass galaxies at low redshift. The constant high mass slope is
found to be about 1. The escape fraction from mergers is constrained at just under 40 per cent, which is considerably higher than the 20
per cent we assumed in \citet{Moster:2013aa}. The stripping parameter $f_\mathrm{s}$ is slightly larger than 10 per cent, which means that satellite
galaxies get stripped to the ICM once its subhalo's mass has reached a tenth of its peak value. The satellite quenching times are found to be
about 4 dynamical times for massive galaxies with $m\ge10^{10}\Msun$, and considerably longer for galaxies with lower mass (e.g. about 10
dynamical times for galaxies with $m=10^{9}\Msun$). We discuss the correlations between the model parameters in Appendix \ref{sec:cov}.
In the right panel of Figure \ref{fig:box} we show the distribution of galaxies in the simulation
box with $150\Mpc$ side length, calculated with our best-fit model. As in the left panel, the
volume has been remapped into a a sheet with $6\Mpc$ thickness and $750\Mpc$ width.
Each dot corresponds to a galaxy, while the size of the dots corresponds to the stellar mass
and the colour indicates the specific star formation rate as given by the colour bar.
Massive quenched (red) galaxies are clustered at the knots of the cosmic web, while active
(blue) galaxies are preferentially located along the filaments.

\begin{figure*}
\includegraphics[width=0.99\textwidth]{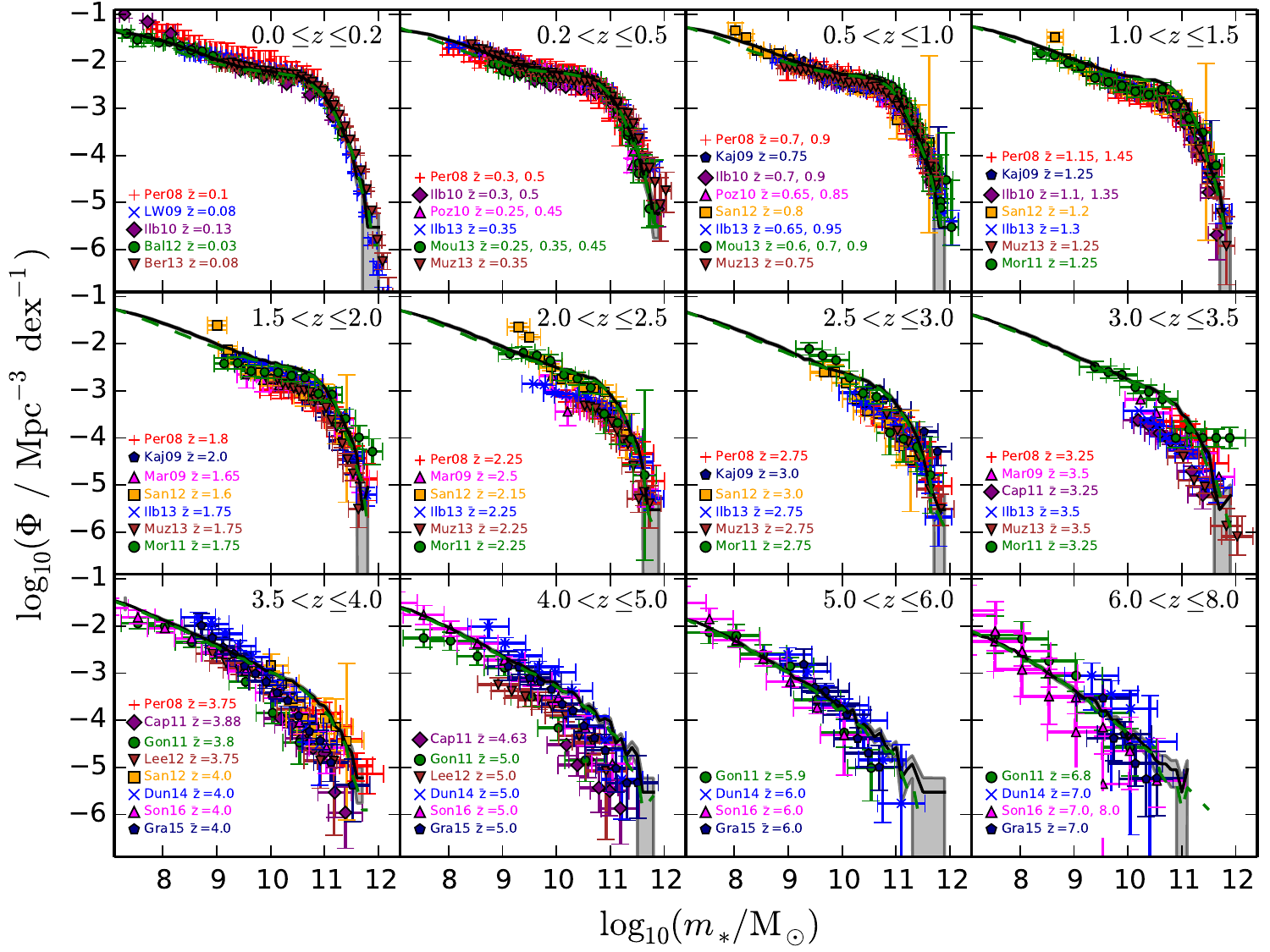}
\caption{Comparison of the model (lines) and the observed stellar mass
  functions (symbols, c.f. Table \ref{tab:obssmf}) for $z=0-8$ as indicated in each panel. The
  model stellar mass functions have been computed with the best fit
  model parameters employing the $150\Mpc$ box (black solid lines)
  and the $200\Mpc$ box (green dashed lines), and the shaded regions
  correspond to poisson noise.
}
\label{fig:smf}
\end{figure*}

We first check how well the best-fit model is able to reproduce the data that has been
used to fit the parameters to judge if the empirical relations we employed are sensible.
In Figure \ref{fig:csfrd} we compare the resulting CSFRD (lines)
to the data that have been used to fit the model (symbols). Symbols of different colours
denote estimates derived with different methods. These data strongly constrain the overall
normalisation of the instantaneous conversion efficiency and its redshift evolution. The
black solid line shows the model results obtained for the box with $150\Mpc$ side length
which has been used to fit the parameters. The green dashed line has been obtained
by running the model with the previously fitted parameters on the box with
$200\Mpc$ side length, which is not only bigger but also better resolved. This
indicates that the model is converged and reproduces the same SFRs
independent of the simulation size and mass resolution. The red/blue
solid lines show the contribution to the CSFRD from galaxies
and their progenitors that are quenched/star-forming at $z=0$. This demonstrates
that the progenitors of local star-forming galaxies played a very minor role in the
peak of cosmic star formation at $z\sim2-3$, while the main contribution to cosmic
star formation beyond $z\ge0.5$ is from galaxies that are now quenched.

The SFRs of galaxies strongly depends on their stellar mass and on redshift.
Figure \ref{fig:ssfr} compares the sSFRs as a function of redshift of
observed galaxies (symbols) to the model results (lines) for four stellar mass bins. For this,
only galaxies have been selected that have the quoted stellar mass at the quoted redshift.
Again, symbols of different colours have been derived with different methods. As these data give the
SFR of low-mass and high-mass galaxies individually, they strongly constrain
the slopes and the characteristic mass of the instantaneous conversion efficiency and their
redshift evolutions. The black solid lines show the best-fit model obtained with the $150\Mpc$
box, while the green dashed lines show the same model run on the $200\Mpc$ box. Also
here, the results do not depend on box size and resolution. The model fits the observed data
very well, including the plateau at high redshift. The SFRs for massive galaxies
at high redshift are considerably lower than the radio observations, but the uncertainty between
different observational methods is relatively large in this range as well.

Integrating the SFRs, taking into account stellar mass loss and merging,
provides us with stellar masses for all galaxies, so that we can compute the stellar mass
functions up to high redshift. We compare the observed SMFs (symbols)
to the model results (lines) in Figure \ref{fig:smf}. These data give strong constraints on
the characteristic mass of the instantaneous conversion efficiency and its redshift evolution.
The solid black lines show the best-fit model for the $150\Mpc$ box and the green dashed
lines show the same model run on the $200\Mpc$ box, indicating that also the stellar masses
are converged. The local SMF is reproduced very accurately, while at
intermediate redshift ($1<z<3$) the model is on the upper end of the observations around
the knee. At high redshift the SMFs are fitted very well. Thus it is possible
to fit both observed SFRs and stellar masses if stellar mass loss and merging
are taken into account. We therefore conclude that there is no tension for the stellar mass
density between integrated SMFs and the integrated CSFRD,
in agreement with previous findings \citep{Moster:2013aa, Behroozi:2013aa}.

\begin{figure*}
\includegraphics[width=0.99\textwidth]{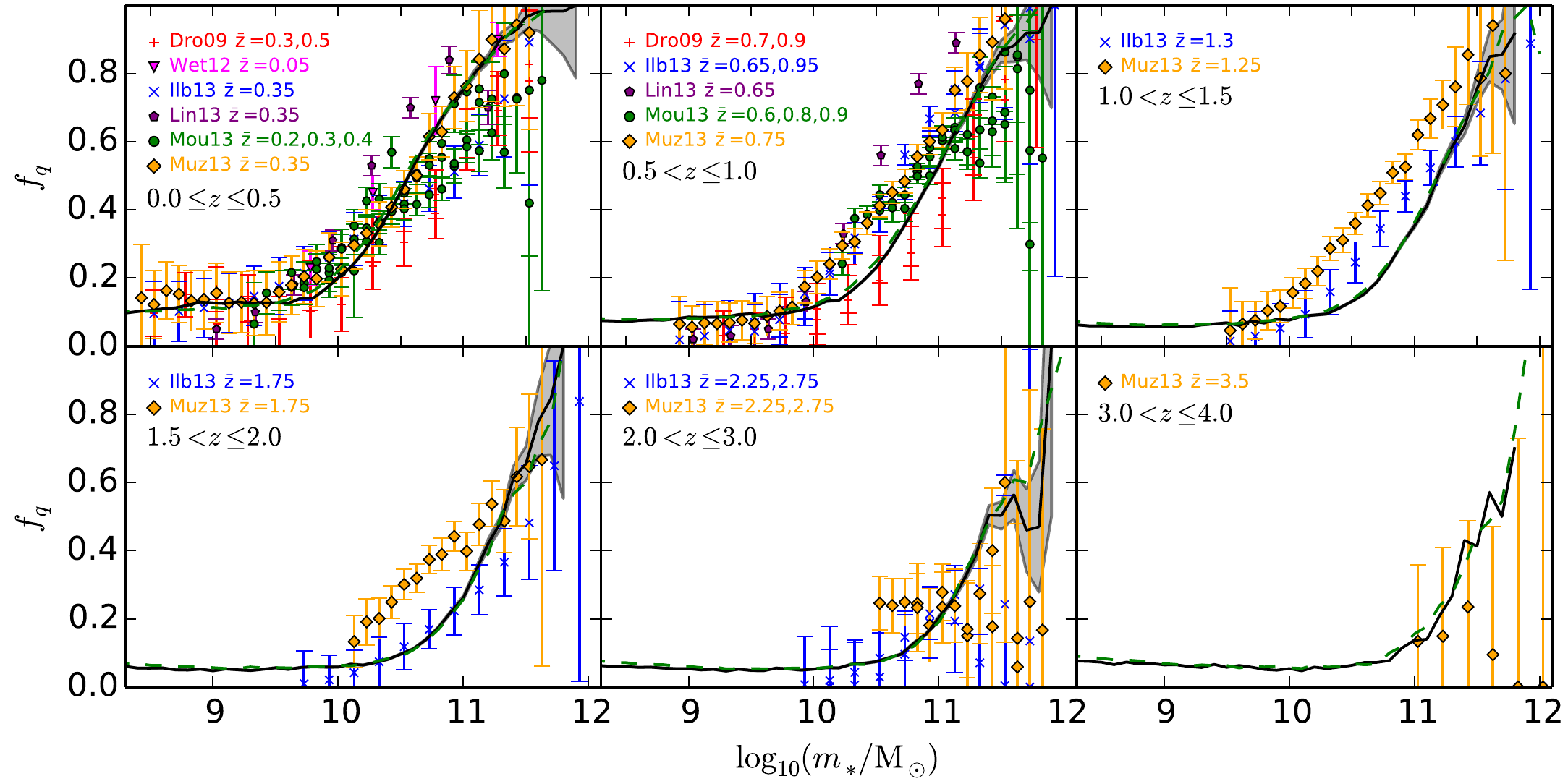}
\caption{Fraction of quenched galaxies as a function of stellar mass
  for six redshift bins from $z=0$ (upper left panel) to $z \sim 4$
  (bottom right panel). The symbols give observational estimates
  (c.f. Table \ref{tab:obsfq}), while the black solid line shows the
  model prediction for the best-fit parameters for the box with a side
  length of $150\Mpc$ and the green dashed line shows the model
  with the same parameters for the $200\Mpc$ box. The shaded
  regions correspond to poisson noise.
}
\label{fig:fq}
\end{figure*}

\begin{figure}
\begin{center}
\includegraphics[width=0.45\textwidth]{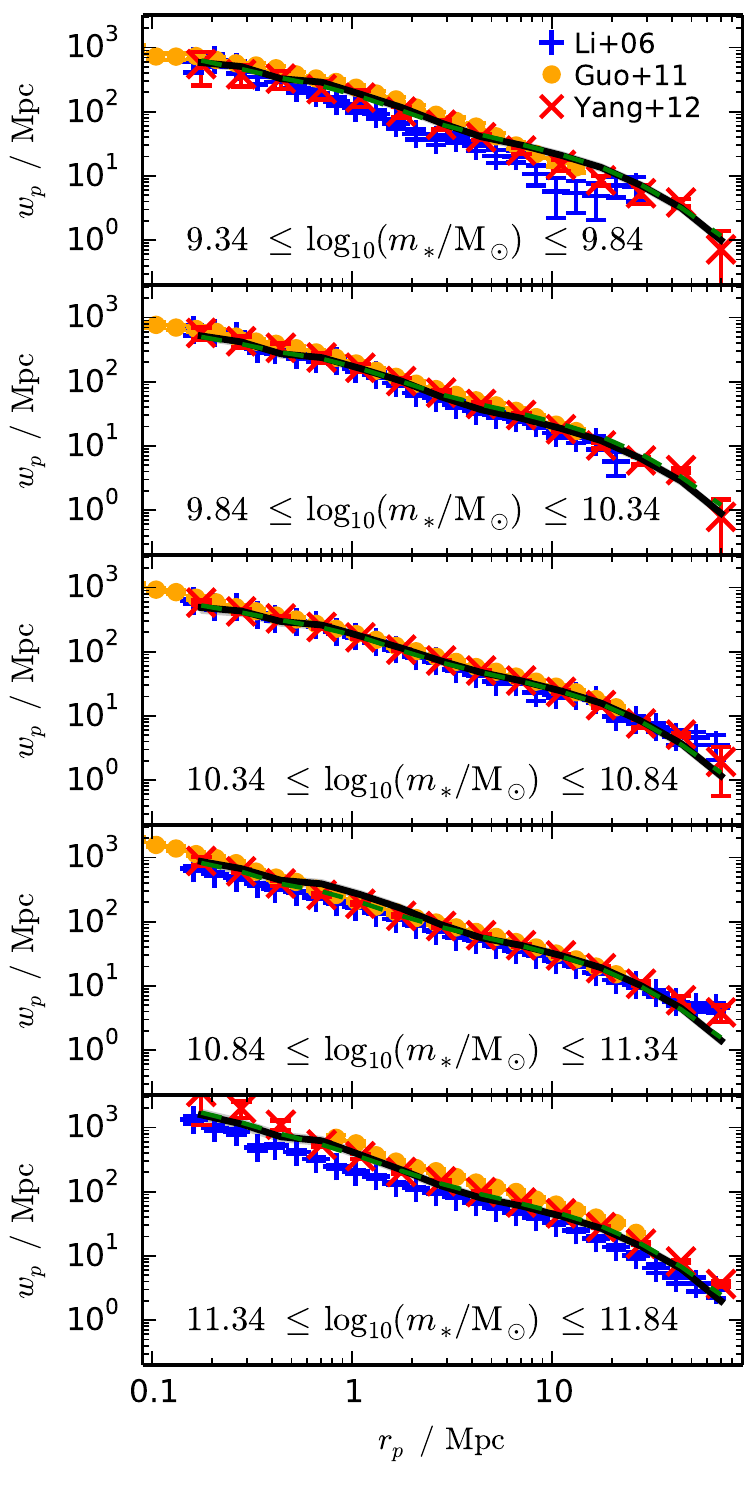}
\caption{Projected galaxy correlation function for five bins of
  increasing stellar mass (from top to bottom). The
  symbols indicate observational estimates \citep{Li:2006aa,Guo:2011aa,Yang:2012aa}, and
  the black and green lines show the model prediction for the best fit
  parameters for the $150\Mpc$ and $200\Mpc$ simulations, respectively.
  The shaded area indicates the $1\sigma$-confidence levels for the model.   
}
\label{fig:wp}
\end{center}
\end{figure}

To constrain the quenching timescale for satellites we have used the fraction of quenched
galaxies as function of stellar mass and redshift. For very short quenching times the quenched
fraction will be too high, while for very long quenching timescales they will be too low. We
compare the observed data (symbols) that have been used in the fit, to the results of the
best-fit model (lines) in Figure \ref{fig:fq}. The solid black lines show the best-fit model for the
$150\Mpc$ box and the green dashed lines show the same model run on the $200\Mpc$ box.
Also for the fraction of quenched galaxies we find that the model has converged. At low
redshift the model provides an excellent fit to the data. At intermediate redshift (around $z\sim1$)
the fraction of quenched galaxies is slightly underpredicted for intermediate mass galaxies
($10^{10}<m_*/\Msun<10^{11}$). This corresponds to the stellar mass and redshift range,
where the model SMF is on the higher side of the data. As the model gives
a very good fit to the observed SFR in this range, we conclude that the
observed SMFs and quenched fractions are well consistent with each other,
but they are in slight disagreement with the observed SFRs. At higher redshift,
the model agrees very well with the observed quenched fractions.

We have used the projected correlation function to constrain the stripping parameter of our
model. If satellite galaxies are stripped very early (i.e. when their subhalo still has a large
fraction of its peak mass), the model correlation function on small scales will be too low
compared to the data, while for late stripping (i.e. the satellite exists until its subhalo has
lost a large fraction of its peak mass) it will be too high. On large scales, beyond the typical
host halo radius for galaxies of a given stellar mass, the projected correlation function is not
affected. In Figure \ref{fig:wp} we show the observed data
that have been used in the fit (symbols), and the results of the best-fit model (lines).
The best-fit model for the $150\Mpc$ box is given by the solid black lines, while the green
dashed lines show the same model run on the $200\Mpc$ box. For both simulations
the model agrees remarkably well with the data. On small scales this is the consequence
of fitting the stripping parameter. However, as on large scales the correlation function is not 
impacted by this, the agreement between data and model is a direct result of assigning
galaxies of a given stellar mass to haloes of the right mass. Galaxy clustering thus directly
follows from halo clustering, which has been found before with simple empirical models, such
as subhalo abundance matching.

\begin{figure*}
\includegraphics[width=0.99\textwidth]{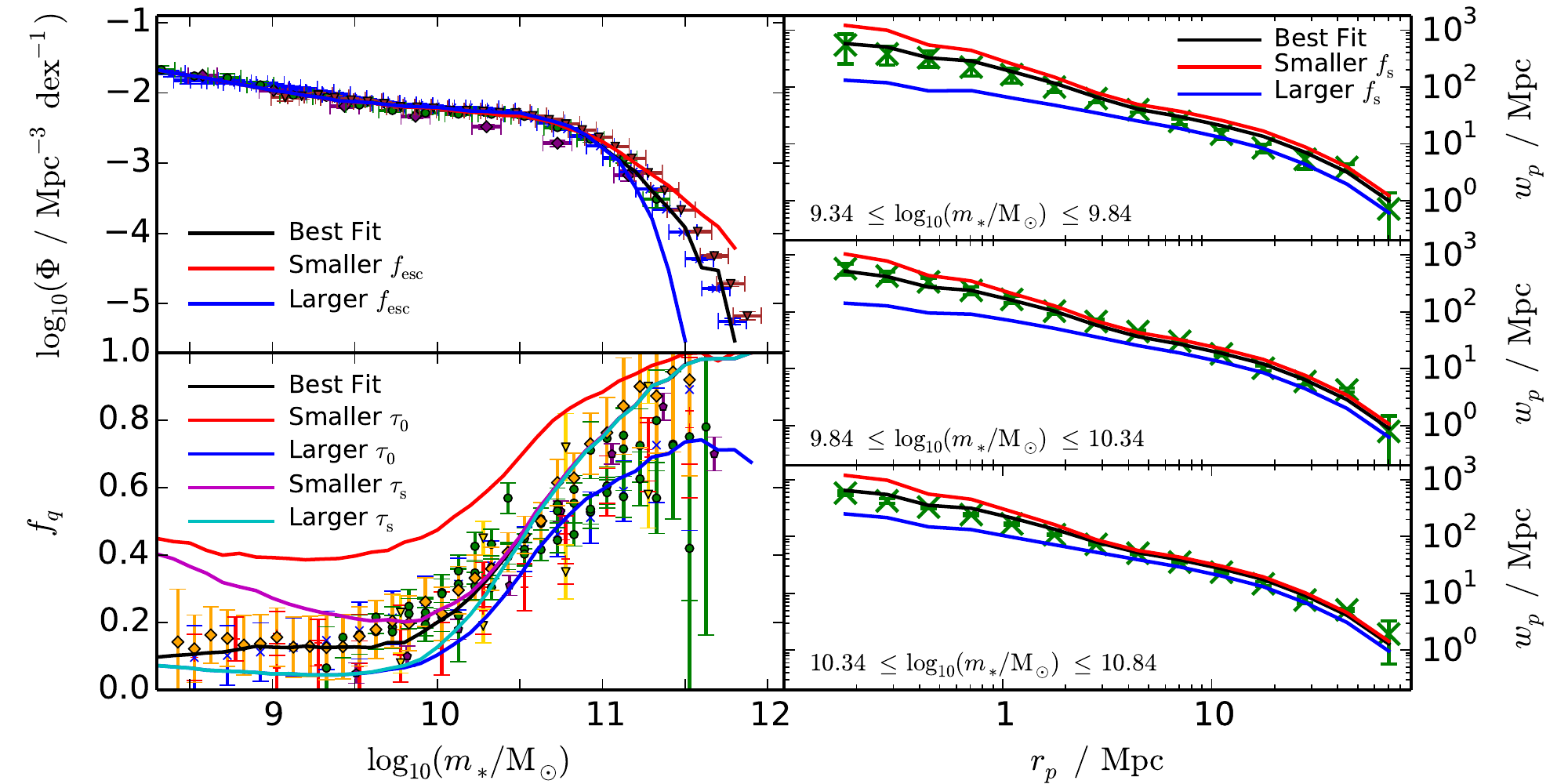}
\caption{Effect of the three parameters that cover the evolution of
  satellite galaxies on different observational predictions. In each
  plot the black line corresponds to the best fit, and the red/blue
  lines show the effect of a smaller/larger value of one parameter,
  while keeping all other parameters fixed. \textit{Top left panel}:
  The stellar mass function at $z=0$ -- a smaller/larger value of
  $f_\mathrm{m}$ leads to more/less massive galaxies compared to the
  observations. \textit{Bottom left panel}: The fraction of quenched
  galaxies at $z=0$ -- a smaller/larger value of $\tau_0$ leads to
  more/less quenched galaxies compared to the
  observations. \textit{Right panels}: The projected correlation
  function for three stellar mass bins at $z=0$ -- a smaller/larger
  value of $f_\mathrm{s}$ leads to more/less small scale clustering
  compared to the observations.}
  \label{fig:satpar}
\end{figure*}

Before we focus on the predictions of the new model and the comparison
to previously published results,
we demonstrate how the different observational data constrain the individual
parameters. While the instantaneous conversion efficiency and its parameters
are fixed by the evolution of the CSFRD (normalisation),
the sSFRs (slopes) and the SMFs (characteristic
halo mass), the satellite parameters are fixed by the fraction of quenched galaxies
(quenching), small-scale clustering (stripping), and the low-redshift evolution of the
massive end of the SMF (merging). In Figure \ref{fig:satpar}, we
illustrate how each satellite parameter is constrained by one specific observation.
In all panels the black lines show the best-fit model and the coloured lines give
the results when the relevant parameter is chosen to have a larger and smaller
value while keeping all other parameters fixed.

The top left panel shows the $z=0$ SMF given different values
for the merging parameter $f_\mathrm{esc}$. A smaller value of $f_\mathrm{esc}$
(red line) leads to more stellar mass ending up in the central galaxy after a merger
(and less mass going into the ICM), so that massive galaxies, where accretion
is a significant growth channel, have even higher stellar masses. This leads to
an over-abundance of massive galaxies compared to the best-fit model and the
observations. On the other hand, a larger value of $f_\mathrm{esc}$ (blue line),
leads to less stellar mass going to the central galaxies and more mass being
expelled into the ICM, such that the galaxies at the massive end grow less
strongly, leading to fewer massive galaxies. As the SFRs of massive
galaxies are directly constrained by observed SFRs, the growth
of the massive end of the SMF directly constrains the merging
parameter leading to $f_\mathrm{esc}=0.388$ such that almost half of the mass
of a satellite escapes into the stellar halo.

The bottom left panel of Figure \ref{fig:satpar} shows the effects on the fraction
of galaxies as function of stellar mass when the quenching parameters are varied.
Keeping all parameters fixed (including the quenching slope $\tau_\mathrm{s}$),
a higher normalisation $\tau_0$ (blue line) leads to longer quenching times for
all satellites. As they can retain star formation for a longer time, this results in lower
quenched fractions compared to the best-fit model (and the data) for all stellar masses.
For a lower normalisation (red line), the satellites become quenched shortly after
their halo reaches peak mass (i.e. when the galaxy becomes a satellite).
Consequently this leads to higher fractions of quenched galaxies for all stellar masses.
The normalisation is thus directly constrained by the quenched fractions,
and we find a best-fit value of $\tau_0=4.282$ which means that in our model, massive
satellites become quenched at about 4 dynamical halo times after their halo reached
its peak mass. At $z=1$ and $z=0$ this corresponds to $\sim4\Gyr$ and
$\sim8\Gyr$, respectively.

A higher value for the quenching slope (cyan line) while keeping all other parameters fixed
(including the normalisation $\tau_0$), leads to longer quenching times for low-mass galaxies.
The fraction of low-mass quenched galaxies therefore is reduced compared to the best-fit
model and the observations. At the massive end this does not change the quenched fraction,
as we have fixed a minimal quenching time of $\tau_0\cdot t_\mathrm{dyn}$ such that the
quenched fraction of massive satellites is only set by the normalisation $\tau_0$. A lower value
for the quenching slope (magenta line) results in shorter quenching times for low-mass galaxies,
increasing the fraction of low-mass quenched galaxies compared to the best-fit model.
Consequently, a flat slope ($\tau_\mathrm{s}=0$), where the quenching time does not depend
on the stellar mass of a satellite, leads to an upturn of the quenched fraction below a stellar mass
of $m_*\approx 10^{10}\Msun$, and to significantly more quenched galaxies than observed.
The slope is thus directly constrained by the quenched fractions at low stellar masses,
and we find a best-fit value of $\tau_\mathrm{s}=0.363$. Thus, a low-mass satellite with
$m_*\approx 10^{8.5}\Msun$ keeps forming stars 3.5 times longer than massive satellites,
corresponding to 14 dynamical halo times after their halo reached its peak mass. While this
equals $\sim30\Gyr$ for satellites of this mass falling in at $z=0$, it only equals $\sim7.8\Gyr$
at $z=2$ so that enough time passes to quench them before $z=0$.

Finally, the right-hand-side panels of Figure \ref{fig:satpar} show the effects on the
projected galaxy correlation function $w_\mathrm{p}(r_\mathrm{p})$ when the stripping
parameter is varied. Each panel shows a different stellar mass bin. A smaller value of
$f_\mathrm{s}$ (red line) means that satellite galaxies are stripped to the main halo
once their subhalo has lost a larger fraction of its peak mass compared to the best-fit
case. Therefore, the satellite has more time to sink closer to the central galaxy due
to dynamical friction. This results in more close pairs and more power on small scales
boosting the one-halo term of the auto-correlation function. Consequently, the projected
correlation function is higher on small scales. On the other hand, a larger value of
$f_\mathrm{s}$ (blue line) means that satellites are stripped when their subhalo has lost
a smaller fraction of its peak mass. They are therefore stripped faster and do not have as
much time to sink closer to the central galaxy resulting in less close pairs and reduced
small-scale power. This leads to a lower projected correlation function on small scales.
The stripping parameter is thus directly constrained by small-scale clustering leading to
$f_\mathrm{s}=0.122$ such that satellites get stripped to the ICM once their subhalo
has been stripped to 12 per cent of its peak mass.

In summary we note that the empirical model fulfils its main purpose:
to follow the stellar content of dark matter haloes over cosmic time
such that all relevant observational data is reproduced. The model
has been specifically designed to do this; for fewer parameters we
are not able to fit all observations simultaneously, while for more
parameters the agreement with the data does not improve significantly
(see appendix \ref{sec:bayes} for more details on the model selection
process). The presented model is therefore the most simple model
that is able to explain the observations, or phrased differently, the
currently available statistical data does not provide any stronger
constraints on galaxy formation than what has been implemented
in this empirical model. When new data becomes available, we
will be able to test if the model is consistent with the new data.
If the model cannot explain these, we can increase the complexity
of the model and in this way use the data to increase our knowledge
about the formation of galaxies. We can also use the empirical
model as a constraint for more detailed hydrodynamical simulations
and in this way test whether our current understanding of the physical
processes that drive galaxy formation is sufficient to explain the
observed data.

\section{Results of the model}
\label{sec:results}

The new empirical method that has been presented in the previous section
was designed to follow the assembly of galaxies based on the assembly of
their dark matter haloes. Its free parameters have been fitted by requiring
that a number of statistical observations be reproduced: SMFs, CSFRDs, sSFRs,
quenched fractions, and small-scale clustering. We now focus on how the stellar
content builds up in detail within different dark matter haloes,
and the resulting SHM ratio for quenched and star-forming galaxies.

\subsection{The evolution of the stellar content}
\label{sec:growth}

\begin{figure*}
\includegraphics[width=0.98\textwidth]{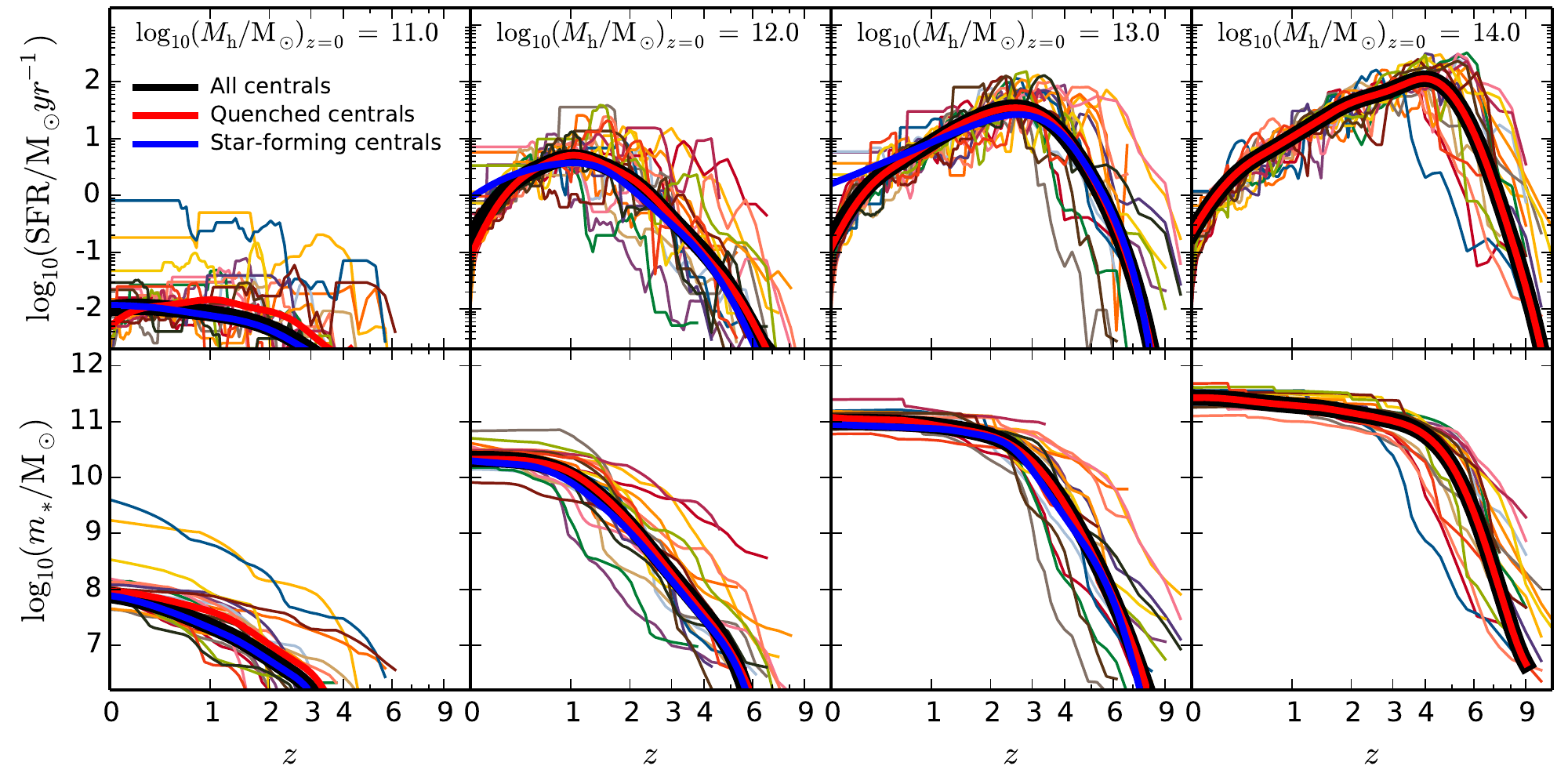}
\caption{Stellar mass build-up in haloes of different mass. {\it Top panels}: 
Star formation rate as a function of redshift for central galaxies in halos of
$10^{11} -10^{14}\Msun$ at $z=0$ (from left to right). Thin lines are tracks
for 20 randomly chosen haloes. Individual tracks can differ significantly in
particular at high redshift and for low present day halo masses. The black,
red, and blue thick lines indicate the ensemble averages for all, quenched
and star-forming centrals, respectively. For higher halo masses star formation
peaks at higher values and higher redshifts.
{\it Bottom panels}: Stellar mass as a function of redshift for the same $z=0$ halo
masses. Lines are the same as in the top panels, i.e. lines of the same colour
represent the same halo.
}
\label{fig:histA}
\end{figure*}

Since the SFR of a galaxy in {\sc Emerge} depends on both
the current mass of its halo (through the efficiency) and its current growth
rate, each galaxy has an individual SFH based on the
growth history of its halo. This means that in haloes that experience most
of their growth early-on, the SFHs also follow this
behaviour with higher SFRs at high redshift compared to
average systems. In the top panels of Figure \ref{fig:histA} we present the resulting
SFHs of central galaxies. Each panel shows the SFR as a
function of redshift for galaxies in haloes of different $z=0$ mass from
$M=10^{11}\Msun$ to $10^{14}\Msun$. The thin lines are individual tracks
for 20 randomly chosen systems for each panel. This shows that although
for any given panel the final halo mass of each system is identical, the
galaxies at the centre of these haloes can have very different SFHs,
especially at high redshift and for low masses.

Even though individual SFRs can differ considerably between
galaxies in haloes with the same $z=0$ mass, we can compute the average
SFHs as function of halo mass. The thick black lines show
the median SFRs at each redshift for all central
galaxies. This confirms the results of \citet{Moster:2013aa} for the dependence of the
average SFHs on halo mass. While for low halo masses the SFRs
peak late at relatively low values, the SFRs of galaxies in massive haloes
peak earlier and at higher values. Typical Milky Way-like galaxies with 
$M(z=0)=10^{12}\Msun$ peak around $z=1$ with SFRs of a few solar masses
per year, while massive systems with $M(z=0)=10^{14}\Msun$ peak around $z=4$
with SFRs of more than $150\Msunpyr$. Individual galaxies however, can peak
significantly earlier or later with much higher or lower SFRs.

While each galaxy in {\sc Emerge} has an individual SFR, we
can still group them into galaxies that are star-forming and quenched at $z=0$,
and investigate how the SFHs differ. The thick blue and red lines
in indicate the median SFHs for central galaxies
that are star-forming and quenched at $z=0$, respectively. For all halo masses, galaxies
that are quenched at $z=0$ had on average a higher peak SFR than galaxies that are
still star forming at $z=0$. This can be explained by the fact that the SFR is directly
connected to the halo growth rate. Galaxies that are quenched today live in haloes
that have low growth rates at low redshift. As by $z=0$ they need to reach the same
virial mass as haloes that have high growth rates at late times, they need to form more
mass at early times resulting in high growth rates at high redshift. Consequently, at early
times the SFRs are higher for the galaxies in these haloes compared to galaxies that have
high SFRs at late times (but low SFRs early-on). Since the peak SFR is reached at high
redshift galaxies with low SFRs at late times had a higher peak SFR.

Integrating the SFHs over cosmic time while taking into account the effects of stellar mass
loss and mergers yields the stellar mass growth histories (eqn. \ref{eqn:mstar}). In the bottom panels of Figure
\ref{fig:histA} we show the resulting growth histories of central galaxies. Each panel shows the stellar mass of
the haloes' central galaxies as a function of redshift for different $z=0$ virial masses. The thin lines again give
individual tracks for the same 20 systems that were presented in the top panels, i.e. lines
of the same colour represent the same halo. In each panel the final halo mass is the same for each halo, but the
stellar mass of their central galaxies can vary significantly between individual systems. This shows that in
{\sc Emerge}, scatter in stellar mass at fixed halo mass is introduced automatically, without the need
to add a random scatter to the average stellar mass as done in standard subhalo abundance matching. This
has several reasons. Integrating the SFR of a halo over time shows that the final stellar mass
just depends on the integrated efficiency. From equation (\ref{eqn:sfrcen}) we get for the final stellar mass:
\begin{equation}
m_*(M) = f_\mathrm{b} \int_0^M \epsilon(M^\prime,z) \, \mathrm{d}M^\prime \, .
\end{equation}
If the instantaneous efficiency is independent of redshift, the integral has the same value for each halo with
the same final mass $M$. However, as the efficiency is redshift dependent, the final stellar mass depends on
which track has been taken through $\epsilon-M$ space (c.f. top panel of Figure \ref{fig:model}). If the
efficiency is higher at high redshift for example, a halo with a lot of early growth will have an enhanced
SFR at high redshift, resulting in a higher total stellar mass compared to a halo that growth late.
Moreover, a different amount of merging satellite galaxies will lead to different final stellar masses
of central galaxies even if the final halo masses are identical.

\begin{figure*}
\includegraphics[width=0.98\textwidth]{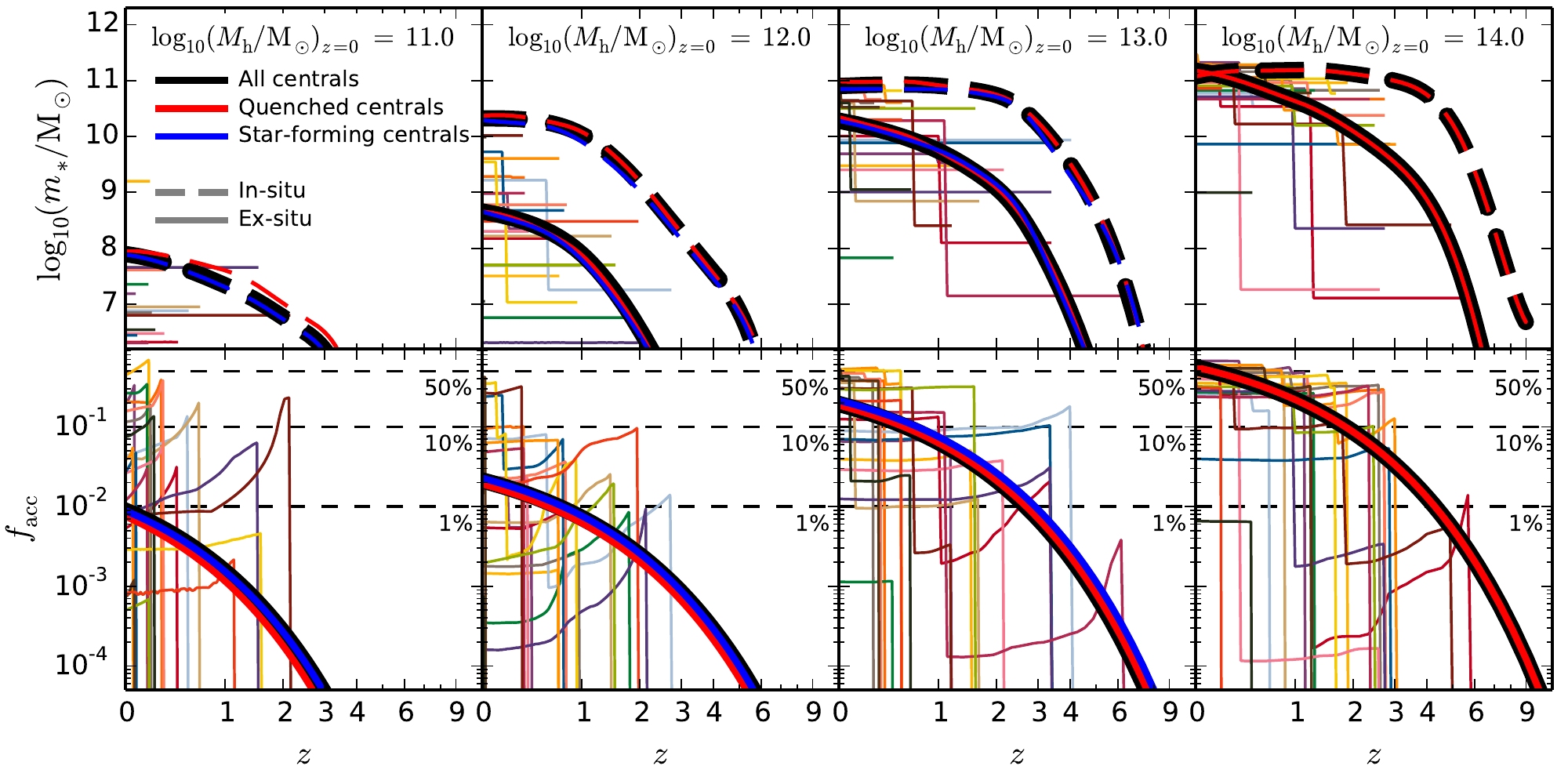}
\caption{Contribution of accreted material to the total stellar mass for haloes of
different mass. {\it Top panels}: Total stellar mass formed in central galaxies by
star formation within the galaxy (in-situ), and by accretion of satellite galaxies (ex-situ).
Thin lines give the accreted stellar mass for 20 randomly chosen systems. In
individual galaxies accreted material is added incrementally. The black, red, and blue
thick lines show the averages for all, quenched and star-forming centrals, respectively.
Dashed and solid lines give the mean amount of stellar mass that has formed in-situ
and ex-situ, respectively. Unlike for individual galaxies, the mean amount of ex-situ
formed stellar mass grows smoothly.
{\it Bottom panels}: Fraction of accreted stellar mass as a function of redshift for the same
$z=0$ halo masses. Tracks for individual galaxies (thin lines) show that mergers are
discrete events. The average fraction increases from $\sim 1$ per cent in $10^{11} \Msun$
haloes to $\sim 50$ per cent in $10^{14} \Msun$ haloes.
}
\label{fig:histB}
\end{figure*}

The median stellar mass growth histories at each redshift for all central galaxies at a given
$z=0$ halo mass are indicated by the thick black lines. Galaxies in massive haloes thus typically grow
by orders of magnitude at early times and then have relatively little growth while galaxies in low mass
halo tend to grow most of their stellar mass late. Still, individual galaxies can oppose this trend, e.g. there
are low-mass galaxy that have most of their mass already in place at $z>4$ and thus have old stellar
populations. Grouping centrals galaxies into systems that are star-forming and quenched at $z=0$
(thick blue and red lines), we find that the final stellar mass at $z=0$ is always larger for quenched
systems compared to systems that are still forming stars. Since passive galaxies formed most of their
stars at high redshift where the conversion efficiency is higher they were able to obtain higher stellar
masses than systems that have formed most of their stars late at lower efficiency.

\begin{figure*}
\includegraphics[width=0.98\textwidth]{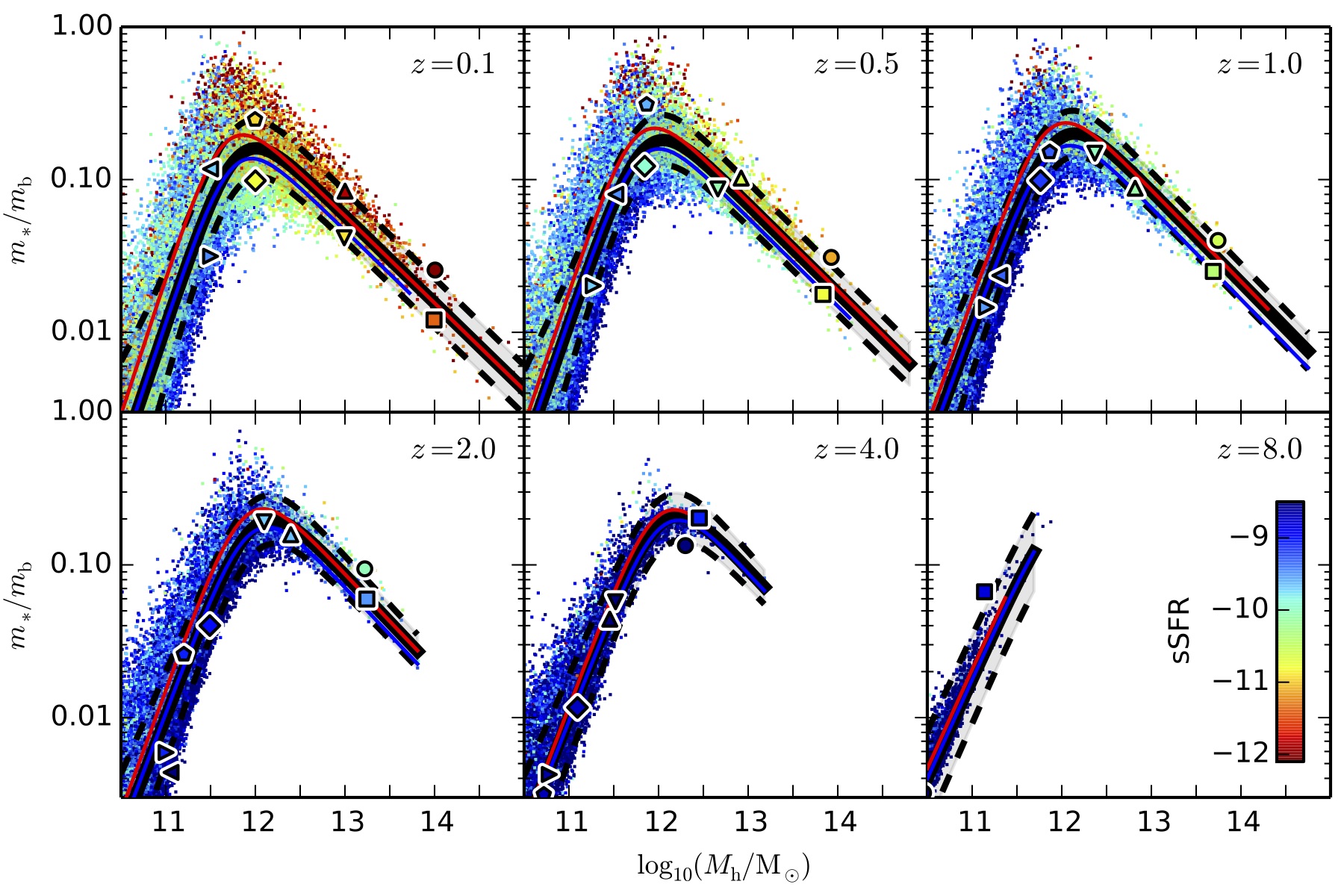}
\caption{Integrated baryon conversion efficiency (ratio between stellar mass $m_*$ and total baryonic mass
$m_\mathrm{b}$) for central galaxies as a function of peak halo mass $M_\mathrm{M}$ at six redshifts from $z=0$
to $z=8$. The colour of each point corresponds to the specific star formation rate ($\mathrm{yr}^{-1}$) of the galaxy as
indicated by the colour bar. The solid black lines show the median conversion efficiency at fixed halo mass and
the dashed black lines indicate the $1\sigma$ scatter. The median conversion efficiencies for quenched and
star-forming galaxies are given by the red and blue lines, respectively. The symbols represent 8 individual
systems that have been selected from the upper and lower $1\sigma$ contours for four halo masses at $z=0.1$.
Identical symbols in different panels represent the same system and the colour represents the specific star
formation rate at this redshift.
}
\label{fig:efficiencyA}
\end{figure*}

As we can trace the evolution of every galaxy, we can investigate how much stellar mass was formed
in-situ, i.e. within each galaxy, and how much was formed in other galaxies that have been accreted.
In the top panels of Figure \ref{fig:histB} we show the stellar mass as a function of redshift for mass
that has formed in-situ (solid lines), and ex-situ (dashed lines), in haloes of different $z=0$ mass. The
thin lines are ex-situ tracks for 20 randomly chosen individual galaxies for each panel. As the accreted material
is added instantly to the central galaxy during mergers the ex-situ formed mass increases incrementally.
The average amount of accreted stellar mass is given by the thick solid lines, and the average stellar mass
that has formed within the galaxy is given by the thick dashed lines. The amount of in-situ formed stellar mass
is generally much larger than the accreted mass. Only in very massive haloes with $M(z=0)\gtrsim10^{14}\Msun$
the final accreted mass exceeds the mass formed within the galaxy. We also note that at late times the
in-situ formed mass decreases with time due to stellar mass loss from dying stars and low SFRs.
Dividing these two components between star-forming and quenched central galaxies, we do not notice
any significant variation. Both groups show very similar amounts of in-situ and ex-situ formed stellar mass.

\begin{table}
 \caption{Fitting function parameters for star formation and accretion histories}
 \label{tab:fittingfunctions}
 \begin{tabular}{@{}crrrrrr@{}}
  \hline
$\log_{10} M$ & $\Psi_1$ & $\Psi_2$ & $\Psi_3$ & $\Psi_4$ & $f_1$ & $f_2$\\
  \hline
  \hline
  \multicolumn{7}{c}{All Galaxies}\\
  \hline
  \hline
11.0 & 73.29 & 0.28 & 1.07 & -1.87 & 0.76 & 1.29\\
12.0 & 3.87 & 5.94 & 1.46 & 4.94 & 1.08 & 0.06\\
13.0 & 2.24 & 3.98 & 1.86 & 11.69 & 1.10 & 0.58\\
14.0 & 1.44 & 3.58 & 1.99 &16.26 & 0.94 & 1.43\\
  \hline
  \hline
  \multicolumn{7}{c}{Quenched Galaxies}\\
  \hline
  \hline
11.0 & 147.50 & 2.50 & 0.89 & -1.99 & 1.87 & 0.05\\
12.0 & 7.13 & 7.43 & 1.43 & 4.88 & 1.12 & 0.06\\
13.0 & 2.58 & 4.14 & 1.86 & 11.72 & 1.09 & 0.55\\
14.0 & 1.44 & 3.58 & 1.99 &16.26 & 0.94 & 1.43\\
  \hline
  \hline
  \multicolumn{7}{c}{Star-forming Galaxies}\\
  \hline
  \hline
11.0 & 63.00 & -0.04 & 1.04 & -2.02 & 1.06 & 0.01\\
12.0 & 0.98 & 3.02 & 1.54 & 5.03 & 1.09 & 0.07\\
13.0 & 0.51 & 1.96 & 2.45 & 15.76 & 1.05 & 0.61\\
  \hline
\end{tabular}
 \medskip\\
  \textbf{Notes:} Columns are halo mass at $z=0$ in \Msun (1), fitting parameters for the star formation histories (2-5), and
  fitting parameters for the accreted fractions (6-7).
\end{table}

To compare the contributions of stellar mass formed in-situ and ex-situ further we present the fraction of
accreted stellar mass as a function of redshift for the same $z=0$ halo masses in the bottom panels of
Figure \ref{fig:histB}. The thin lines show the tracks for the same 20 randomly selected individual galaxies
as in the top panels. As mergers are discrete events, the accreted fraction increases instantaneously and
then decreases again smoothly due to in-situ star formation. This process can be repeated several times,
especially in massive haloes. The average fraction of accreted stellar mass is given by the thick black lines.
We find that in low mass haloes the amount of accreted stellar mass is negligible. For a halo with 
$M(z=0)=10^{11}\Msun$ only $\sim 1$ per cent of the total stellar mass at $z=0$ has been accreted. This
fraction increases with halo mass but is still just $\sim 2$ per cent at $z=0$ for a typical Milky Way-like
galaxy with $M(z=0)=10^{12}\Msun$. Only in massive haloes and at late times, the fraction of accreted
stellar mass becomes significant. At $z=0$ an average halo with a mass of $M(z=0)=10^{13}\Msun$ has
an accreted fraction of $\sim 20$ per cent, and an average halo with $M(z=0)=10^{14}\Msun$ even has
accreted more than 50 per cent of its total mass. Interestingly though, individual galaxies can have drastically
different accreted fractions. There are several galaxies in low mass haloes with accreted fractions of more
than 10 per cent and up to 50 per cent, while in massive haloes we find a few galaxies that have accreted
less than1 per cent of their total stellar mass. We also group the galaxies into star-forming and quenched
systems (blue and red lines), but we find no significant difference between them.

We find that the SFHs $\Psi(z)$ and the accreted fractions $f_\mathrm{acc}(z)$ for central
galaxies in haloes with a given $z=0$ virial mass can be well approximated by the following fitting functions:
\begin{eqnarray}
\log \Psi(z) &=& -\log\left[ \Psi_1 (z+1)^{-\Psi_2} + \mathrm{e}^{\Psi_3(z+1)-\Psi_4} \right]  ,\\
f_\mathrm{acc}(z) &=& f_2 \exp\left[-f_1 (z+1)\right] \, .
\end{eqnarray}
The fitting parameters for all, star-forming, and quenched centrals as function of their $z=0$ virial mass
are presented in Table \ref{tab:fittingfunctions}.

\subsection{The integrated conversion efficiency} 
\label{sec:shm}

The new empirical model describing the growth of galaxies in dark matter haloes is based on the
instantaneous baryon conversion efficiency $\epsilon(M,z)$. Together with the growth rate of
dark matter haloes it was used to derive SFRs for galaxies at each time, which
were then integrated to get stellar masses. As we have shown that this leads to a range of stellar
masses at a fixed halo mass, this will result in scatter in the SHM relation $m(M)$
and consequently in the integrated baryon conversion efficiency, which is defined as
$\epsilon_\mathrm{int}(M) = m/m_\mathrm{b} = m/(f_\mathrm{b}M)$, where $m_\mathrm{b}$ is the
total baryonic mass in a halo and $f_\mathrm{b}$ is the universal baryon fraction. In Figure
\ref{fig:efficiencyA} we plot the integrated conversion efficiency as a function of peak halo mass for each
individual central galaxy in the simulation box. Each panel corresponds to a different redshift from $z=0.1$
(top left) to $z=8$ (bottom right). The colour of each point gives the sSFR of each galaxy ranging
from star-forming (blue) to quenched (red) as indicated by the colour bar. The solid lines show the
average conversion efficiency and the dashed lines indicate the standard deviation ($1\sigma$
scatter).

The resulting average integrated conversion efficiencies for centrals are in very good agreement with previous
results for the SHM ratio from subhalo abundance matching. At $z=0.1$ the maximum
integrated conversion efficiency is 17 per cent -- slightly lower than what has been found in previous
empirical models -- and it increases to 20 per cent at $z=4$.
The corresponding halo mass at the peak is $\log_{10}(M/\Msun)=12$
at $z=0.1$, increasing to $\log_{10}(M/\Msun)=12.25$ at $z=4$. The low-mass slope is quite steep at
low redshift and becomes shallower at high redshift. The high-mass slope is shallower and does
not depend on redshift. We find that the majority of galaxies in massive haloes at $z=0.1$ are
quenched, while at low halo masses most galaxies are star-forming. However, there are also
active galaxies in massive haloes if the halo has a large growth rate at late times, and passive
galaxies in low-mass haloes, typically in haloes that have stopped growing. At high redshift, all central
galaxies are actively forming stars.

\begin{figure}
\includegraphics[width=0.48\textwidth]{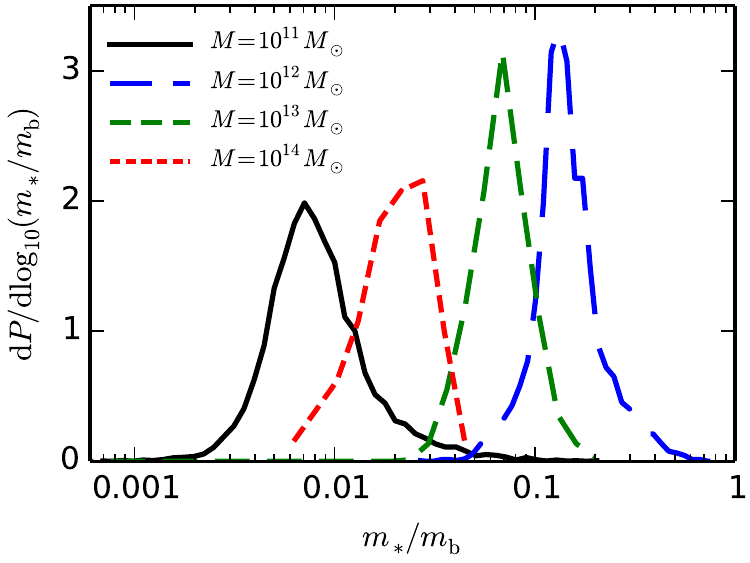}
\caption{Probability density function of the integrated baryon conversion efficiency $m/m_\mathrm{b}$
at $z=0.1$. Each line is for a different fixed peak halo mass: $M=10^{11}\Msun$ (black), $10^{12}\Msun$
(blue), $10^{13}\Msun$ (green), $10^{14}\Msun$ (red).
}
\label{fig:effslice}
\end{figure}

\begin{table}
 \caption{Fitting function parameters for the integrated baryon conversion efficiency and scatter}
 \label{tab:shm}
 \begin{tabular}{@{}rrrrrrrr@{}}
  \hline
$z$ & $M_1$ & $\epsilon_\mathrm{N}$ & $\beta$ & $\gamma$ & $M_\sigma$ & $\sigma_0$ & $\alpha$\\
  \hline
  \hline
  \multicolumn{8}{c}{All Centrals}\\
  \hline
  \hline
0.1 & 11.80 & 0.14 & 1.75 & 0.57 & 10.80 & 0.16 & 1.00\\
0.5 & 11.85 & 0.16 & 1.70 & 0.58 & 10.70 & 0.14 & 0.90\\
1.0 & 11.95 & 0.18 & 1.60 & 0.60 & 10.60 & 0.12 & 0.75\\
2.0 & 12.00 & 0.18 & 1.55 & 0.62 & 10.50 & 0.10 & 0.50\\
4.0 & 12.05 & 0.19 & 1.50 & 0.64 & 10.40 & 0.08 & 0.40\\
8.0 & 12.10 & 0.24 & 1.30 & 0.64 & 10.30 & 0.02 & 0.10\\
  \hline
  \hline
  \multicolumn{8}{c}{Quenched Centrals}\\
  \hline
  \hline
0.1 & 11.65 & 0.17 & 1.80 & 0.57 & 10.00 & 0.14 & 0.55\\
0.5 & 11.75 & 0.19 & 1.75 & 0.58 &  9.90 & 0.12 & 0.45\\
1.0 & 11.85 & 0.21 & 1.65 & 0.60 &  9.80 & 0.08 & 0.40\\
2.0 & 11.90 & 0.21 & 1.60 & 0.62 &  9.70 & 0.07 & 0.35\\
4.0 & 12.00 & 0.21 & 1.55 & 0.64 & 9.60 & 0.06 & 0.30\\
8.0 & 12.10 & 0.28 & 1.30 & 0.64 & 9.50 & 0.04 & 0.20\\
  \hline
  \hline
  \multicolumn{8}{c}{Star-forming Centrals}\\
  \hline
  \hline
0.1 & 11.75 & 0.12 & 1.75 & 0.57 & 10.35 & 0.20 & 1.10\\
0.5 & 11.80 & 0.14 & 1.70 & 0.58 & 10.25 & 0.10 & 0.50\\
1.0 & 11.90 & 0.15 & 1.60 & 0.60 & 10.15 & 0.08 & 0.45\\
2.0 & 11.95 & 0.16 & 1.55 & 0.62 & 10.05 & 0.05 & 0.35\\
4.0 & 12.05 & 0.18 & 1.50 & 0.64 & 9.95 & 0.03 & 0.30\\
8.0 & 12.10 & 0.24 & 1.30 & 1.64 & 9.85 & 0.02 & 0.10\\
  \hline
  \hline
  \multicolumn{8}{c}{All Galaxies}\\
  \hline
  \hline
0.1 & 11.78 & 0.15 & 1.78 & 0.57 & 10.85 & 0.16 & 1.00\\
0.5 & 11.86 & 0.18 & 1.67 & 0.58 & 10.80 & 0.14 & 0.75\\
1.0 & 11.98 & 0.19 & 1.53 & 0.59 & 10.75 & 0.12 & 0.60\\
2.0 & 11.99 & 0.19 & 1.46 & 0.59 & 10.70 & 0.10 & 0.45\\
4.0 & 12.07 & 0.20 & 1.36 & 0.60 & 10.60 & 0.06 & 0.35\\
8.0 & 12.10 & 0.24 & 1.30 & 0.60 & 10.40 & 0.02 & 0.30\\
  \hline
\end{tabular}
 \medskip\\
  \textbf{Notes:} Masses are in \Msun. Parameters are given for the relations as function of
  peak halo mass through history.
\end{table}

The symbols in each panel of Figure \ref{fig:efficiencyA} give the conversion efficiency of 8 individual
systems that have been selected from the upper and lower $1\sigma$ contours for 4 halo masses at $z=0.1$.
Identical symbols in different panels show these systems at higher redshift and the colour indicates
their sSFR. Interestingly, galaxies that reside at the upper or lower $1\sigma$ level at $z=0.1$ have not
been there throughout their evolution, but have moved there from different, typically more average, efficiencies.
Massive systems were even at opposite $1\sigma$ level at high redshift: the high efficiency system with
$\log_{10}(M/\Msun)=14$ and low sSFR at $z=0.1$ (circle) started as a low efficiency system with high sSFR
at $z\gtrsim4$, while the low efficiency system with the same halo mass and higher sSFR at $z=0.1$ (square)
had a very high efficiency and lower sSFR during its early evolution. This can be understood as a result of the
connection between SFR and halo growth rate. A system with a high halo growth rate early-on has a high SFR
as well, but in order to reach the same halo mass at low redshift as other systems, the growth rate at late times
and consequently the SFR must be low. Since the instantaneous conversion efficiency is higher at high redshift,
systems with a high growth and SFRs at early times form more stars than systems with the same
final halo mass but low growth and SFRs at high redshift.

Separating central galaxies into star-forming and quenched systems at any given redshift shows that the average
integrated baryon conversion efficiency of quenched galaxies (red lines) is higher than that of star-forming galaxies
(blue lines). Consequently, this means that at fixed halo mass, passive galaxies have a higher stellar mass than
active galaxies. This trend is most pronounced at low redshift. We can understand this behaviour as a result of linking
SFR and halo growth rate, as well. Passive galaxies live in haloes that have little growth now but experienced high
growth at early times, and thus had high SFRs then. As the instantaneous conversion efficiency was higher at high
redshift, these galaxies were able to form more stars compared to other galaxies with the same final halo mass. 

Naively, this result seems to contradict the finding of weak lensing studies \citep[e.g.][]{Mandelbaum:2006aa,
Mandelbaum:2016aa} that at fixed stellar mass the average mass of haloes harbouring passive galaxies
is higher than that of haloes hosting active galaxies. However, one cannot simply invert the average $m(M)$
relation to obtain the $M(m)$ relation, because of the scatter. As a consequence of the Eddington bias and the larger
fraction of active galaxies in low-mass systems, the average halo mass at fixed stellar mass is higher for quenched
systems. We will explore this in more detail in future work.

Another interesting feature of the relation between halo mass and the integrated conversion efficiency is the scatter at
fixed halo mass, which arises naturally as a consequence of different haloes taking different paths through the
$\epsilon(M,z)$ diagram. From the distribution of dots in Figure \ref{fig:efficiencyA} one can get the impression that
there is an large amount of scatter for low halo masses, especially at low redshift. However, the standard deviation
(dashed black lines) is only a little larger at low masses than at high masses. To investigate this further we show the
probability density function of the integrated conversion efficiency for galaxies in four different halo masses at $z=0.1$
in Figure \ref{fig:effslice}. The distributions of the efficiency are quite narrow in massive haloes and somewhat broader
in low mass haloes. Moreover, while in massive haloes the distributions are cut off at $\gtrsim2\sigma$, there is a longer
tail in the low mass haloes, so that a few of these haloes host massive galaxies. This explains the slightly lower
efficiencies compared to the one derived with subhalo abundance matching. Due to the form of the SMF, this increases
the number of galaxies with higher stellar mass (Eddington bias). Consequently, the overall integrated efficiency needs
to be slightly lower to accommodate for the higher number of more massive galaxies. To confirm this, we have performed
a simple abundance matching experiment: both the halo and the stellar masses of our model catalogue including also
subhaloes and satellite galaxies have been rank ordered independently and then matched one-to-one. The resulting
efficiency is higher at all halo masses, and we find a peak efficiency at $z=0.1$ of 19 per cent, in good agreement with
previous results from abundance matching. However, we stress that if the self-consistent scatter is taken into account 
this is reduced to a peak efficiency of 17 per cent.

\begin{figure}
\includegraphics[width=0.48\textwidth]{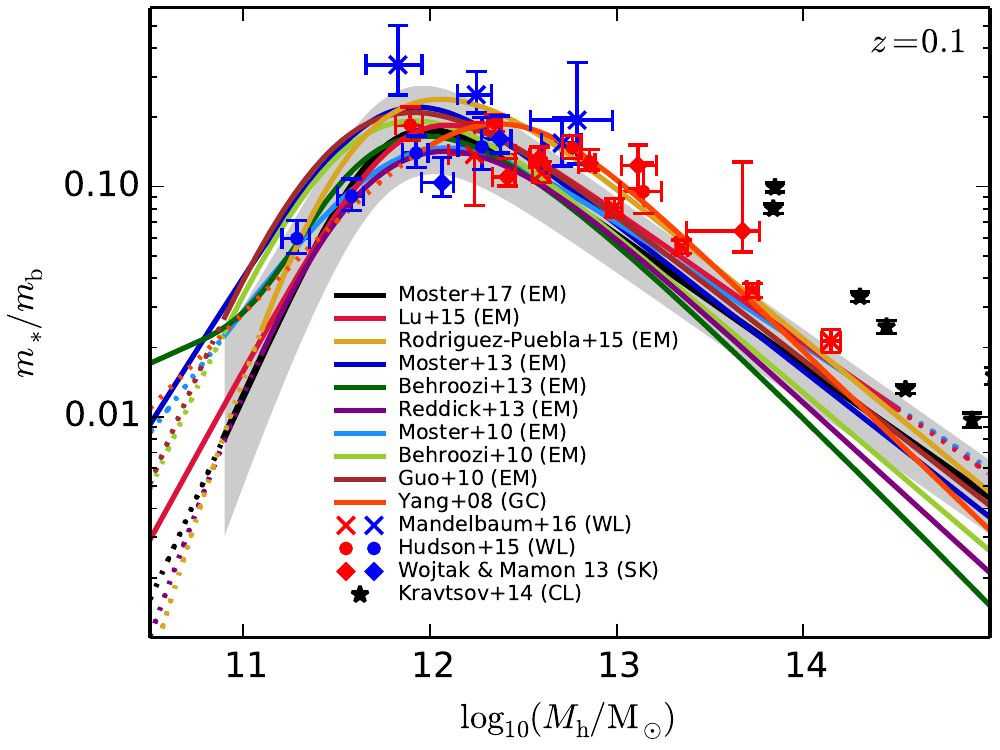}
\caption{Comparison of the average conversion efficiency of all galaxies at $z=0.1$ to previously published
results. This includes empirical models \citep{Moster:2010aa,Behroozi:2010aa,Guo:2010aa,Moster:2013aa,
Behroozi:2013aa,Reddick:2013aa,Lu:2015aa,Rodriguez-Puebla:2015aa}, a galaxy group catalogue
\citep*{Yang:2008aa}, X-ray observations of clusters \citep{Kravtsov:2014aa}, and results for active and passive
galaxies (blue and red symbols) from weak lensing \citep{Mandelbaum:2016aa,Hudson:2015aa} and from
satellite kinematics \citep{Wojtak:2013aa}. The shaded region corresponds to the $1\sigma$ confidence levels
of our model. 
}
\label{fig:shm}
\end{figure}

We find that at a given redshift the integrated baryon conversion efficiency can be well approximated by a
double-power-law (eqn. \ref{eqn:epsilon}). In Table \ref{tab:shm} we present the values of the parameters
for all centrals, quenched and star-forming centrals, and all galaxies at 6 different redshifts. Note that the
parameters have been obtained with respect to the peak mass a halo had up to the given redshift.
Furthermore we find that the logarithmic scatter (in dex) can be well approximated by
\begin{equation}
\sigma = \sigma_0 + \log_{10}\left[\left(\frac{M}{M_\sigma}\right)^{-\alpha}+1\right] \, .
\label{eqn:sigma}
\end{equation}
The fitted values for these parameters are also shown in Table \ref{tab:shm}.

We show a comparison of our model result for the integrated baryon conversion efficiency at $z=0.1$ to
previously published stuff in Figure \ref{fig:shm}. All stellar and halo masses have been converted to our
definitions. Overall, there is a good agreement between the different methods. However, there are some
notable differences. The conversion efficiency of our model at the low-mass end is lower than the results
that have been obtained with subhalo abundance matching \citep{Moster:2010aa,Behroozi:2010aa,
Guo:2010aa,Moster:2013aa,Behroozi:2010aa}. This can be understood as the consequence of the scatter
in the relation. While in abundance matching a constant log-normal scatter for all halo masses is typically
assumed, the scatter in our new model results from different formation histories of the haloes, which leads
to an increased scatter with a tail towards higher stellar masses in low-mass haloes (c.f. Figure \ref{fig:effslice}).
As explained above, this results in a lower efficiency, especially for low-mass haloes. Interestingly, the
empirical model by \citet{Lu:2015aa}, which does not add scatter artificially, also predicts lower conversion
efficiencies for low halo masses.

At the massive end, our model agrees very well with other empirical models, though it predicts comparably
high conversion efficiencies. Still, compared to direct methods, all empirical models predict rather low
efficiencies. One reason for this discrepancy is that weak lensing \citep{Hudson:2015aa,Mandelbaum:2016aa}
and satellite kinematics studies \citep{Wojtak:2013aa} measure halo mass for galaxy populations with
fixed stellar mass. Because of the scatter in the relation it cannot simply be inverted. Another reason is the
derivation of the stellar mass. Studies specifically targeting massive systems \citep*[e.g.][]{Kravtsov:2014aa}
integrate the surface brightness up to large radii or use a fitting function, while for the SMFs used
in the empirical models typically Petrosian magnitudes are used. At the massive end this can lead to a
discrepancy as a significant fraction of the light can be outside the aperture. However, it is unclear how far
the surface brightness should be integrated and how much light should be included to derive a galaxy's
stellar mass. To some degree it is ambiguous if the light (or mass) belongs to the galaxy or if it is part of
the ICM. It is more important that if a comparison is made, the definitions should be the same, which is
unfortunately not always easy to achieve. We refer to section \ref{sec:smfz0} for more discussion on this topic.

\begin{figure}
\includegraphics[width=0.48\textwidth]{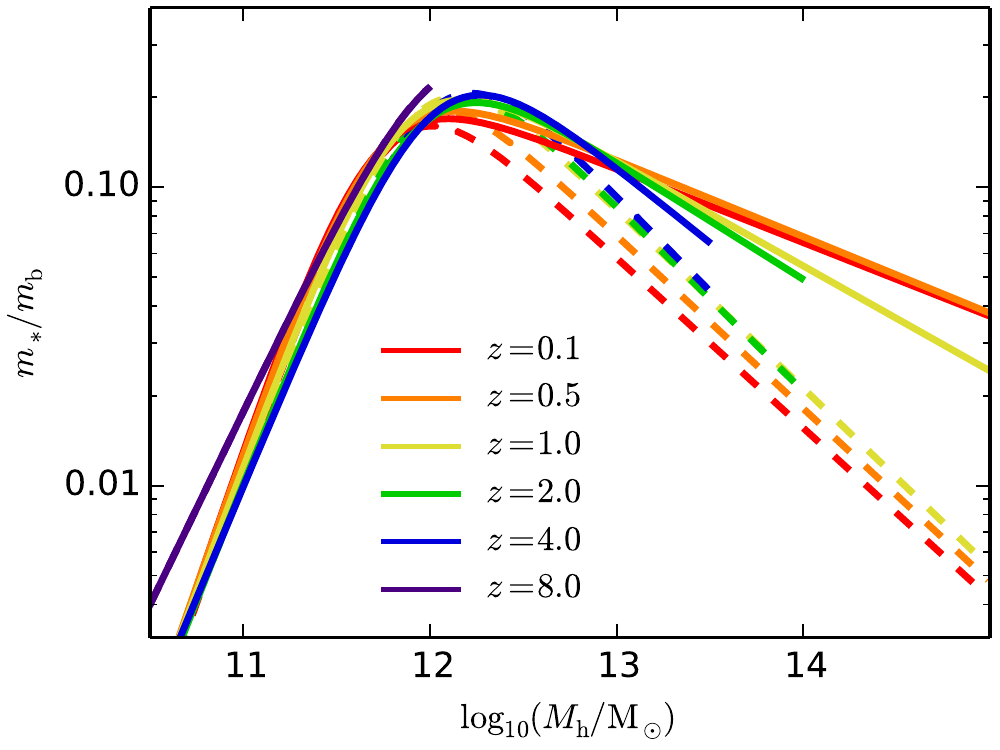}
\caption{The average conversion efficiency of main haloes from $z=0.1$ to $z=8$. The dashed lines
only include the stellar mass of the central galaxy ($m_*=m_\mathrm{c}$), while the solid lines
include the total stellar mass of the central galaxy and the ICM ($m_*=m_\mathrm{c}+m_\mathrm{ICM}$).
}
\label{fig:icm}
\end{figure}

Are more unambiguous quantity is the total amount of stellar mass that can be associated with a main halo,
i.e. the stellar mass in the central galaxy $m_\mathrm{c}$ plus the ICM $m_\mathrm{ICM}$ (but without
the mass in satellites). In our model the central galaxy can grow through star formation and mergers, and
the ICM can grow through ejected stars in mergers, tidal stripping of satellites, and the infall of a subhalo
with its own ICM (which is then transferred to the main halo). In Figure \ref{fig:icm}, we plot the conversion
efficiency including only the mass of the central galaxy (dashed lines), and including the mass of the central
galaxy plus the ICM (solid lines) from $z=0.1$ to $z=8$. We note that the lines can cross, because the halo
mass at the given redshift has been used to calculate the conversion efficiency (and not the $z=0$ mass).
While the ICM adds very little to the total mass at high redshift and for low halo masses, it dominates the stellar
mass budget of massive main haloes at low redshift. For a halo with $M=10^{15}\Msun$ the ICM is larger than
the mass of the central galaxy by a factor of almost 8. We note that over 80 per cent of the ICM in these haloes
forms from the tidal disruption of satellites and is distributed throughout the halo. The contribution of ejected
stars from merging satellites is therefore rather small and even if all those stars would be associated with
the central galaxy ($f_\mathrm{esc}=0$), this would boost its mass only by a factor up to 2.

\subsection{Clustering of star-forming and quenched galaxies}
\label{sec:wpc}

\begin{figure}
\includegraphics[width=0.45\textwidth]{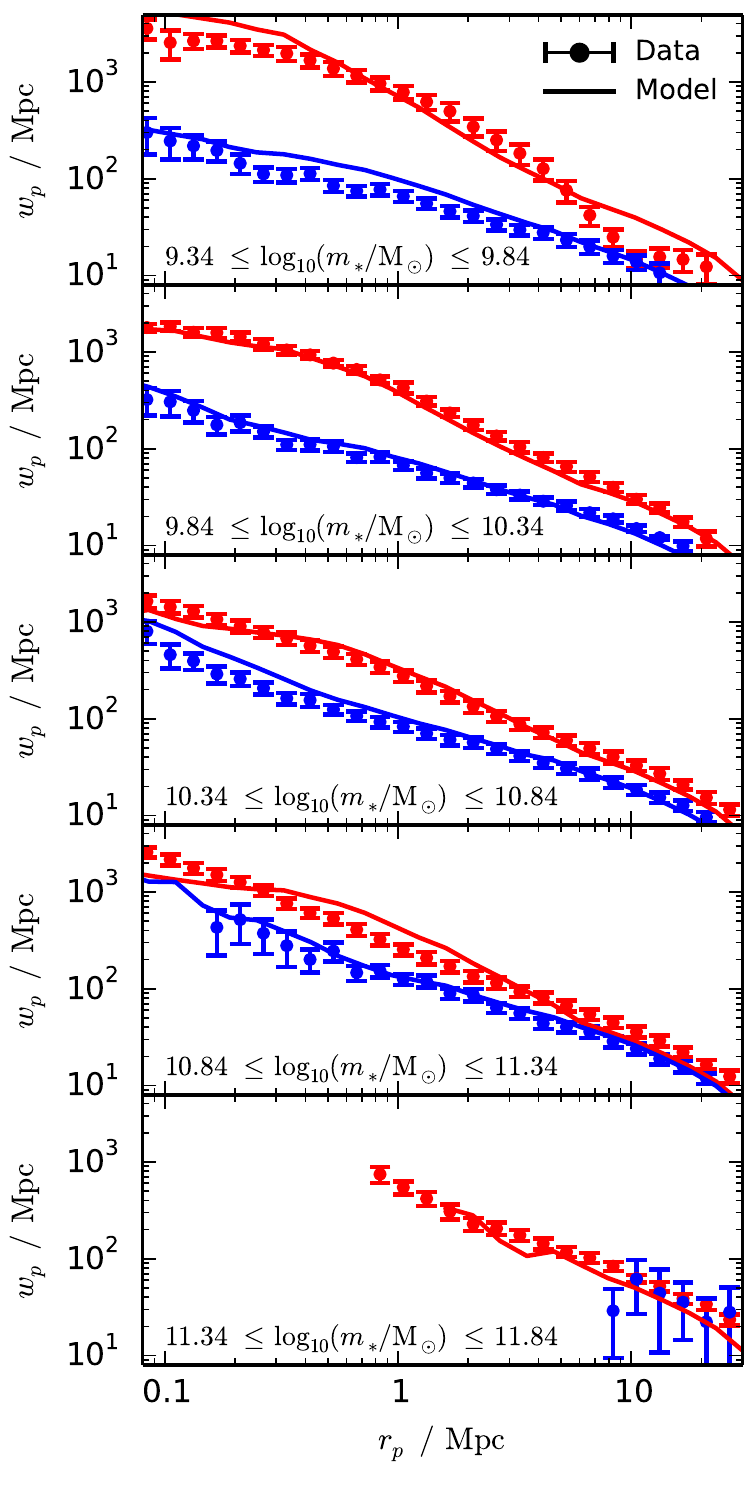}
\caption{Projected galaxy correlation function for star-forming (blue) and quenched (red) galaxies in
five bins of increasing stellar mass (from top to bottom). The symbols represent the observational estimates
\citep{Guo:2011aa}, and the lines show the model prediction for the best fit parameters for the $150\Mpc$
box.
}
\label{fig:wpc}
\end{figure}

We have used the projected auto-correlation functions of galaxies in different stellar mass bins to
constrain our model. As the tidal stripping of satellite galaxies reduces the number of pairs at small
scales, the clustering provides a measure for how effective the stripping can be. On large scales,
galaxy clustering is only a consequence of halo clustering, so that it can be used to test if galaxies
with given properties form in the right dark matter haloes. We have shown that on large scales
the projected correlation function computed with our model agrees very well with the observed
data (Figure \ref{fig:wp}). Here, the correlation results from pairs that reside in different haloes
and are predominantly centrals. This indicates that the relation between central galaxies and main
haloes is described well by our model.

Additional insight into the connection between galaxies and dark matter haloes can be obtained by
studying the clustering of star-forming and quenched galaxies. This can reveal whether the modelled quenching
processes lead to the correct distribution of galaxies in haloes. As the SFR depends on the halo growth rate
and the instantaneous conversion efficiency, a galaxy can be quenched because it runs out of fuel
(environmental quenching), or because the efficiency of the feedback becomes very high as the halo grows
(mass quenching). As halo clustering at fixed mass is stronger for haloes with low growth rates, and passive
galaxies are located in these haloes because of environmental quenching, the clustering of quenched galaxies
is stronger. On small scales the correlation function is dominated by pairs in which at least one galaxy is a satellite,
so the clustering of active and passive galaxies is strongly affected by how satellites are quenched.
Because of mass quenching, active and passive central galaxies of a given stellar mass live in haloes with different
masses, and consequently have a different correlation function at large scales where the signal is dominated
by pairs of centrals. The large-scale clustering of star-forming and quenched galaxies therefore provides a strong
constraint on mass quenching and consequently on the instantaneous conversion efficiency.

We can asses if the effects of the quenching processes are captured well in the model by comparing the
clustering of star-forming and quenched galaxies in the model to observational data. In Figure \ref{fig:wpc},
we plot the projected galaxy correlation function computed with our model (lines) and the observations
by \citep{Guo:2011aa} (symbols) in five stellar mass bins for star-forming (blue) and quenched (red) galaxies.
The observed galaxies have been divided with a $g-r$ colour cut, while the model galaxies were separated
with a cut in the sSFR. The agreement between model and data is good. Passive galaxies are more
clustered than active galaxies on all scales. Especially at small scales, the clustering of quenched galaxies is
enhanced compared to star-forming galaxies, showing that the adopted satellite quenching model captures the
environmental effects quite well. Also at large scale the clustering of passive galaxies is enhanced demonstrating
that the mass quenching leads to active and passive galaxies being located in the correct haloes.

\section{Discussion}
\label{sec:disc}

\begin{figure*}
\includegraphics[width=0.99\textwidth]{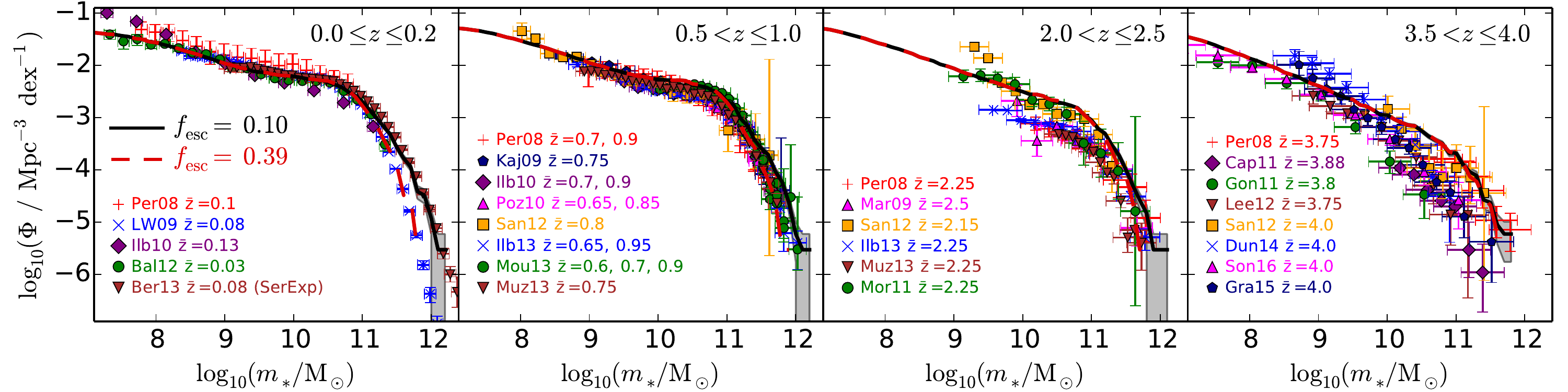}
\caption{The stellar mass functions up to $z=4$ for different values
  of $f_\mathrm{m}$. At $z=0$ the Bernardi et al. 2013 SMF based on
  the SerExp profile has been plotted. For a value of
  $f_\mathrm{m}=0.1$ the model prediction is in agreement with the
  different $z=0$ SMF, while at higher redshift the difference is
  negligible. }
\label{fig:serexp}
\end{figure*}

Our new empirical model {\sc Emerge} links the formation of galaxies to the evolution of dark matter haloes.
The resulting average build-up of the stellar component is in good agreement with previous empirical results,
which have typically been used as a benchmark to test hydrodynamical simulations. For instance, the integrated
conversion efficiency of simulated galaxies is often compared to the empirical findings to judge whether the
simulation has formed the correct amount of stellar mass in a certain halo. The result is then often used to
tune the unconstrained parameters of the relevant baryonic physics, most notably the efficiency of the supernova
feedback. This test can also be applied to infer if a previously unconsidered process can release the tension. For
example, \citet{Stinson:2013aa} have compared the SFHs of simulated galaxies to the empirical average SFHs
by \citet{Moster:2013aa}, and found that supernova feedback alone can reduce star formation enough to match
the integrated conversion efficiency at $z=0$, but the simulations still form too many stars before $z\sim2$. They
then invoke `early feedback', which mimics an ultraviolet ionisation source and provides pressure from the radiation
of massive young stars, and show that the correct average SFHs can be reproduced.

The limitation of previous empirical models was always the ability to only derive average galaxy properties for a
given halo mass, while the scatter in the relations was typically introduced artificially, and did not reflect the formation
history of the halo. It was therefore always unclear, if a particular simulated galaxy was simply an outlier to the relation,
or was in disagreement with the empirical findings. In the first case the conclusion would be that the modelled baryonic
physics can explain the observations, while in the second case one would modify the model or include new physics.
Since {\sc Emerge} computes the formation of galaxies in individual dark matter haloes in an empirical
fashion that automatically reproduces a large number of observational constraints, it provides an ideal testbed for
hydrodynamical simulations. It is possible to run the model on the halo merger tree provided by a dark matter
simulation. If the hydrodynamical simulation with the same initial conditions is in good agreement with the empirical
results for this specific system, the baryonic physics in the simulation is modelled such that statistical observations
like SMFs can be reproduced for a larger simulation volume. In the case where the hydrodynamic simulation results
in a large amount of stellar mass, while the empirical model predicts an average or lower amount of stars, then the
efficiency of the feedback may not be modelled correctly and will lead to a disagreement with the SMF for a larger
simulation volume.

\subsection{The conversion efficiency in massive haloes}
\label{sec:smfz0}

The characteristic shape of the instantaneous and integrated baryon conversion efficiencies results from the interplay
of the different physical processes that prevent the infalling gas in a dark matter halo from cooling and forming stars.
The contribution of each process strongly depends on the mass of the halo. While in low-mass haloes, feedback from
stars (e.g. supernova-driven winds or radiation pressure) can expel large amounts of gas from haloes with low escape
velocities, in more massive haloes the gas cannot escape as easily and falls back to the centre. However, AGN feedback
can heat the gas in the halo and prevent it from cooling and falling back to the centre. At the massive end, this feedback
can thus dominate the conversion efficiency. Comparing the SHM ratio in simulations of massive galaxies to
empirical constraints has therefore been a primary way to test models for AGN feedback and to determine its efficiency.
The question whether AGN feedback is the dominant mechanism at high masses and how strong it has to can thus be
inferred from observed galaxy properties.

The main observational input for the empirical constraints has been the SMF. For low-redshift galaxies the massive end
is well sampled by the SMF presented by \citet{Li:2009aa}, who use Petrosian magnitudes to compute stellar masses. In this
work we used the updated SMF in \citet{Guo:2010aa} which are based on \texttt{cmodel} magnitudes. These have been derived by
fitting an exponential and a de Vaucouleurs profile to the photometry and using the best fit to compute the total magnitude. In
this way the light that falls outside the Petrosian aperture can be captured. Recently, \citet{Bernardi:2013aa} presented a SMF
which is based on a fit of a Sersic profile or a two-component profile (Sersic and exponential) to the observed surface brightness.
As these profiles are more realistic for objects that neither have a de Vaucouleurs nor a pure exponential profile, the authors
claim their stellar masses provide a better estimate of the true mass. Since the profiles they adopt typically return more of the
light in the outskirts, the derived stellar masses are generally larger then masses based on Petrosian or \texttt{cmodel} magnitudes.

Using the \citet{Bernardi:2013aa} SMFs, \citet{Kravtsov:2014aa} apply the subhalo abundance matching method and find that the
SHM ratio is significantly higher than previously found with SMFs based on Petrosian or \texttt{cmodel} magnitudes.
While they acknowledge that empirical models predict the total stellar mass in the galaxy and the ICM to be of the same order as for
their result, they conclude that the overall efficiency of star formation in massive halos is considerably less suppressed than previously
thought, and that feedback in massive halos should be weaker than assumed in most of the current simulations. However, we stress
that this conclusion heavily depends on what is defined as galaxy mass and what is defined as ICM, and can therefore be misleading.
It is well established that the mass from accreted satellites is mainly located at the outskirts of massive galaxies \citep{Oser:2012aa,
Hilz:2013aa}. It is therefore not unambiguous if this material should be ranked among the central galaxy or the ICM, even if the light
profile is continuous as there may be overlap. If extended profiles are used, more of this accreted mass is taken into account and
assigned to the central galaxy.

This can be illustrated with our model. Figure \ref{fig:serexp} shows the SMF from $z\sim0.1$ to
$z\sim4$ (left to right panels). The symbols are the same data as used to fit our model, with the exception of the SMF of
\citet{Bernardi:2013aa}. Where before we have used their SMF based on \texttt{cmodel} magnitudes, the left panel shows the SMF
based on their two-component (Sersic and exponential) fit. The red dashed lines correspond to our best-fit model with a fraction of
stars in satellites that escape to the ICM during a merger of $f_\mathrm{esc}=0.39$. The black lines shows the same model but with
an escape fraction of 10 per cent. While beyond $z>0.5$ the two models give almost identical results, the massive end of the $z\sim0.1$
SMF is very different. The best-fit model reproduces the \citet{Li:2009aa} SMF (as fitted), while the model with $f_\mathrm{esc}=0.1$ has
a much shallower slope and reproduces the \citet{Bernardi:2013aa} SMF. This shows that while in both models the total amount of stars
formed is equal, the SMF and thus the integrated baryon conversion efficiency can be boosted if more of the material is assigned to the
galaxy instead of the ICM. In our model, the escape fraction $f_\mathrm{esc}$ regulates this assignment. The SFR is independent of this
choice, and is additionally constrained by the sSFR measurements.

In this sense, the instantaneous efficiency is given by the measured sSFR,
while the escape fraction is fixed by the growth of the massive end of the SMF which is mainly though accretion. So while the escape fraction
and the integrated efficiency depend on the assumed surface brightness profile and whether the mass is assigned to the galaxy or the ICM,
the instantaneous efficiency provides a strong constraint for the feedback processes. As we have shown, the SFHs of massive galaxies are
quenched strongly after $z\sim4$ falling from $\sim100\Msunpyr$ to $<1\Msunpyr$ at $z=0$. This implies that strong AGN feedback may still
be necessary to prevent the gas from cooling and forming stars.

If the integrated conversion efficiency in simulations is compared to empirical constraints, the definition of the surface brightness profiles does
matter, i.e. what belongs to the galaxy and what belongs to the halo. In this case, it is important to ensure that the comparison between
simulation and observation is as fair as possible. The ideal comparison would be to create mock images of the simulation with a radiative transfer
code such as {\sc Sunrise} \citep{Jonsson:2006aa} or {\sc Grasil-3D} \citep{Dominguez-Tenreiro:2014aa}, and to analyse these images in the
exact same way as the observed data. If this is not possible, then the stellar mass in the simulation should be computed as close as possible
to the observations. If the simulation is compared to SMFs based on Petrosian or \texttt{cmodel} magnitudes, then only the stellar mass within
a few scale radii should be counted. For a comparison to SMFs based on profiles that capture the light out to large radii, then it is better to count
all stellar mass within a large radius up to the virial radius, subtracting all stars bound to satellites.

\subsection{The conversion efficiency in low-mass haloes}

The empirically determined SHM ratio has also often been used at the low mass end to provide a constraint for hydrodynamical
simulations, and to test or tune the effects of stellar feedback, such as supernova-driven winds, radiation pressure, and cosmic rays. The integrated
conversion efficiencies computed by different empirical models agree very well within the derived uncertainties up to the minimum mass that is
used in the model. The minimum stellar mass for which observational constraints are available (typically from the SMF) is $m_\mathrm{min}\sim
10^{7.5}\Msun$, with a corresponding halo mass of $M_\mathrm{min}\sim10^{10.5}\Msun$. Below this minimum mass that has been used to
fit the model, the empirical models do not provide any constraint. Still, it has become common to extrapolate the SHM mass relation
down to very low masses. Since different empirical models can use different fitting functions this extrapolation can lead to orders of magnitude
differences in the stellar mass at fixed halo mass, even if the models agree very well up to the minimum mass. In practice, the empirical model
is used that fits the hydrodynamical results best when extrapolated, and success is then claimed for the simulations.

We strongly caution against this practice, as extrapolating the conversion efficiency has several pitfalls. First, reionisation suppresses galaxy formation
in haloes that have a low mass early-on \citep{Efstathiou:1992aa}, which can lead to dark subhaloes that do not host a galaxy \citep{Sawala:2015aa}.
This can lead to the mean SHM relation bending over with respect to the interpolation. Secondly, it is not clear that galaxies at these
mass scales follow a tight SHM relation, as observations of their stellar populations show `bursty' star formation histories. As our model shows, 
this can lead to a large scatter at low halo masses, and we have demonstrated that an increased scatter leads to a modified mean SHM relation.
Thirdly, it is unlikely that the physical processes that govern galaxy formation at intermediate mass ($10^{11}<M/\Msun<10^{12}$), are also the
dominant processes at lower halo masses. As the processes can have different efficiencies, the low-mass slope of the SHM relation may change
at the halo mass, where the efficiencies of two dominant processes cross, and currently it is difficult to find this transition mass. Finally, the main
justification why simple empirical models such as abundance matching can be used to constrain the conversion efficiency is not because a
monotonic SHM relation is an obvious choice, but because the models are able to correctly reproduce observations that have not been used
to calibrate the model, such as galaxy clustering. This has been shown to be successful above the minimum stellar mass that has been used
in the fit, but not yet below that mass. Until such an independent validation of the model can be performed, the results have to be regarded as
tentative at best.

\subsection{Differences to a semi-analytic model}

The new empirical model presented here follows dark matter halo trees through cosmic time and populates
them with galaxies according to a set of simple parameterised prescriptions where the parameters are adjusted
to fit observed data. In this regard the question arises how this model differs from a SAM. To shed some light
on this issue we first need to recall the purpose and history of SAMs which is excellently illustrated in the 
review by \citet{Baugh:2006aa}. When full hydrodynamical simulations were still unfeasible, the only possibility
to model the formation of galaxies was to treat all gas physics semi-analytically \citep{White:1978aa,White:1991aa}.
However, as many of the physical processes are poorly understood, several of the adopted prescriptions need to be
parameterised. As these parameters are directly connected to a physical process their values often cannot be chosen
completely freely, but they need to comply with physical priors, typically from direct observations or more detailed simulations.
The star-formation efficiency at low redshift is strongly constrained by the observed star-formation-relation
\citep{Schmidt:1959aa,Kennicutt:1998aa}, for example. Within these limits the parameters are then typically chosen
to reproduce a subset of statistical observations.

Following the general philosophy of ab initio models, SAMs can then be employed to learn about galaxy formation.
For a given model with certain values chosen for the parameters, the model is run and the results are compared to
observations. Varying the parameter values within physical limits, this is repeated until the best agreement is found.
If some subsets of the observations cannot be reproduced the model is changed, either by altering the parameterisations
or by including new physical processes. A very insightful example is the inability of earlier SAMs to reproduce the bright
end of the luminosity function. This eventually lead to the inclusion of feedback from AGN into the models
showing that the data can be reproduced in this way. Compared to hydrodynamical simulations, SAMs have the great
advantage that model prescriptions can easily be varied or disabled. Therefore, SAMs are very useful to to gain a better
understanding of how the different physical processes impact a particular observation. Moreover, once all physical
processes driving galaxy formation are fully understood, future SAMs that parameterise these processes will likely be
the final way to describe galaxy formation.

Empirical models on the other hand avoid directly modelling the physics of the baryon cycle, i.e. the different states of the
baryons (hot gas, cold gas, and stars) and the physical processes that regulate the transition between these states, but
rather employ empirical relations between galaxy and halo properties, marginalising over the baryon cycle.
This can technically be done similarly to SAMs by populating halo merger trees with galaxies according to parameterised
relations. However, the motivation determining the choice of the parameterised models is different. The parameters in an
empirical model are not directly related to a physical process or quantity, so there are no physical priors on them. Therefore
all model parameters can be fitted to reproduce the data with statistical methods such as an MCMC algorithm without concerns
about the parameter values. Moreover, model selection criteria can easily be applied to find the simplest model that agrees with
the data. In SAMs this is more complicated because of observed physical priors, e.g. outflow velocities.
However, as empirical models do not explicitly implement physical processes it is difficult to use them directly  to
learn about the physics of galaxy formation, i.e. one cannot test the impact of certain physical processes on the galaxy
population (empirical models can nevertheless provide strong constraints that can be used to study the physics). Still, this does
not imply that empirical models are `unphysical', unless they violate some law of physics -- they simply do not model the
physics of the baryon cycle explicitly, but rather provide a minimal set of assumptions to reproduce the observed data.

Lately, several studies have used statistical methods such as MCMC methods to explore the parameter space
of SAMs \citep[e.g.][]{Henriques:2009aa,Lu:2011aa,Lu:2012aa,Henriques:2013aa}. This is a very useful exercise as it provides
many insights into the model, such as degeneracies between parameters or unconstrained parameters. This also provides
uncertainties on model predictions and can thus help considerably with interpretation of the results. However, this does not
imply that the models normalised in this way represent all the underlying physics, nor that they are unique. Rather they are a simple
physically motivated representation that can reproduce the set of observations. For instance, if a SAM provides a very good fit
to the low-mass end of the SMF, it is not guaranteed that the fitted supernova feedback parameters (e.g. the efficiency)
represent the underlying physics. It cannot be ruled out that other processes that may not have been considered, such as
cosmic ray feedback \citep[e.g.][]{Pfrommer:2007aa}, may be equally or even more relevant, and that their effects have
just been mimicked by the supernova feedback model. This can make the interpretation of MCMC-fitted SAMs difficult.
Unless there are strong priors on the physical parameters, a model fitting these parameters is no longer based on first
principles and effectively becomes empirical.

Given these differences, one could argue for employing a state-of-the-art SAM and fit all possible model parameters
to reproduce observations. Even if the model effectively is empirical when fitted with a statistical method, it may still be able to
reproduce the observations as well as a model that uses empirical relations between halo and galaxy properties.
Galaxies with realistic properties can thus be followed through cosmic time if a limited physical interpretation of the model
is accepted. However, there a some complications with this approach. First, as SAMs typically have a very large number of
parameters it becomes difficult to sample the whole parameter space. 
Secondly, as SAMs directly model the physics of the baryon cycle, the parameters may degenerate and it becomes
critical to find observational data that can break the degeneracies \citep[e.g.][]{Henriques:2015aa}.
Empirical models can be designed to minimise parameter degeneracies, as they do not aim to follow all components of the baryon cycle.
Thirdly, because the relations in empirical models are not explicitly motivated by baryonic physics, it is straight-forward to
use model selection statistics to optimise the model, which is much more difficult in a SAM. This means that if the goal is simply to track galaxies
through cosmic time while having galaxy populations that agree as well as possible with observed galaxies, then
Occam's razor can favor empirical models.

\section{Summary}
\label{sec:sum}

In this paper, we present the novel empirical galaxy formation model {\sc Emerge}, which follows the evolution of galaxies in individual dark matter haloes through
cosmic time. First, halo merger trees are extracted from cosmological $N$-body simulations and the growth rate is calculated for each halo at every redshift.
Secondly, the SFR of the galaxy in each halo is determined as the product of the halo growth rate, which specifies how much material becomes available, and the instantaneous baryon conversion efficiency, which specifies how efficiently this material can be converted into stars. It captures the effects of all baryonic physics
that governs galaxy formation, depends on halo mass and redshift, and has a peak around $10^{12}\Msun$. The stellar mass of each galaxy is computed by
integrating the SFHs taking into account mass loss from dying stars. Once a halo stops growing, i.e. when it starts to fall onto a larger halo and becomes a subhalo,
its galaxy continues to form stars for a specified amount of time and is then rapidly quenched. If the halo has lost a significant fraction of its peak mass, the stars in
the galaxy become unbound and are stripped to the ICM. When a satellite has lost its kinetic energy due to dynamic friction, it merges with the central galaxy and
ejects a fraction of its stars to the ICM. 

We constrain our model with several sets of observed data. The conversion efficiency is constrained by SMFs, sSFRs, and the CSFRD up to $z\sim10$.
The satellite quenching timescale is constrained by the fraction of quenched galaxies as function of stellar mass up to $z=4$, and the stellar stripping is
constrained by small-scale galaxy clustering. The fraction of stars ejected to the ICM is determined with the low-redshift evolution of the massive end of the
SMF and the sSFR of massive galaxies. We fit all model parameters with an MCMC ensemble sampler by requiring that all observed data be reproduced
simultaneously. The adopted empirical relations are as flexible as possible. We increase the complexity of the model stepwise if the data require it, which
is assessed by a number of different model selection statistics. The result is thus the simplest model that is in agreement with the data.

For our best-fit model, we find that the characteristic halo mass where the instantaneous conversion efficiency peaks, decreases from $1.1\times10^{12}\Msun$
at high redshift to $2.2\times10^{11}\Msun$ at $z=0$. The peak efficiency at high redshift is 70 per cent and decreases to less than one per cent towards
low redshift. We find a steep low-mass slope for the efficiency of 1.3 at high redshift that steepens to 3.3 towards low redshift, such that star formation is
strongly suppressed in low-mass haloes. The high-mass slope is $\sim1$, independent of redshift, so as haloes become more massive the galaxy gets
quenched. We find that almost 40 per cent of all stars in satellite galaxies get ejected to the ISM during a merger, and that satellite galaxies are tidally
disrupted once the mass of their subhalo has dropped to 10 per cent of its peak value. Massive satellites with $m\ge10^{10}\Msun$ keep forming stars for
about 4 dynamical halo times ($r_\mathrm{vir}/v_\mathrm{vir}$) after their halo has stopped growing. In lower-mass galaxies this quenching timescale
is considerably longer, such as 10 dynamical halo times for satellites with $m=10^{9}\Msun$.

Using our best-fit model, we study the SFHs of individual galaxies, and find that even if the final halo masses are identical, galaxies can have very different
SFHs. The average SFHs and accretion rates as a function of $z=0$ halo mass in our model agree very well with previous findings. The SFRs of central galaxies in
massive haloes typically peak at high redshift, after which the galaxies become increasingly quenched, e.g. the SFR of a galaxy in a halo with a $z=0$ mass
of $10^{14}\Msun$ peaks around $z=4$. In low-mass haloes the SFRs peak much later, e.g. in a halo with a $z=0$ mass of $10^{12}\Msun$ the SFR peaks
around $z=1$. However, individual galaxies can deviate significantly from these overall trends. The average peak SFRs of central galaxies that are quenched at
$z=0$ is slightly higher than that of their star-forming counterparts, as quenched galaxies generally form most of their mass at high redshift, where the peak
is located. We find that the fraction of ex-situ formed (accreted) stars to be insignificant in low mass haloes, while in massive haloes these stars can dominate
the total stellar mass of central galaxies, e.g. the accreted fraction of galaxies in haloes with $M=10^{14}\Msun$ is 55 per cent at $z=0$. We do not notice any 
difference between quenched and star-forming galaxies for the accreted fraction.

Our model predicts the integrated baryon conversion efficiency $m/(f_\mathrm{b}M)$, i.e. the SHM ratio, for each individual system in contrast to previous
models that only predicted the average SHM ratio at a given halo mass. The stellar mass of each galaxy depends on the formation history of the halo, such
that scatter in the SHM relation is automatically produced. The average peak conversion efficiency at $z=0.1$ is 17 per cent at a halo mass of
$M\sim10^{12}\Msun$. As quenched galaxies formed most of their mass at high redshift where the conversion efficiency is higher, they have a larger SHM
ratio than star-forming galaxies. We find a scatter of 0.16 dex at the massive end, which is in good agreement with observational estimates from satellite
kinematics, while at the low-mass end the scatter is larger. This increased scatter causes the overall average SHM to be slightly lower than previously found.
Still, the overall agreement with other empirical models is very good within the model uncertainties. We further find good agreement with direct measurements
of the SHM ratio from weak lensing and satellite kinematics, while X-ray measurements indicate higher SHM ratios. Taking into account also the ICM, we find
that the SHM ratio is strongly enhanced in massive systems at low redshift. For example, a halo with $M=10^{15}\Msun$ at $z=0.1$ has a conversion efficiency
of 0.004 when only the mass in the central galaxy is taken into account, while if also the ICM is considered, which is larger by a factor of 8, the conversion
efficiency increases to 0.037.

Computing galaxy clustering for star-forming and quenched galaxies we find a very good agreement with observational constraints. Passive galaxies are more
strongly clustered at all scales, while at small scales this effect is even enhanced. Quenched galaxies live in haloes that have low growth rates, and these haloes
are more strongly clustered, so quenched galaxies are generally more clustered as well. Moreover, satellite galaxies get environmentally quenched, so at small
scales, where the clustering is dominated by pairs in which at least one galaxy is a satellite and these are preferentially quenched, the clustering of passive
galaxies is boosted. The good overall agreement indicates a realistic assignment of galaxies to haloes.

\section*{Acknowledgements} 

We thank all authors who provide their data in electronic form.
We are also grateful to 
Peter Behroozi,
Andreas Burkert,
Darren Croton,
Benedikt Diemer,
George Efstathiou,
Martin Haehnelt,
Andrew Hearin,
Bruno Henriques,
Ben Hoyle,
Houjun Mo,
Jerry Ostriker,
Debora Sijacki,
Rachel Somerville,
Jeremy Tinker,
and
Frank van den Bosch
for enlightening discussions.
The cosmological simulations used in this work were carried out at
the Odin Cluster at the Max Planck Computing and Data Facility in Garching,
and the Darwin Supercomputer of the University of Cambridge High Performance Computing Service.
BPM acknowledges
an Emmy Noether grant funded by the Deutsche Forschungsgemeinschaft (DFG, German Research Foundation) -- MO 2979/1-1.
TN acknowledges support from the DFG Cluster of Excellence ``Origin and Structure of the Universe''.


\bibliographystyle{mnras}
\bibliography{astro}

\appendix

\section{Model selection}
\label{sec:bayes}

The philosophy of the empirical model {\sc Emerge} is to construct the simplest self-consistent galaxy evolution model
that is able to explain all available observed statistical data without being restricted by our limited understanding of
baryonic physics. The model is thus designed to allow for a very broad range of possibilities such that all observational
constraints can be fulfilled. We start with very simple models to describe the instantaneous efficiency and the processes
for the satellite galaxies containing a minimal number of parameters, and increase the complexity and the number of
parameters in a stepwise manner, as the data require it. We assess this with a number of different model selection
criteria that are described below; for more details on the methods see \citet{Liddle:2007aa}.

\begin{table}
 \caption{Summary of the tested models}
 \label{tab:models}
 \begin{tabular}{@{}cccccccc@{}}
  \hline
  ~ Model ~&~ $M_\mathrm{z}$ ~&~ $\epsilon_\mathrm{z}$ ~&~ $\beta_\mathrm{z}$ ~&~ $\gamma_\mathrm{z}$ ~&~ $\tau_\mathrm{s}$ ~&~ $f_\mathrm{s}$ ~&~ $\tau_\mathrm{d}$ ~\\
  \hline
  1 & 0 & 0 & 0 & 0 & 0 & $M_\mathrm{p}$ & 0\\
  2 & F & F & 0 & 0 & 0 & $M_\mathrm{p}$ & 0\\
  3 & F & F & F & 0 & 0 & $M_\mathrm{p}$ & 0\\
  4 & F & F & F & F & 0 & $M_\mathrm{p}$ & 0\\
  5 & F & F & F & 0 & F & $M_\mathrm{p}$ & 0\\
  6 & F & F & F & 0 & F & $M_\mathrm{p}$ & F\\
  7 & F & F & F & 0 & F & $m_*$ & 0\\
  \hline
  \end{tabular}
 \medskip\\
  \textbf{Notes:} Columns are model number (1), $z$-evolution of $M_1$ (2), $N$ (3), $\beta$ (4), and $\gamma$ (5), slope of quenching time (6), threshold used
  for stripping (7), and star formation decay time-scale (8). A 0 indicates a parameter is set to $0$, F states that the parameter is free, and $M_\mathrm{p}$ signifies
  that the stripping threshold is taken with respect to the halo peak mass, while for $m_*$ it is taken with respect to the present stellar mass.
\end{table}

\begin{table*}
 \caption{Model selection results}
 \label{tab:selectionresults}
 \begin{tabular}{@{}cccccccccc@{}}
  \hline
  ~~ Model ~~&~~ $\chi^2_\mathrm{min}$ ~~&~~ $\chi^2_\mathrm{mean}$ ~~&~~ $N_\mathrm{p}$ ~~&~~ $p_\mathrm{D}$ ~~&~~ AIC ~~&~~ BIC ~~&~~ DIC ~~&~~ $-2\ln Z$ ~~&~~ $-2\ln(Z/Z_5)$ ~~\\
  \hline
  1 & 2761.65 & 2768.94 & 7 & 7.29 & 2775.71 & 2814.09 & 2776.24 & 2859.32 & 990.67\\
  2 & 1939.60 & 1948.75 & 9 & 9.16 & 1957.70 & 2007.04 & 1957.91 & 2023.47 & 154.82\\
  3 & 1932.65 & 1943.08 & 10 & 10.42 & 1952.78 & 2007.01 & 1953.50 & 2020.29 & 151.64\\
  4 & 1932.65 & 1943.77 & 11 & 11.12 & 1954.80 & 2015.07 & 1954.88 & 2045.40 & 176.75\\
  5 & 1789.41 & 1800.42 & 11 & 11.01 & 1811.56 & 1871.83 & 1811.44 & 1868.65 & 0.0\\
  6 & 1789.41 & 1801.55 & 12 & 12.14 & 1813.59 & 1879.33 & 1813.69 & 1908.77 & 40.12\\
  7 & 1845.19 & 1856.16 & 11 & 10.97 & 1867.34 & 1927.62 & 1867.13 & 1917.29 & 48.64\\
  \hline
  \end{tabular}
 \medskip\\
  \textbf{Notes:} Columns are model number (1), minimum deviance $\chi^2_\mathrm{min}$ (2), mean deviance $\chi^2_\mathrm{mean}$ (3), number of free parameters $N_\mathrm{p}$ (4),
  effective number of parameters $p_\mathrm{D}$ (5), Akaike information criterion (6), Bayesian information criterion (7), deviance information criterion (8), twice the logarithmic
  marginal likelihood $Z$ (9), and twice the logarithmic Bayes factor w.r.t. the best model (10).
\end{table*}

\subsection{Model selection criteria}

While models with a large number of free parameters can achieve a good fit more easily, the high complexity reduces
the predictiveness of the model. Therefore, any model selection aims to balance the quality of the fit to observational
data against the complexity of the model. This acts as Occam's razor and prefers simpler models if the fits are similar.
In practice, model selection statistics attach a number to each model, which is based on the quality of the fit and penalised
for larger numbers of parameters.

A simple statistic that can be applied easily to almost any model is the Akaike Information Criterion \citep[AIC,][]{Akaike:1974aa},
as it only requires the maximum likelihood a model can achieve $\mathcal{L}_\mathrm{max}$, rather than the full likelihood
surface. It is defined as
\begin{equation}
\label{eqn:AIC}
\mathrm{AIC} = -2 \ln \mathcal{L}_\mathrm{max} + 2k \; ,
\end{equation}
where $k$ is the number of model parameters. The best model is the one that minimises the AIC. For small sample sizes,
a corrected AIC has been introduced by \citet{Sugiura:1978aa}:
\begin{equation}
\label{eqn:AICc}
\mathrm{AIC_c} = \mathrm{AIC} + \frac{2k(k+1)}{N-k-1} \; ,
\end{equation}
where $N$ is the number of data points used in the fit. This strengthens the penalty for $N/k$ being only a few. The Bayesian
Information Criterion \citep[BIC,][]{Schwarz:1978aa} is an approximation of the Bayes factor, and is defined as
\begin{equation}
\label{eqn:BIC}
\mathrm{BIC} = -2 \ln \mathcal{L}_\mathrm{max} + k \ln N \; ,
\end{equation}
assuming that the data points are independent and the parameters are not degenerate.

The Deviance Information Criterion \citep[DIC,][]{Spiegelhalter:2002aa} combines elements from Bayesian statistics and
information theory. Unlike the AIC and BIC it accounts for situations where parameters are unconstrained by data. It requires
knowledge of the full likelihood surface, but can be computed from a posterior sample generated by a MCMC method.
First the Bayesian complexity is introduced:
\begin{equation}
\label{eqn:pd}
p_D = \langle\chi^2(\vec\theta)\rangle - \chi^2_\mathrm{min} \; ,
\end{equation}
where the chevrons $\langle\rangle$ indicate the average over the posterior distribution. It corresponds to an effective
number of model parameters indicating the number of parameters actually constrained by the data. The DIC is then
defined as
\begin{equation}
\label{eqn:DIC}
\mathrm{DIC} = -2 \ln \mathcal{L}_\mathrm{max} + 2 p_D \; ,
\end{equation}
so that it does not penalise the inclusion of parameters that are unconstrained by the data.

In Bayesian statistics the Bayesian Evidence (also referred to as model likelihood or marginal likelihood) is the probability
of the data $D$ given the model $\mathcal{M}$, and it can be computed as
\begin{equation}
\label{eqn:evidence}
Z = P(D|\mathcal{M}) = \int \mathrm{d}\vec\theta \, P(\vec\theta|\mathcal{M}) \, P(D|\vec\theta,\mathcal{M}) \; ,
\end{equation}
where $P(\vec\theta|\mathcal{M})$ is the prior distribution and $P(D|\vec\theta,\mathcal{M})$ is given by the likelihood function
$\mathcal{L}(\vec\theta)$. With a prior probability $P(\mathcal{M}_i)$ of a model $\mathcal{M}_i$, the probability
of this model given the data is specified by Bayes Theorem: $P(\mathcal{M}_i,D)=P(\mathcal{M}_i) P(D|M)/P(D)$. Thus, the
posterior odds of model $\mathcal{M}_i$ relative to model $\mathcal{M}_j$ become
\begin{equation}
\label{eqn:bayes}
\frac{P(\mathcal{M}_i|D)}{P(\mathcal{M}_j|D)}
= \frac{P(\mathcal{M}_i)}{P(\mathcal{M}_j)} \, \frac{P(D|\mathcal{M}_i)}{P(D|\mathcal{M}_j)}
= \frac{P(\mathcal{M}_i)}{P(\mathcal{M}_j)} \, \frac{Z_i}{Z_j}\; ,
\end{equation}
and we can interpret the Evidence $Z$ as the support for a model given the data. In the absence of information about the model
priors, a ratio of $P(\mathcal{M}_i)/P(\mathcal{M}_j)=1$ is adopted, and the relative probability of the models given the data is
specified by the Bayes Factor $\mathcal{B}=Z_i/Z_j$. This property can be used for model selection and considers a combination
of data fit and predictiveness. Moreover, the Evidence does not penalise parameters that are unconstrained by the data. Thus,
models that fit the data by varying few parameters are favoured. Models with many unconstrained parameters therefore do not
need to be discarded, only the unconstrained parameters.

Calculating the Evidence by integrating over parameter space is difficult in practice, however. Here, we use the method presented
by \citet{Weinberg:2012aa} and \citet*{Weinberg:2013aa} to derive the Evidence from the posterior distributions computed with
our adopted MCMC algorithm. We first define a define a region of high posterior probability employing a volume peeling strategy
and then integrate the sample in this region numerically using a volume tessellation algorithm. This integration is performed for
both the Riemann and the Lebesgue variants, and we have verified that both techniques result in the same Evidence within a few
percent for all models, i.e. the values for $\ln Z$ differ by 0.2 at most.

\subsection{Tested models}

We start with a very simple model, test how well this model can fit the data, and then increase the complexity stepwise.
For all models, we assume that at any redshift the instantaneous baryon conversion efficiency $\epsilon$ has a double power
law dependence on halo mass, as specified by equation (\ref{eqn:epsilon}). The four parameters may vary with redshift and
depend linearly on the scale factor $a$:
\begin{align}
\log_{10} M_1(z)& = M_0 + M_\mathrm{z}(1-a) = M_0 + M_\mathrm{z}\frac{z}{z+1} \; ,\label{eqn:m1}\\
\epsilon_\mathrm{N}(z)& = \epsilon_0 + \epsilon_\mathrm{z}(1-a) = \epsilon_0 + \epsilon_\mathrm{z}\frac{z}{z+1} \; ,\label{eqn:epsN}\\
\beta(z)& = \beta_0 + \beta_\mathrm{z}(1-a) = \beta_0 + \beta_\mathrm{z}\frac{z}{z+1} \; ,\label{eqn:beta}\\
\gamma(z)& = \gamma_0 + \gamma_\mathrm{z}(1-a) = \gamma_0 + \gamma_\mathrm{z}\frac{z}{z+1}\; .\label{eqn:gamma}
\end{align}
In model 1, we assume that the four parameters of $\epsilon(M)$ are constant through cosmic time, i.e. the efficiency does
not evolve with redshift, and we set $M_\mathrm{z}=\epsilon_\mathrm{z}=\beta_\mathrm{z}=\gamma_\mathrm{z}=0$.
Moreover we set the slope of the quenching timescale $\tau_\mathrm{s}$ to zero, so that the
quenching time does not depend on stellar mass and is equal for all satellites. After the quenching time has elapsed,
the SFR is set to zero instantly. The satellite is stripped to the halo once the mass of the associated subhalo has fallen
below a fraction of its peak mass $M_\mathrm{p}$.

In model 2 we allow for a redshift evolution of the characteristic halo mass $M_1$ and the normalisation $\epsilon_\mathrm{N}$
of the conversion efficiency as specified by equations, i.e. $M_\mathrm{z}$ and $\epsilon_\mathrm{z}$ are now free parameters,
while everything else is identical to model 1. In model 3, we also let the low-mass slope $\beta$ evolve, i.e. $\beta_\mathrm{z}$
becomes a free parameter as well. All four parameters, including the high-mass slope $\gamma$ are allowed to evolve with
redshift in model 4, i.e. $\gamma_\mathrm{z}$ is non-zero, too. The quenching and stripping in models 2 to 4 are identical to
model 1.

Model 5 is identical to model 3 ($M_\mathrm{z}$, $\epsilon_\mathrm{z}$, and $\beta_\mathrm{z}$ are free while
$\gamma_\mathrm{z}=0$), but now we let the quenching timescale depend on stellar mass according to equation
(\ref{eqn:satquenching}), so that $\tau_\mathrm{s}$ is allowed to vary freely. In model 6, we also let the SFR decline
exponentially on a timescale $\tau_\mathrm{d}$ after the quenching timescale has elapsed. Finally, model 7 is
identical to model 5, but the galaxy is stripped to the halo once its subhalo has reached a mass that is a specific factor of the
galaxy's stellar mass (typically a factor of a few), instead of a fraction of its peak mass.

Table \ref{tab:models} summarises all models we have tested. For each model we first found the most likely parameters
using the {\sc Hybrid} method and then ran the ensemble MCMC as described in section \ref{sec:modelfitting} using
the same number of ensembles, walkers, and steps.

\begin{figure*}
\includegraphics[width=0.95\textwidth]{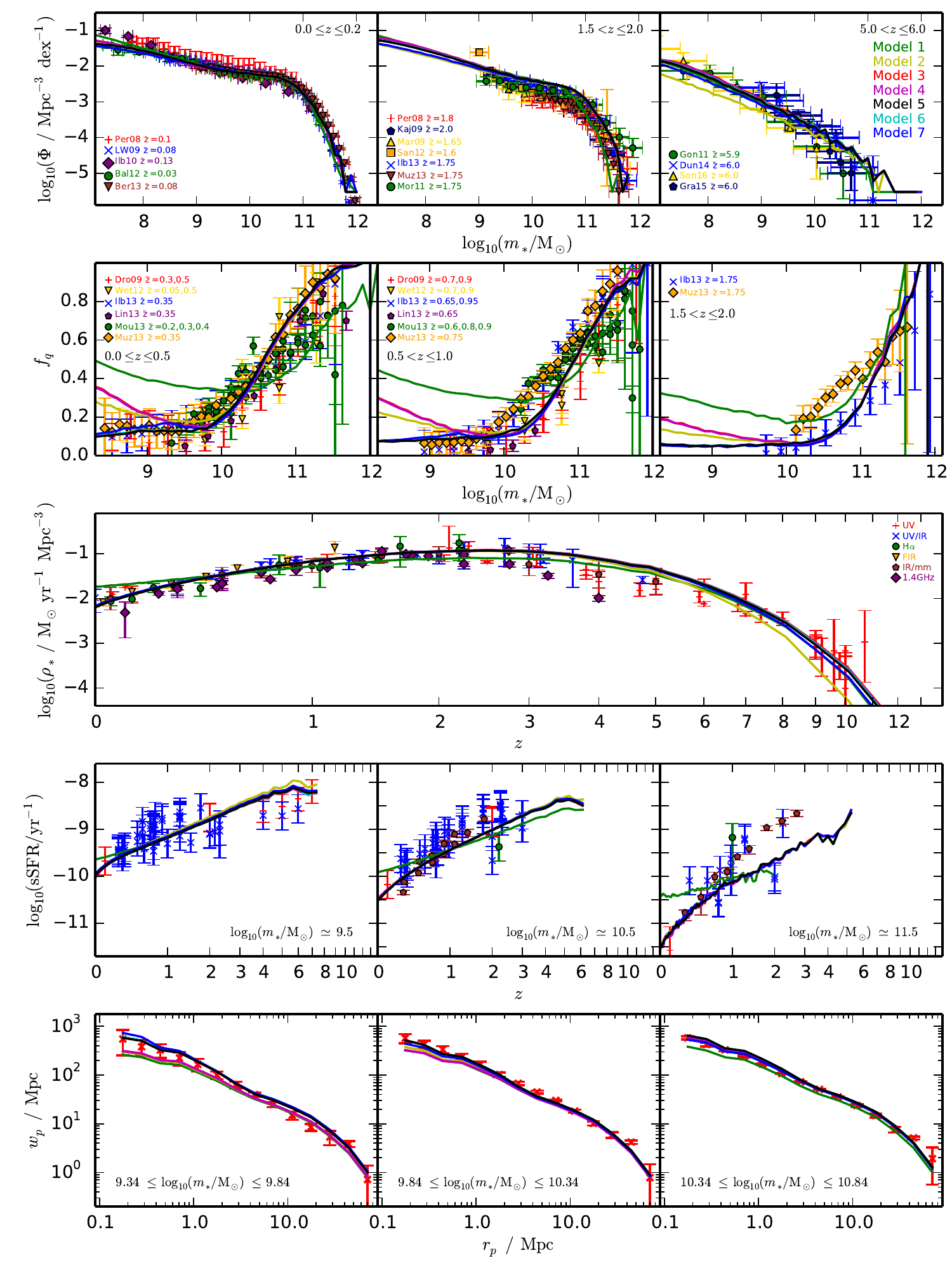}
\caption{Comparison of different models using their best-fit parameters. Each line corresponds to the global best-fit result
of a specific model, while the symbols represent the observed data.
See table \ref{tab:models} for more details about the different models. The top panels show the
SMF in three redshift bins, the panels in the second row show the quenched fractions in three redshift bins, the panel
in the third row shows the CSFRD, the panels in the fourth row show the sSFRs in three stellar mass bins, and the bottom
panels show the $z=0$ correlation functions in three stellar mass bins.
}
\label{fig:bayes}
\end{figure*}

\subsection{Model selection results}

We use the posterior distributions from the Markov chains to compute the information criteria and Bayesian
Evidences. The results are summarised in Table \ref{tab:selectionresults}. We present the best-fit results for
a subset of the SMF, CSFRD, sSFR, quenched fraction and clustering data for all models in Figure \ref{fig:bayes}.

The simplest model, which has no redshift evolution of the efficiency (model 1), provides a rather poor fit to the data.
The massive end of the SMF is underpredicted at intermediate redshift, the fraction of quenched galaxies is too high
at the low-mass end and too low at the massive end, the CSFRD at low redshift is overpredicted as are the sSFRs
of massive galaxies, and the small-scale clustering of all galaxies is underpredicted. Consequently, all information
criteria have a very high value and the Evidence is very low.

Adding two more free parameters, $M_\mathrm{z}$ and $\epsilon_\mathrm{z}$ (model 2), improves all model
selection statistics significantly as the quality of the fit has increased. The SMFs can be reproduced much better,
except for very high redshift, the fraction of quenched galaxies is only too high at very low masses, the CSFRD is
slightly underpredicted at high redshift, and the small-scale clustering of low-mass galaxies is too low. The Evidence
increases dramatically with a logarithmic Bayes factor of 418 with respect to model 1, so that model 2 cleary outperforms
model 1.

Letting also the low-mass slope $\beta$ evolve with redshift (model 3), improves the fit and thus the model selection
statistics further. Now all SMFs, sSFRs, and the CSFRD can be reproduced very well. Still the small-scale clustering
of low-mass galaxies is too low and the fraction of quenched low-mass galaxies is too high. With a logarithmic Bayes
factor of 1.6, model 3 is preferred over model 2. Allowing the high-mass slope $\gamma$ to evolve with redshift as well
(model 4), does not further improve the fit, and thus the model selection statistics become worse, as the model is
penalised for the additional free parameter. The logarithmic Bayes factor of model 3 with respect to model 4 is 12.5,
so that model 4 does not constitute an improvement, and model 3 is still favoured.

Model 5 is identical to model 3, but adds one more free parameter, letting the quenching timescale depend on stellar mass,
specifically low-mass galaxies can now have longer quenching timescales and form stars longer than in model 3. While
the SMFs, sSFRs, and the CSFRD are still reproduced very well, this new addition significantly improves the fit to the
quenched fractions of low-mass galaxies, as low-mass satellites keep forming stars longer. Moreover, the small-scale
clustering of low-mass galaxies is improved. As the quality of the fit is much higher than for model 3, all model selection
statistics are better for model 5, and the logarithmic Bayes factor with respect to model 3 is 75.8, so that model 5 is clearly
favoured.

Adding another free parameter by letting the SFR decline exponentially after the quenching timescale, does not improve
the quality of the fit. In fact, the decline timescale $\tau_\mathrm{d}$ is constrained to zero, so that the data require the
quenching to be rapid. Since the fit is not improved and constrained parameter is added, the Evidence drops such that
the logarithmic Bayes factor of model 5 with respect to model 6 is 20. Finally, model 7 is identical to model 5, but the
stripping method has been changed, and the subhalo mass is compared to the present stellar mass instead of the peak
subhalo mass. While this retains the same number of parameters as model 5, the quality of the fit slightly decreases, as
the small-scale clustering is not reproduced as well as for model 5. While this effect is relatively small, the best-fit $\chi^2$
is larger by 56 for model 7, and the logarithmic Bayes Factor of model 5 with respect to model 7 is 24. Therefore, the
preferred model of our analysis is model 5, which has been presented in this paper.

\section{Correlations between parameters}
\label{sec:cov}

\begin{figure*}
\includegraphics[width=0.95\textwidth]{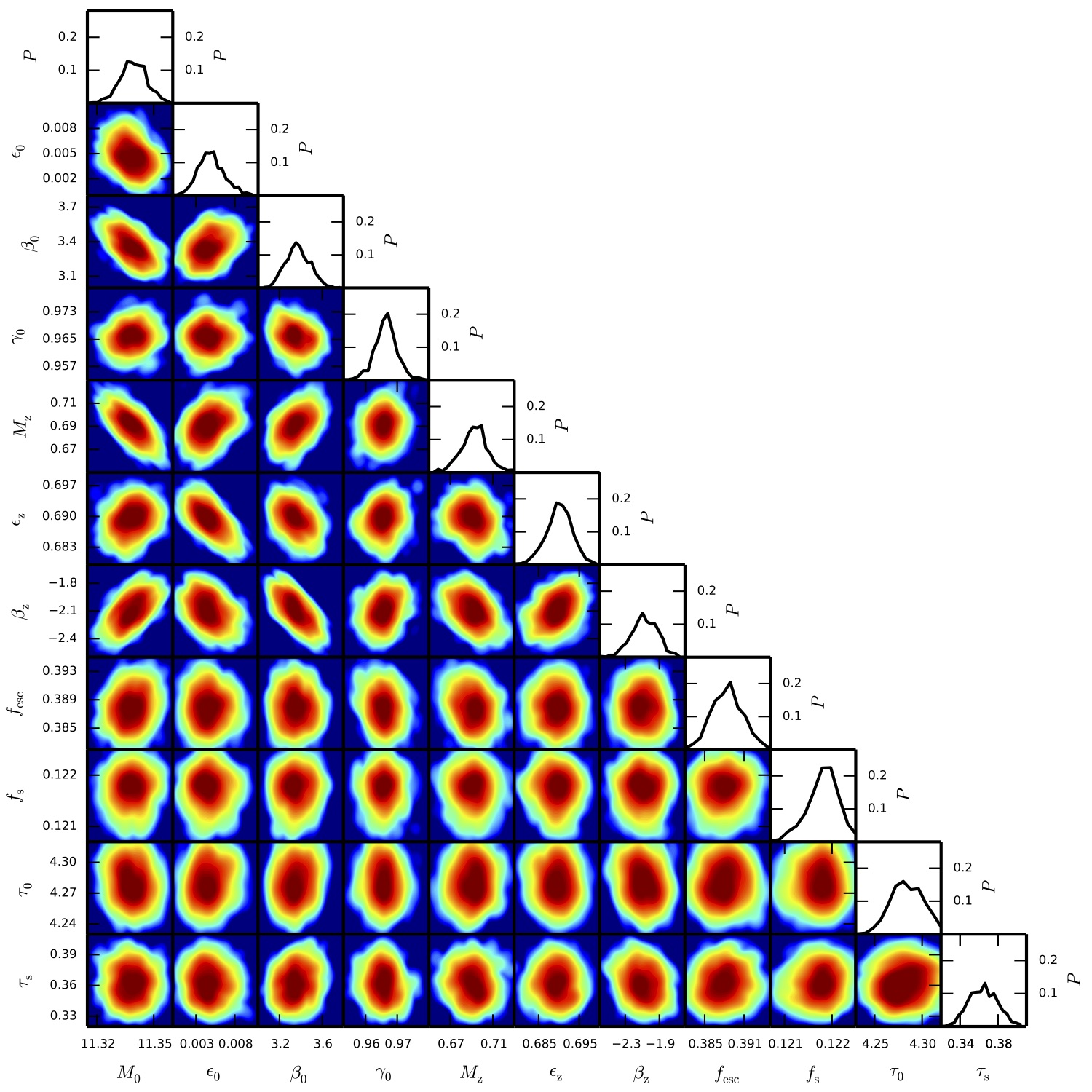}
\caption{Correlation between model parameters, shown by projecting the likelihood surface obtained with
the MCMC algorithm to 2-dimensional surfaces for each parameter pair. Dark red colours represent locations
with a high likelihood, while dark blue colours identify low-likelihood regions.}
\label{fig:cov}
\end{figure*}

All model parameters have been fitted using an MCMC algorithm. This does not only provide the best-fit
values and their uncertainties, but from the full likelihood surface we can also determine the correlations
of the model parameters. The likelihood surface projected to 2-dimensional surfaces for each parameter
pair and the probability density functions for each parameter are shown in Figure \ref{fig:cov}. 

While most parameters are uncorrelated (especially the ones governing the behaviour of satellite galaxies),
we can identify a number of interesting correlations. For the characteristic mass $M_1$, the normalisation
$\epsilon_\mathrm{N}$, and the low-mass slope $\beta$ of the conversion efficiency, the $z=0$ values
are anti-correlated with the respective evolution slopes. This means that they can be either (relatively)
high/low at high/low redshift, or reverse, as long as their sum is roughly the same. For example, for a lower
normalisation at low redshift, the normalisation at high redshift needs to be higher to compensate, so that
the final amount of stellar mass is the same and the SMF is reproduced. Of course, other constraints
specifically constrain the high redshift parameters, such as the sSFR for low-mass galaxies at high redshift
specifically constrains $\beta_\mathrm{z}$. However, as these data have a non-negligible uncertainty, the
high redshift parameters can vary, and as the final amount of stellar mass is rather well constrained, the
low redshift parameters then adapt leading to correlations.

We can also identify weaker correlations between other parameters, such as a slight anti-correlation between
the low-mass slope at low redshift $\beta_0$ and the characteristic mass at low redshift $M_0$, and
consequently a correlation between the low-mass slope at high redshift $\beta_\mathrm{z}$ and $M_0$.
The first of these can be understood easily, as a lower value of $M_0$ leads to higher efficiencies for
low-mass galaxies (the efficiency curve is simply shifted to lower halo masses), so that the low-mass slope
$\beta$ increases and becomes steeper leading to again lowered efficiencies to compensate. The second
correlation is just a consequence of the first correlation and the anti-correlation between $\beta_0$ and
$\beta_\mathrm{z}$. Similarly, $M_\mathrm{z}$ is correlated with $\beta_0$, but anti-correlated with
$\beta_\mathrm{z}$.

\label{lastpage}

\end{document}